\documentclass[a4paper,10pt]{article}
\usepackage{amsmath}
\usepackage{amssymb}
\usepackage{amsthm}
\usepackage{graphicx}
\usepackage[caption=false]{subfig}
\usepackage{enumerate}
\usepackage{float}
\usepackage{booktabs}

\usepackage{xcolor,comment}

\usepackage{authblk}

\usepackage{hyperref}
\theoremstyle{definition}

\title{Heterogeneity of the Attractor of the Lorenz '96 Model: Lyapunov Analysis, Unstable Periodic Orbits, and Shadowing Properties}
\author[1,2]{Chiara Cecilia Maiocchi}
 \author[1,2,3,*]{Valerio Lucarini}
  \author[4]{Andrey Gritsun}
 \author[5,6,7]{Yuzuru Sato}

\affil[1]{Department of Mathematics and Statistics, University of Reading, Reading, UK}
\affil[2]{Centre for the Mathematics of Planet Earth, University of Reading, Reading,  UK}
  \affil[3]{School of Systems Science/Institute of Nonequilibrium Systems, Beijing Normal University,  Beijing, China}
\affil[4]{Marchuk Institute of Numerical Mathematics, Russian Academy of Sciences, Moscow, RF}
 \affil[5]{RIES, Hokkaido University, N12 W7, Kita-ku, Sapporo 060-0812, Japan}
\affil[6]{Department of Mathematics, Hokkaido University, Sapporo, Japan}
\affil[7]{London Mathematical Laboratory, 8 Margravine Gardens, London, W6 8RH, UK}

  \affil[*]{\small Corresponding author. Email address: \texttt{v.lucarini@reading.ac.uk}}

\begin{document}

 \date{\today} 

\maketitle

\begin{abstract}
It is well known that the predictability of weather and climate is strongly state-dependent. Special, easily recognisable, and extremely relevant atmospheric states like blockings are associated with anomalous instability. This reflects the general property that the attractors of chaotic dynamical systems can feature considerable heterogeneity in terms of dynamical properties, and specifically, of their instability. The attractor of a chaotic dynamical systems is densely populated by unstable periodic orbits that can be used to approximate any forward trajectory through the so-called shadowing. Dynamical heterogeneity can lead to the presence of unstable periodic orbits with different number of unstable dimensions. This phenomenon - unstable dimensions variability -  has considerable implications in terms of the structural stability of the system and of the possibility to model accurately its behaviour through numerical models. As a step in the direction of better understanding the properties of  high-dimensional chaotic systems, here we provide an extensive numerical investigation of the variability of the dynamical properties across the attractor of the  much studied Lorenz '96 model. By combining the Lyapunov analysis of the tangent space with the study of the shadowing of the chaotic trajectory performed by a very large set of unstable periodic orbits, we show that the observed  variability in the number of unstable dimensions, which is a serious breakdown of hyperbolicity, is associated with the presence of a substantial number of finite-time Lyapunov exponents that fluctuate about zero also when very long averaging times are considered. The transition between regions of the attractor with different degrees of instability is associated with a significant drop of the quality of the shadowing. By performing a coarse graining based on the shadowing unstable periodic orbits, we are able to characterise the slow fluctuations of the system between regions featuring, on the average, anomalously high and anomalously low instability. In turn, such regions are associated, respectively, with states of anomalously high and low energy, thus providing a clear link between the microscopic and thermodynamical properties of the system.

\end{abstract}

\section{Introduction}

In Chapter 4 of his book Science and M\'ethode \cite{Poincare1908}, Poincar\'e proposed for the first time the concept of sensitive dependence of the evolution of a system on its initial conditions, making also explicit reference to the relevance of this issue in the context of Meteorology \cite{Ruelle1998}. After a long hiatus, the theoretical and practical relevance of the phenomenon of sensitive dependence on initial conditions and its compatibility with the presence of orbits contained in a compact set  became apparent arguably through the seminal contributions by Lorenz \cite{lorenz_1963}, Ruelle and Takens \cite{Ruelle1971}, and Li and Yorke \cite{Yorke1975}. Since then, there has been a great effort in creating sophisticated mathematical frameworks for chaotic systems able to include, at the same time, phenomenology of practical relevance in science and engineering. In what follows, we provide a short summary of some key developments. 

At first, uniform hyperbolicity was conjectured to be the default condition for systems featuring sensitive dependence on initial conditions \cite{hammerlindl_2019, smale_1967, anosov1967geodesic}.  Uniformly hyperbolic systems feature uniform asymptotic contraction and expansion rates for the derivative of the flow, and allow for the possibility of exponential divergence of initially infinitesimally nearby orbit to be compatible with a dynamics occurring in a compact manifold. This idea stems from the fact that hyperbolic dynamics is robust with respect to perturbations \cite{katok1995}. More precisely, Smale and others, conjectured that any chaotic dynamical systems could be transformed into an hyperbolic system by applying an appropriate perturbation. As a result of this conjecture, it would be possible to identify any chaotic behaviour emerging in a mathematical model of a physical process with a uniformly hyperbolic system. Uniformly hyperbolic systems are structurally stable \cite{Ruelle_1989} and amenable to analysis via linear response theory \cite{ruelle_1998} but are, unfortunately, \textit{very atypical} \cite{Bonatti_2011}.

\subsection{Beyond Uniform Hyperbolicity}
Pesin theory had the great advantage of relaxing the definition of hyperbolicity by removing the requirement of being uniform \cite{Pesin1977} and focuses on studying the rates of expansion or decay (and corresponding modes) of perturbations with respect to a background trajectory, which are defined in terms of the properties of the so-called tangent linear operator. For nonuniformly hyperbolic systems, such asymptotic rates of expansion or contraction - the Lyapunov exponents (LEs) \cite{Benettin1980} are bounded away from zero, except for the direction associated with the flow (in the case of continuous time systems)\cite{Young2013}. The sum of the first $p$ LEs (ordered from the largest to the smallest) describe the asymptotic rate of growth (or decay) of the $p-$volume of infinitesimal $p-$parallelepiped, and the sum of all the LEs gives the time-averaged growth rate of the  phase space volume, which is negative - contractive - for dissipative systems \cite{eckmann_1985}. The presence of a positive largest LE can be taken as a proof that a system has sensitive dependence on initial conditions. The notion of LEs, which are global quantities, has been extended via their local version, constructed by considering finite time horizon (finite-time LEs - FTLEs) \cite{Nese1989,Abarbanel1991,Gallez1991}, and by considering finite scale, rather than infinitesimal, perturbations with respect to the background trajectory (finite size LEs - FSLEs) \cite{Aurell1997}. In the last few decades, the so-called Lyapunov analysis has grown into a  portfolio of very powerful methods for studying rather general complex systems \cite{Cencini2013,pikovsky2016lyapunov}. 

A separate way to relax the notion of uniform hyperbolicity entails introducing a nontrivial centre manifold in the tangent space where expansion or contraction are extremely slow, thus removing hyperbolicity. The so-defined partial hyperbolic systems can be further generalized by allowing for nonuniformity outside the centre manifold: the continuous-time nonuniform partially hyperbolic systems feature more than one zero Lyapunov exponents \cite{barreira2007}.

A very influential attempt at creating a powerful paradigm of chaos beyond uniform hyperbolicity has been presented by Bonatti et al. \cite{bonatti_2004}, 
According to such a paradigm, chaos can originate from different mechanisms, such as heterodimensional cycles or homoclinic tangencies. As compared to uniformly hyperbolic systems, these more general systems are less understood and
advances have been mainly made on discrete dynamical systems, because of the simpler structure of the phase space \cite{hammerlindl_2019}, hence numerical approaches play an important role in shedding light on the global organisation of phase space \cite{zhang2012find}.



\subsection{Unstable Periodic Orbits}
In order to explain the variability of the local properties of the attractor of a chaotic system, the investigation of the tangent space via Lyapunov analysis can be complemented by a different strategy based on the study of special periodic solutions, the so-called Unstable Periodic Orbits (UPOs) \cite{cvitanovic_1988,cvitanovic_1991,cvitanovic_2005}. 
In fact, UPOs, true nonlinear modes of the flow, provide a rigid structure hidden in the chaos of the dynamics. For uniformly hyperbolic systems without continuous symmetries, UPOs are dense in the attractor \cite{Gaspard2005}. This implies that it is always possible to find a periodic orbit arbitrarily near to a chaotic trajectory, allowing for a reconstruction of the trajectory up to any arbitrary accuracy. We can then think of the chaotic trajectory as being continuously scattered from one neighbourhood of an UPO to another, because of their instability.
In general, even though the shadowing property has not yet been formalised for more general system, it is widely assumed. Evidence of turbulent trajectories being shadowed by UPOs are present in forced two-dimensional flows \cite{kazantsev1998unstable, chandler_2013, lucas2015recurrent}, isotropic turbulence \cite{van2006periodic}, plane Couette flow (\cite{krygier2021exact, cvitanovic2010geometry, kreilos2012periodic}), Kolmogorov flow in two \cite{suri2020capturing} and three dimensions \cite{yalniz_2020}. 
UPOs can be used to approximate any chaotic trajectory with an arbitrary accuracy.  While for Axiom A  a rigorous theory that allows to reconstruct statistical properties of the system as sum with well-defined weights over the UPOs has been developed \cite{cvitanovic_2005}, extensions of this approach have been proposed for more general  chaotic systems \cite{Dhamala1999}. 
UPOs have been successfully applied for decomposing and extracting information from the dynamical structure of chaotic flows in many different contexts \cite{cvitanovic_2005}. 
Kawhara and Kida showed in their seminal work that one UPO only in the attractor of a numerical simulation of plane Couette flow manages to capture in a surprisingly accurate way the turbulence statistics \cite{kawahara_2001}.
Chandler and Kerswell \cite{chandler_2013} identified $50$ UPOs of a turbulent fluid at a moderate Reynolds number and used them to reproduce the energy and dissipation probability density functions of the system as dynamical averages over the orbit. More recently Yalniz and Budanur \cite{yalniz_2020} proposed a coarse grained state space decomposition of the dynamics in terms of a few periodic orbits for both the three dimensional Rossler flow and a discretisation of the Kuramoto-Sivashinksky equation. Page at al. \cite{page2022recurrent} provided evidence that the statistics of a fully developed turbulent flow can be reconstructed in terms of a set of UPOs. 

Very interestingly, the investigation of the UPOs of a chaotic system allows one also to identify violations of hyperbolicity in a relatively simple manner. Indeed, if one detects e.g. two UPOs immersed in the attractor of the system that feature a different number of positive LEs, hyperbolicity is broken through the mechanism of so-called unstable dimensions variability (UDV), which establishes the presence of a fundamental heterogeneity in the attractor of the system and hinders the 
existence of an actual trajectory of the systems that stays uniformly close to a numerical one for long time intervals \cite{Lai1997}. UDV is typically associated with the presence of large fluctuations of certain FTLEs between positive and negative values \cite{Sauer1997,Sauer2002,Pereira2007}.

\subsection{Unstable Dimension Variability in Geophysical Fluids}
The study of the tangent space is a key aspect of the science and technology related to geophysical fluids \cite{Kalnay2003,Ghil2020}. In this context, it is well known that the predictability of a system, far from being in any sense uniform, is dramatically state-dependent: certain regions of the attractor feature larger instability than others \cite{Palmer2000,Slingo2011}, and this has great impact on data assimilation strategies \cite{Carrassi2018,Wu2020}. In turn, the skill of  data assimilation exercises can be used to infer the instability of the underlying system \cite{Chen2021}. The  state-dependent predictability results into substantial fluctuations {\color{black}
in the  value of the individual FTLEs \cite{Nese1993,Yoden1995,Nicolis1995,Vannitsem1997}. These fluctuations can be accurately quantified, when considering sufficiently long time horizons, using large deviation laws \cite{de2018exploring}, in agreement with the general theory presented in \cite{Pazo2013,Laffargue2013}. One also finds that the number of positive FTLEs fluctuates across the attractor, which is } clearly indicative of a violation of the condition hyperbolicity and is associated with the UDV mentioned above \cite{lucarini_2020}. The UDV can be particularly problematic for the efficiency of otherwise very powerful data assimilation schemes \cite{Chen2021}.  

The use of UPOs for studying geophysical flows was introduced by Gritsun \cite{gritsun_2008,gritsun_2013}, who proposed using an expansion over UPOs to reconstruct the statistics of a simple atmospheric model based on the barotropic vorticity equation of the sphere. 
Later, Lucarini and Gritsun \cite{lucarini_2020} used UPOs for clarifying the nature of blocking events in a baroclinic model of the atmosphere. Blockings are rare and persistent large scale deviations in the mid-latitudes from the approximately zonal flow. They are most commonly found in either the Atlantic or in the Pacific sector of the Northern Hemisphere.  Specifically, they found that the atmospheric model was characterised by very large variability in the number of unstable dimensions (UDs), thus suggesting  that the dynamics of the atmosphere is far from being hyperbolic. Additionally, it was found that blocked states are associated with conditions of higher instability of the atmosphere, in basic agreement with a separate line of evidence obtained through Lyapunov analysis \cite{Schubert2016,Kwasniok2022} and through the study of recurrent patterns of the atmosphere based on EVT \cite{Faranda2017}. 

\subsection{This Work: Concept and Main Results}
In a previous work \cite{maiocchi_2022} we have studied the  classical version of the 3-dimensional (3D) Lorenz 1963 (L63) model \cite{lorenz_1963} using a rather extensive set of UPOs, following \cite{barrio_2015}, and covering up to period 14 in symbolic dynamics. We have been able to investigate accurately the process of ranked shadowing and elucidate how the statistics of occupation of individual UPOs and of their respective neighbourhood and of the transitions between them can be used to construct a finite-state Markov chain able to represent accurately the statistical and dynamical properties of the system, including its almost-invariant sets \cite{froyland_2014}. The attractor of the L63 model is extremely heterogeneous in terms of predictability, and features specific regions where return of skill is observed \cite{Smith1999}. The detected UPOs do differ in terms of their dynamical characteristics, and specifically in the value of the first LE, thus providing a global counterpart of the heterogeneity of the properties of the tangent space. Nonetheless, in a 3D chaotic flow by construction all UPOs feature one positive, one negative, and one vanishing LE. Hence, if we want to investigate the heterogeneity of the attractor of a chaotic flow and possibly relate it to the presence of variability of the UDs number, one needs to consider a higher dimensional system.

With this work we would like to characterise and explain the heterogeneity of the attractor of the very popular Lorenz '96 (L96) model \cite{lorenz_1996,Lorenz2005} in a chaotic regime.
We show that the system features clear signature of UDV and that this is accompanied by a substantial number of FTLEs whose value fluctuates about zero also when very long averaging times are considered. By combining the information derived from the analysis of an extensive set of UPOs with Lyapunov analysis, we find that anomalously unstable UPOs preferentially populate regions of the attractor where, applying Lyapunov analysis to the tangent space, one gets, coherently, anomalously high instability indicators. This bridges the gap between global and local properties of the system. 

{\color{black}Detecting UPOs in high dimensional, highly chaotic systems is well-known to be extremely challenging \cite{miller2000finding,cvitanovic_2005,gritsun_2008}. We show that, while we are able to identify over $3\times10^5$ UPOs up to period $T\approx22$, the longer-period UPOs are a) vastly underrepresented and b) significantly skewed towards low instability. Such longer-period UPOs occupy preferentially a specific region of the phase space. This provides further supports to the heterogeneity of the attractor in terms of instability and clarifies why restricting our analysis to more thoroughly detected, shorter-period UPOs leads to significant loss of information on the system.} 

We then propose two finite state Markov chain representations of the dynamics based on the shadowing process of the orbit performed by the UPOs. In  first instance, UPOs are grouped in states according to their number of UDs and transitions are recorded at the point when the closest shadowing UPO changes and the new shadowing UPO has a different UDs number. In second instance, we consider a much larger space, where each state corresponds to the trajectory being shadowed by a UPO chosen among a suitable defined subset of the whole database. In both cases, by studying the subdominant eigenvectors and eigenvalues of the stochastic matrix, we are able to characterise the relatively slow fluctuations of the systems from regions with prevalence of anomalously unstable and anomalously stable UPOs, and from regions of typical vs atypical UPOs. We also provide a thermodynamical, energetic interpretation of the results, thus linking microscopic and macroscopic properties of the system.  

The paper is structured as follows. Section \ref{L96model} provides a description of the L96 model and of some of its basic properties in the configuration chosen for this study. In Section \ref{SUPOs} we present the database of detected UPOs and discuss their accuracy in reproducing the dynamics of the system. We also present evidence of the heterogeneity of the attractors in terms of stability properties.  In Section \ref{Local} we supplement the UPOs-based analysis with the Lyapunov analysis. We link local and global properties on the attractor, investigate the transitions between regions of the attractor characterised by different stability, and we emphasize how the breakdown of hyperbolicity and the associated UDV emerges according to these two viewpoints. The statistical angle on the problem is then discussed in Sect. \ref{statistics}, where we study the relaxation of a generic initial ensemble to the invariant measure by extracting two separate finite-state Markov chain from the dynamics. The relaxation can be seen as a mixing associated with transitions between regions of the attractor featuring anomalously high and anomalously low stability, respectively. Finally, in Section \ref{conclusions} we discuss the main results of our study and present perspectives of future investigations. In Appendix \ref{appendix} we report the main mathematical concepts used throughout the paper for the benefit of the reader.

\section{The Lorenz '96 Model: A Toy Model for Spatio-Temporal Chaos}\label{L96model}
The L96 model, while not corresponding to a truncated version of any known fluid dynamical system, was developed as a prototype for the midlatitude atmosphere, with the scope of investigating problems of predictability in weather forecasting \cite{lorenz_1996,Lorenz2005}.
Each variable of the model corresponds to an atmospheric quantity of interest at a discrete location on a periodic lattice, representing a latitude circle on the sphere. The variables are spatially coupled, and their equation of motion include nonlinear (quadratic) terms to simulate advection, linear terms representing dissipation and constant terms representing external forcing. 
While the model only shares only such basic characteristics with more complete geophysical fluid dynamical models, it has emerged as an important testbed for different applications, including the study of 
bifurcations \cite{vanKekem2018PhysD,vanKekem2018NPG}, of  parametrizations \cite{Wilks2005,Arnold2013,Vissio2018}, of data-driven and machine learning techniques, \cite{Chattopadhyay2020,Gagne2020,Gelbrecht2021}, of extreme events \cite{Blender2013,Sterk2017,Hu2019}, of data assimilation schemes \cite{Trevisan2004,Brajard2020}, of  ensemble forecasting techniques \cite{Wilks2006,Duan2016}, to develop new tools for investigating predictability \cite{Hallerberg2010,Carlu2019}, and for addressing basic issues in mechanics and statistical mechanics \cite{AbramovM2008,Lucarini2011,Lucarini2012,Gallavotti2014,Vissio2020}. The evolution equations of the model are:
\begin{equation}\label{L96_eq}
     \dot{X_j} = (X_{j+1}-X_{j-2})X_{j-1}-\alpha X_j+F,\quad j=1,\ldots,J
\end{equation}
where
\begin{equation}
    X_{-1} = X_{J-1}, \quad X_0=X_J, \quad X_{J+1}=X_1.
\end{equation}
impose the periodicity conditions, $F \in \mathbb{R}^+$ is a constant forcing, and {\color{black}$\alpha\in \mathbb{R}^+$} modulates the dissipation. The three free parameters of the model are $J$, $F$, and $\alpha$. For large values of $F/\alpha$ and $J$ the model exhibits extensive chaos \cite{Gallavotti2014}. In the inviscid case - when the dissipation and the forcing are removed by setting $\alpha=F=0$ - the so-called energy $E=1/2\sum_{j=1}^{{\color{black}J}} X_j^2$ is conserved and one can recognise a quasi-symplectic structure in the dynamics \cite{Vissio2020}. The chaotic behaviour of the systems is associated with the presence of unstable waves that grow, accumulate energy, and then decay, with ensuing dissipation  \cite{lorenz_1996,Lorenz2005}.

\begin{figure}
  \subfloat[]{
\includegraphics[width=8cm]{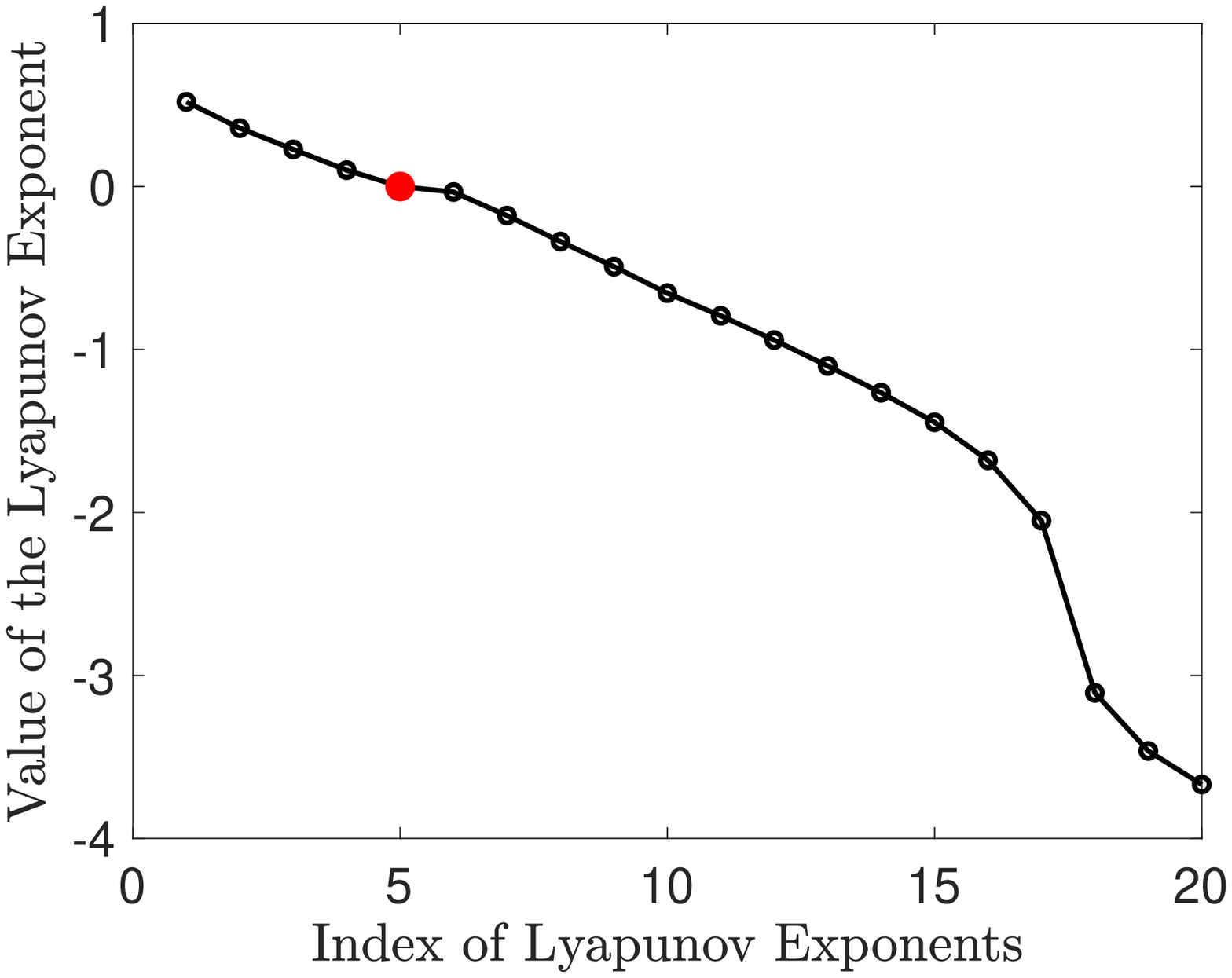}
\label{global_LE}}
  \subfloat[]{
\includegraphics[width=8cm]{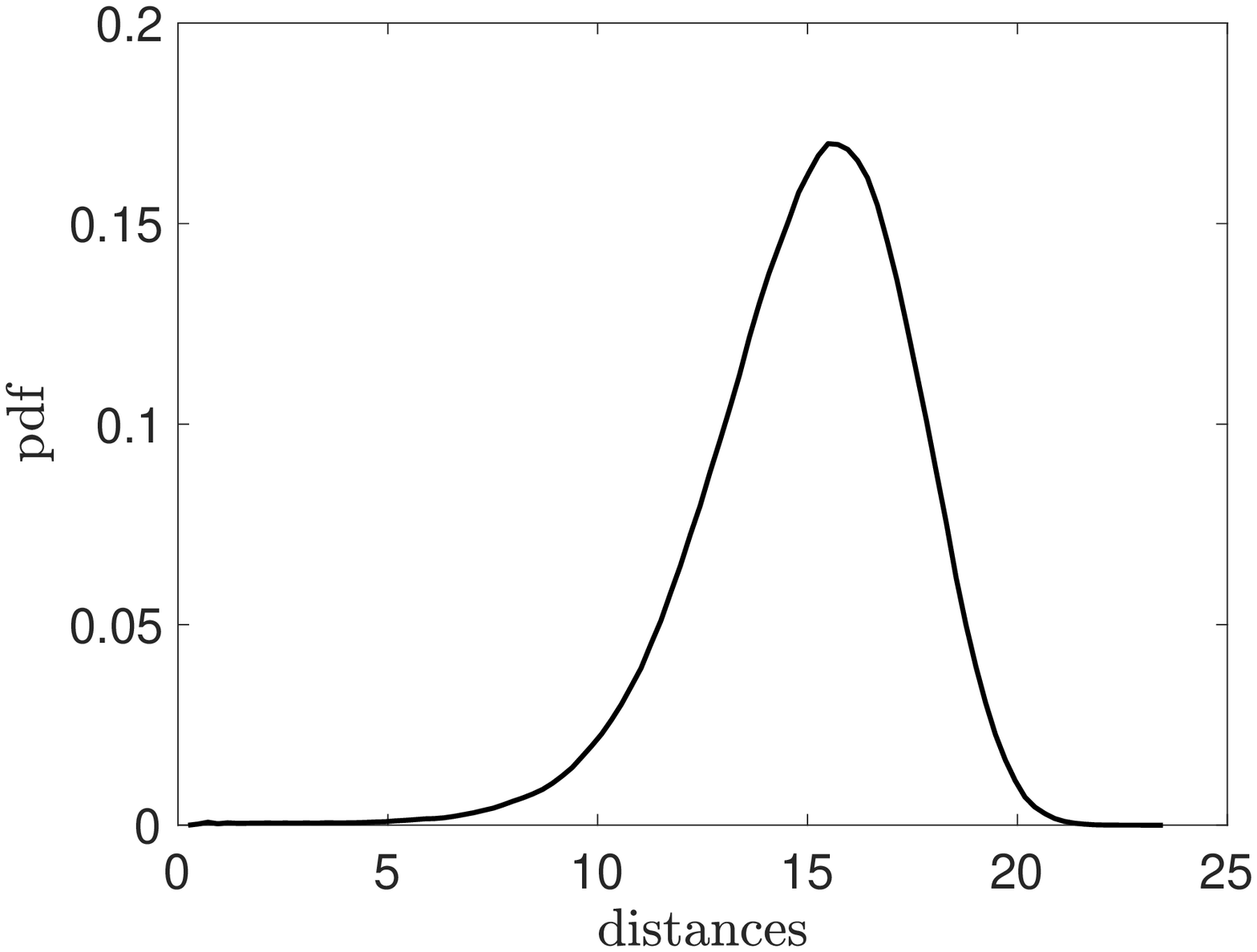}\label{typical_distances}}
\caption{Panel \protect\subref{global_LE} Lyapunov Exponents of the system; $\lambda_5=0$, which corresponds to the direction of the flow, is highlighted in red. Panel  \protect\subref{typical_distances}: Distribution of the  distances between points of the attractor}.
\end{figure}

In this work we have considered $J=20$, $F=5$, and $\alpha=1$, which leads to the onset of a chaotic steady state for the system; {\color{black}see some relevant definitions in App. \ref{chaos}.} In all the simulations the model is integrated with a Runge-Kutta second order midpoint time scheme with fixed time step $dt=0.01$. The choice of this (suboptimal but sufficiently accurate) integrator is motivated by the UPOs detecting algorithm used. For such choice of the parameters the model is well within the chaotic regime. It features $n=4$ positive LEs  with the leading one being $\lambda_1$ $\approx$ $0.54$ (Fig. \ref{global_LE}). The fifth LE vanishes and corresponds to the direction of the flow. The characteristic Lyapunov time of the system is then $\tau_1=1/\lambda_1 \approx 1.85$, the Kolmogorov-Sinai entropy can be approximated as $h_{KS}=\sum_{\lambda_i>0}\lambda_i\approx 1.23$ and the Kaplan-Yorke dimension of the attractor is $d_{KY}\approx 9.25$. {\color{black}See App \ref{finiteLyap} for clarifications on the mathematical terminology.}

The diameter of the attractor is approximately $25$, with point-to-point distances distributed as shown in Fig \ref{typical_distances}. The relatively high dimensionality of the attractor is apparent from the very low prevalence of nearby points \cite{Grassberger1983}. The mean speed over the attractor is $\approx 38$, meaning that on average the trajectory spans a distance of 0.38 for a time-step of $dt=0.01$.

\section{Study of the Attractor via Unstable Periodic Orbits}\label{SUPOs}

\subsection{Database of the Unstable Periodic Orbits}

The numerical extraction of UPOs from a chaotic system {\color{black}- see definitions and some essential information in App. \ref{UPOsdef} -}  is one of the greatest challenges in the applications of periodic orbit theory \cite{saiki_2007, barrio2015database, gritsun_2008}. The problem of finding UPOs can be reduced to the solution of the periodicity condition $S^Tx_0=x_0$, where $S^T$ is the evolution operator associated with the flow given in Eq. \ref{L96_eq} that acts for a time $T$  and $x_0$ is an initial condition on the attractor. Hence, one must solve a system of nonlinear equations with respect to $x_0$ and the period $T$ of the UPO. Even for simple nonlinear systems this represents a complex numerical problem, with a computational cost that grows exponentially with the dimension of the system. The choice of the algorithm and initial condition is thus  crucial (see \cite{gritsun_2008} for  details).  
The equations are symmetric with respect to a cyclic permutation of the variables, so that each time an orbit is detected, the other $19$ can be automatically obtained by simply considering all the possible cyclic permutations of variables. In this work we constructed a database of 15019 fundamental UPOs (i.e. none of these orbits can be obtained from another orbit of the database through cyclic symmetry) immersed in the attractor with period ranging from a minimum of $\approx1.5$ ($\approx0.8/\lambda_1$) to a maximum of $\approx22.8$ ($\approx12.3/\lambda_1$). 
This corresponds to lengths ranging from $\approx2$ to $\approx35$ diameters of the attractor.


\begin{figure}
\subfloat[]{%
      \includegraphics[width=8cm]{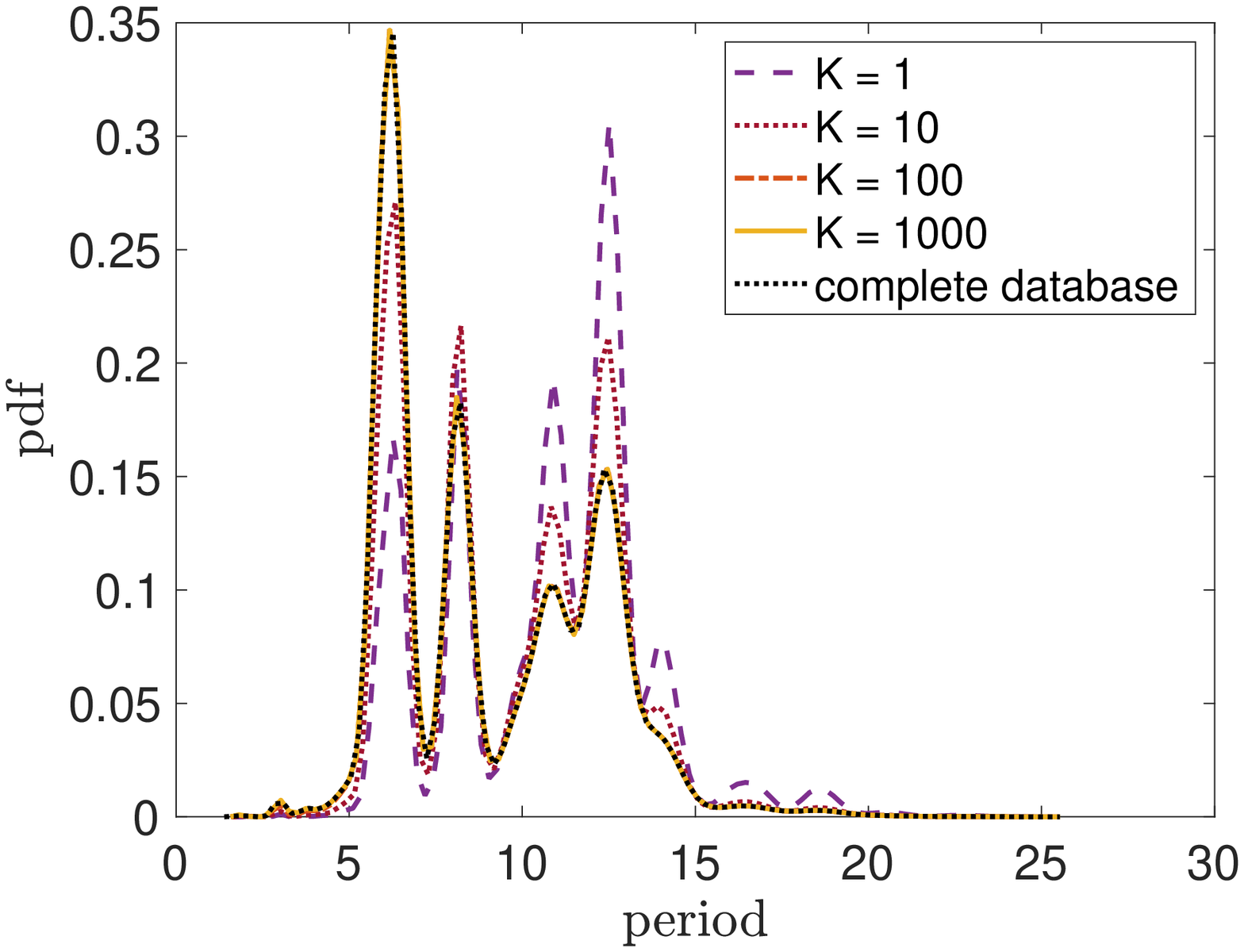}
           \label{distribution_periods_shad_UPOs}
    }
    \subfloat[]{%
      \includegraphics[width=8cm]{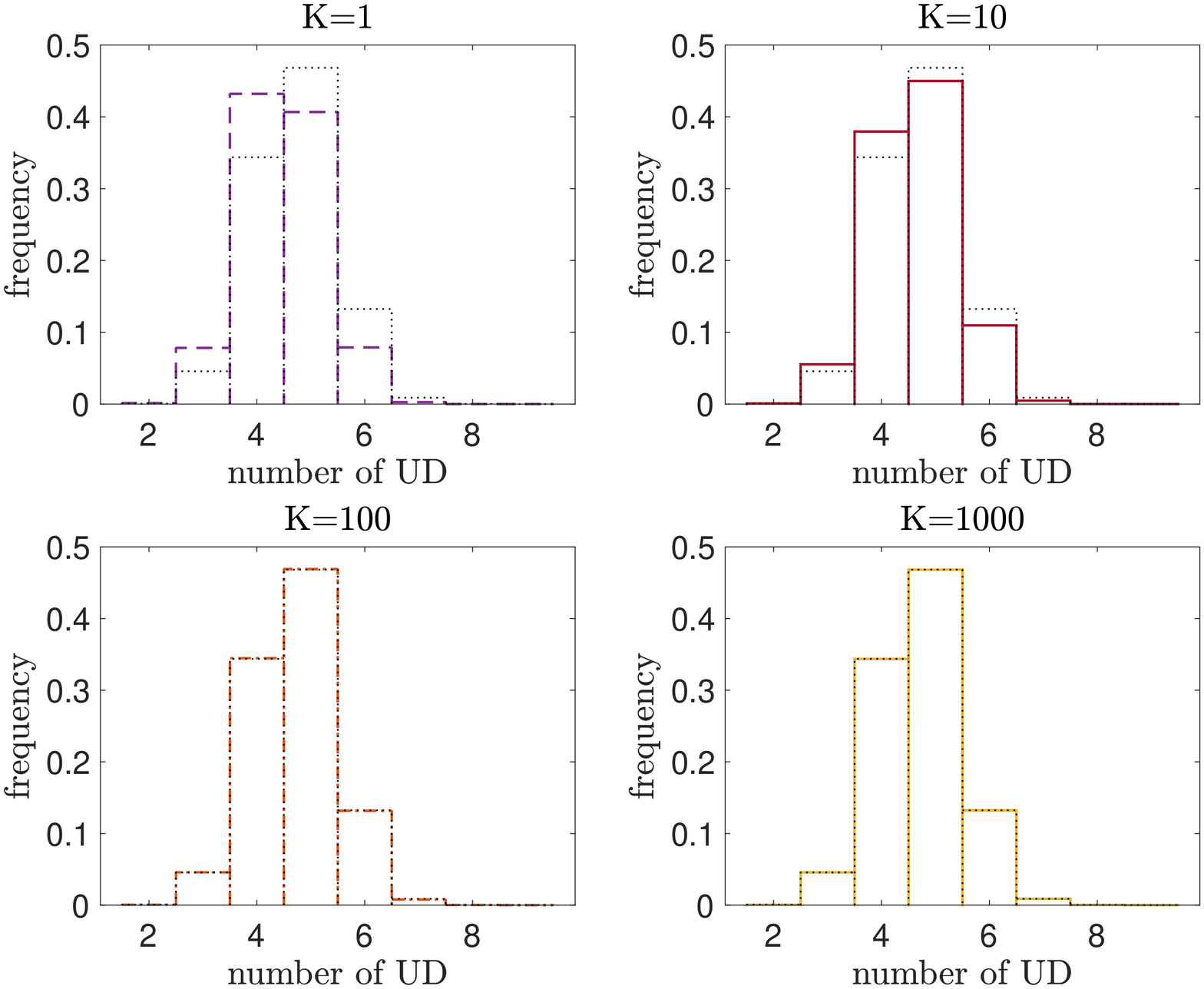}
           \label{positive_LE}
    }\\
     \subfloat[]{%
   \includegraphics[width=8cm]{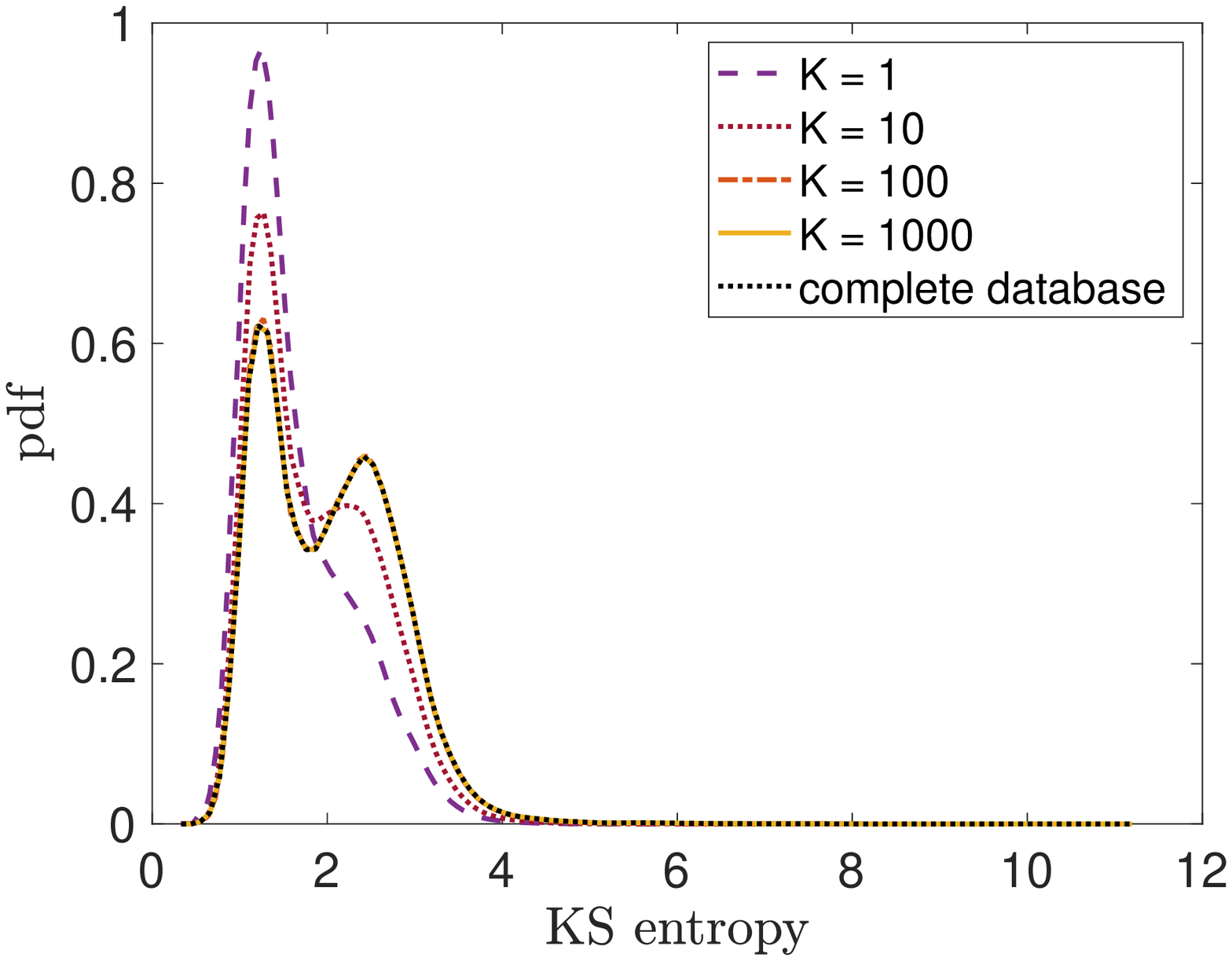}
              \label{KSentropy}
    }
     \subfloat[]{%
   \includegraphics[width=8cm]{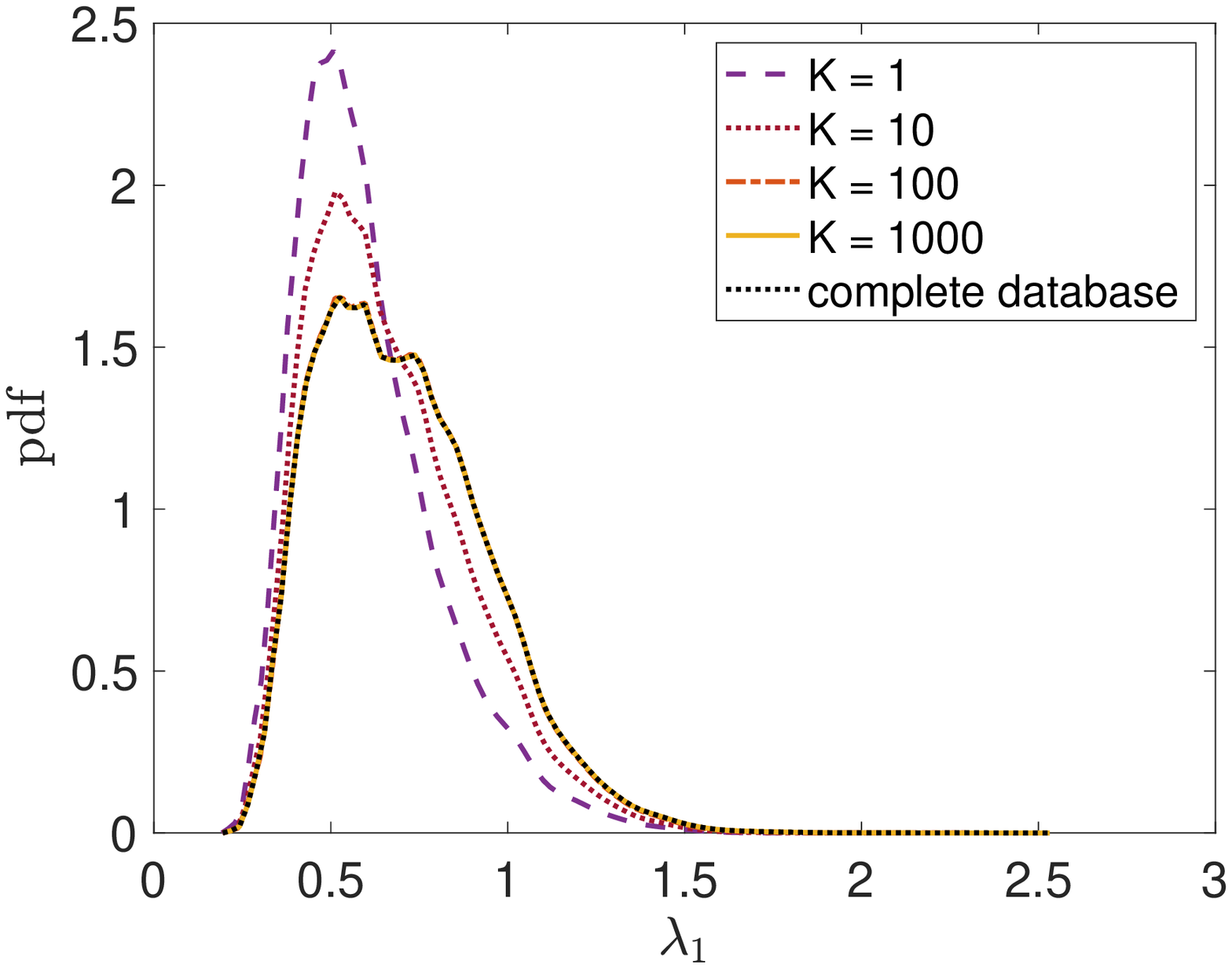}
              \label{first_LE_density}} 
       \caption{Heterogeneity of the instability properties of the shadowing UPOs. Each distribution represent the properties of the UPOs that shadows the trajectory at least once in the first $K = 1, 10, 100, 1000$ tiers (see Sect. \ref{section: shadowing}) and the complete database. 
       Panel \protect\subref{distribution_periods_shad_UPOs}:Distribution of the periods of the shadowing UPOs that shadows {\color{black}- see Sect. \ref{section: shadowing} - }the trajectory at least once in the first $K = 1, 10, 100, 1000$ tiers.
       Panel \protect\subref{positive_LE}: Frequency of the number of UD of the shadowing UPOs for different values of $K$. Panel \protect\subref{KSentropy}: Distribution of the KS entropy of the UPOs. Panel \protect\subref{first_LE_density}: Distribution of $\lambda_1$ across the UPOs of the database. {\color{black}The color code is the same for all panels.}}
  \label{database} 
\end{figure}

 \begin{figure}
       \subfloat[]{%
   \includegraphics[width=8cm]{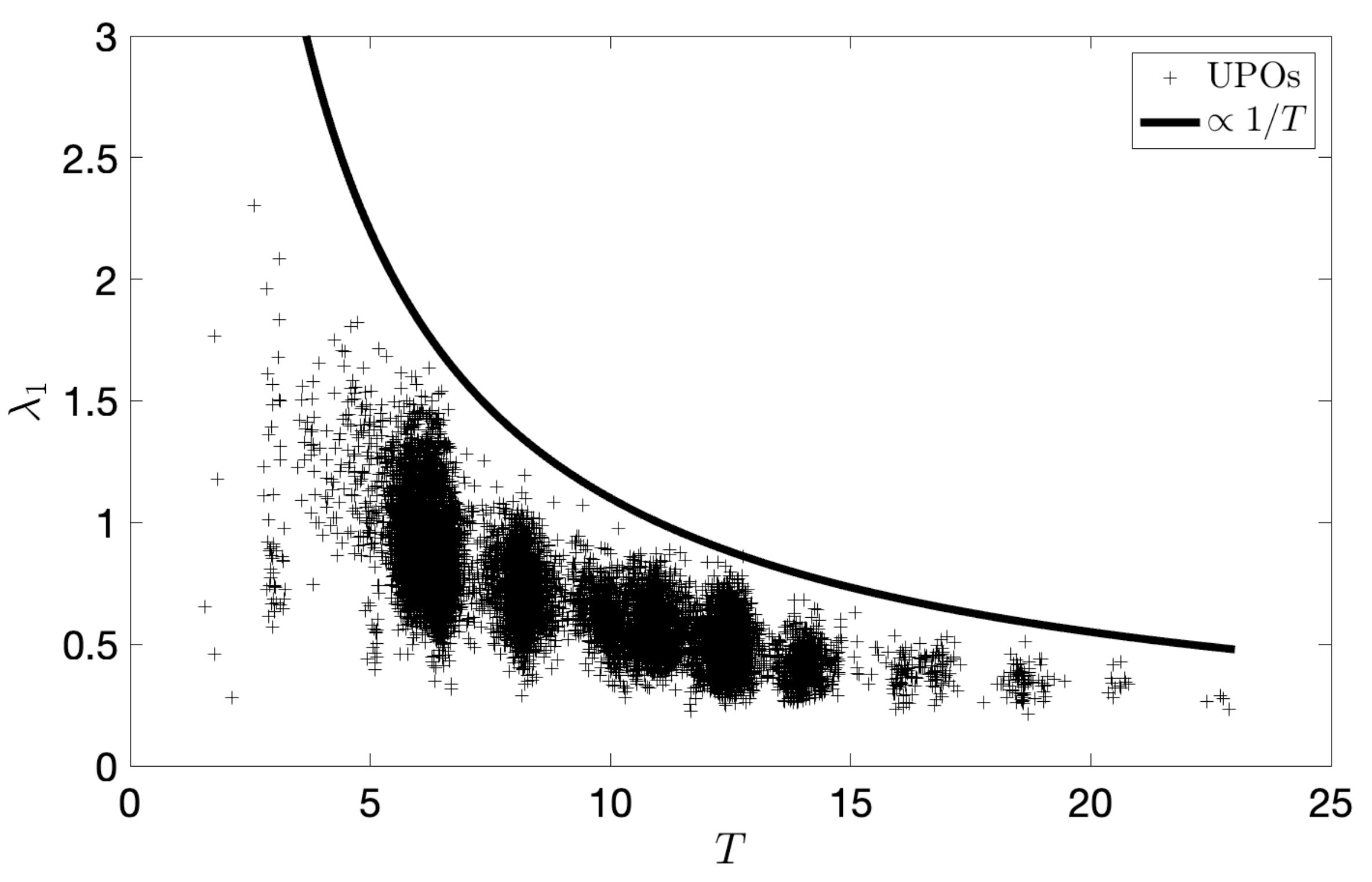}
              \label{scatter_lambda1}      } 
                     \subfloat[]{%
   \includegraphics[width=8cm]{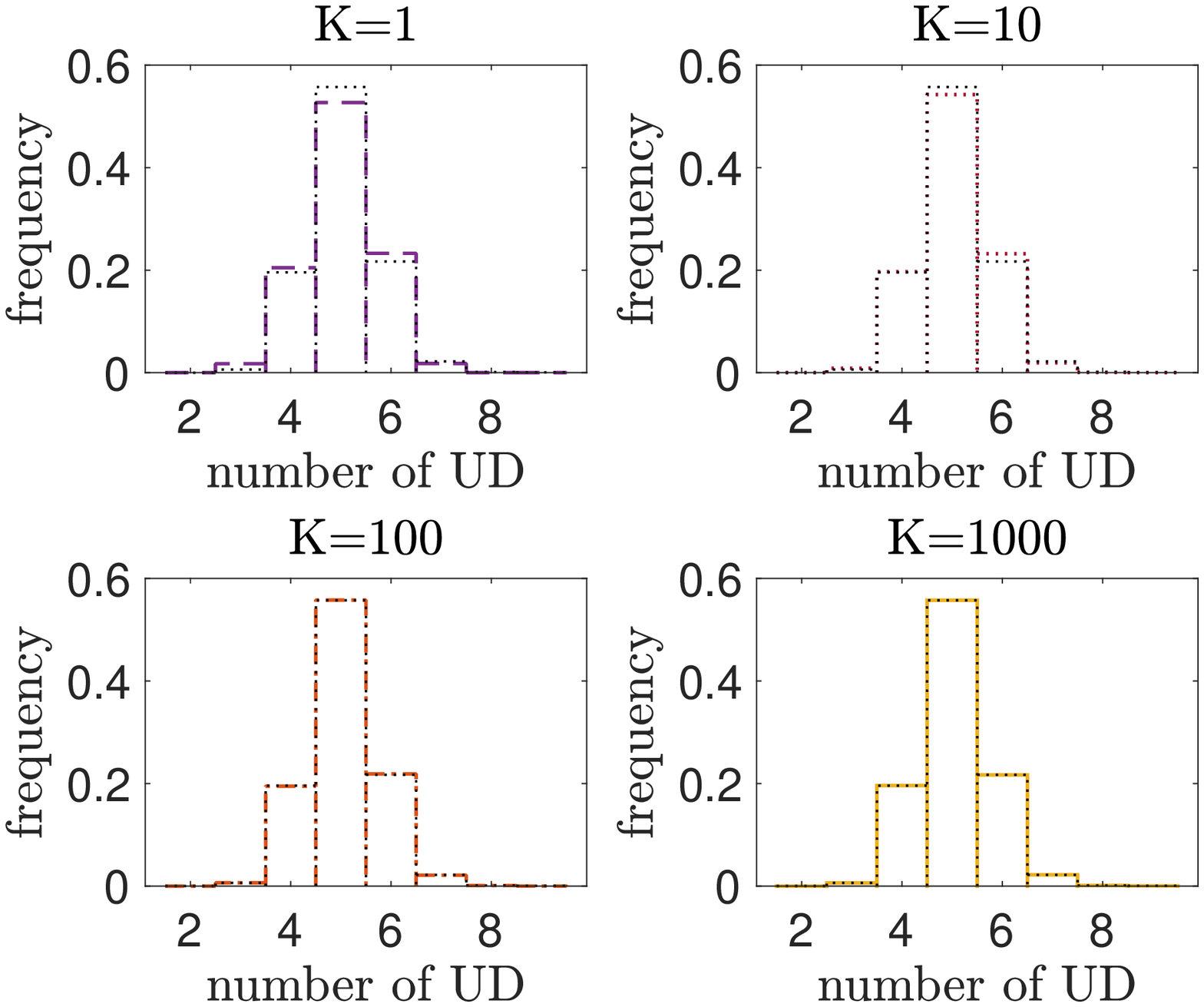}
              \label{positive_LEsmallT}      } \\
                          \subfloat[]{%
   \includegraphics[width=8cm]{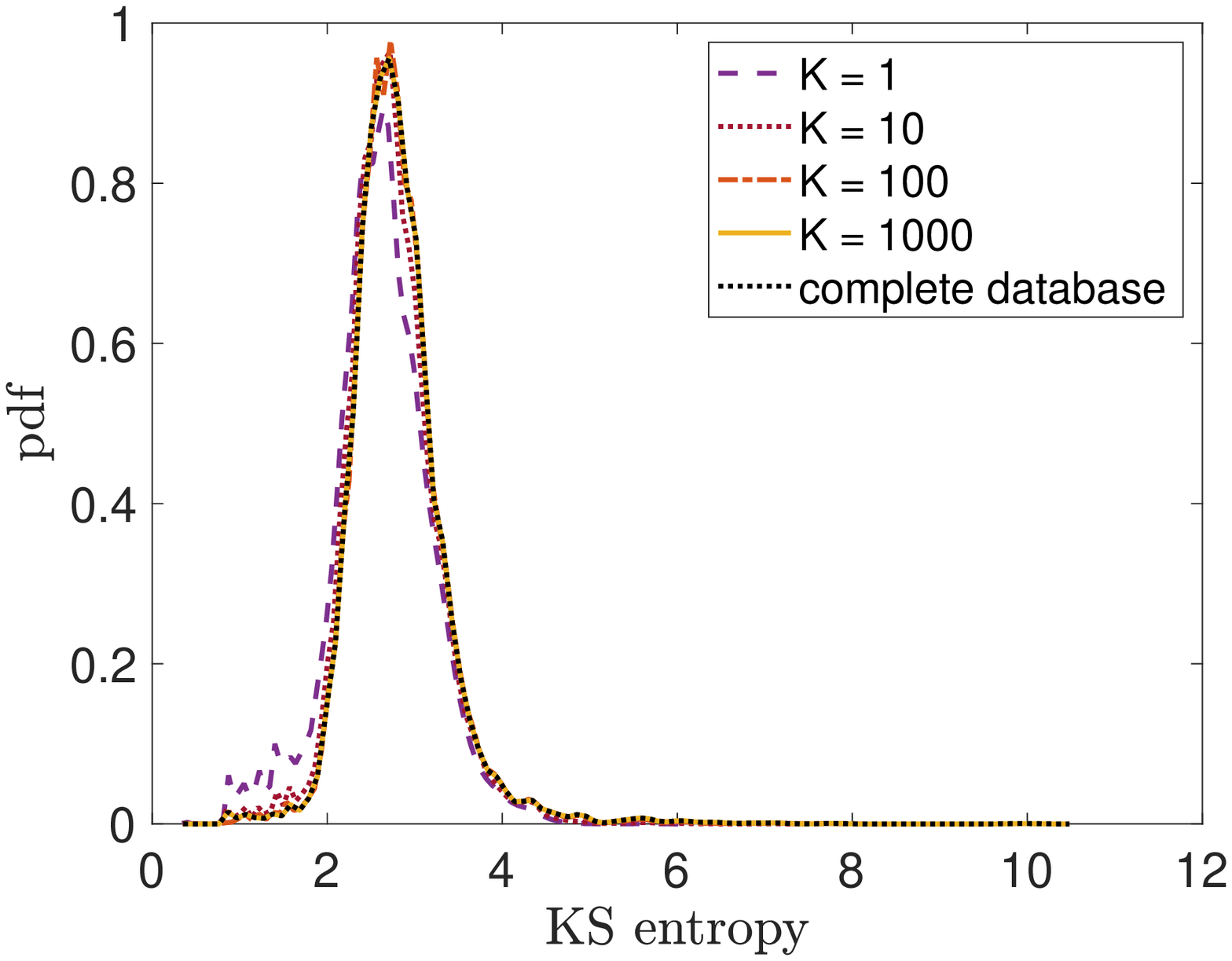}
              \label{KSentropysmallT}      }                 
                     \subfloat[]{%
   \includegraphics[width=8cm]{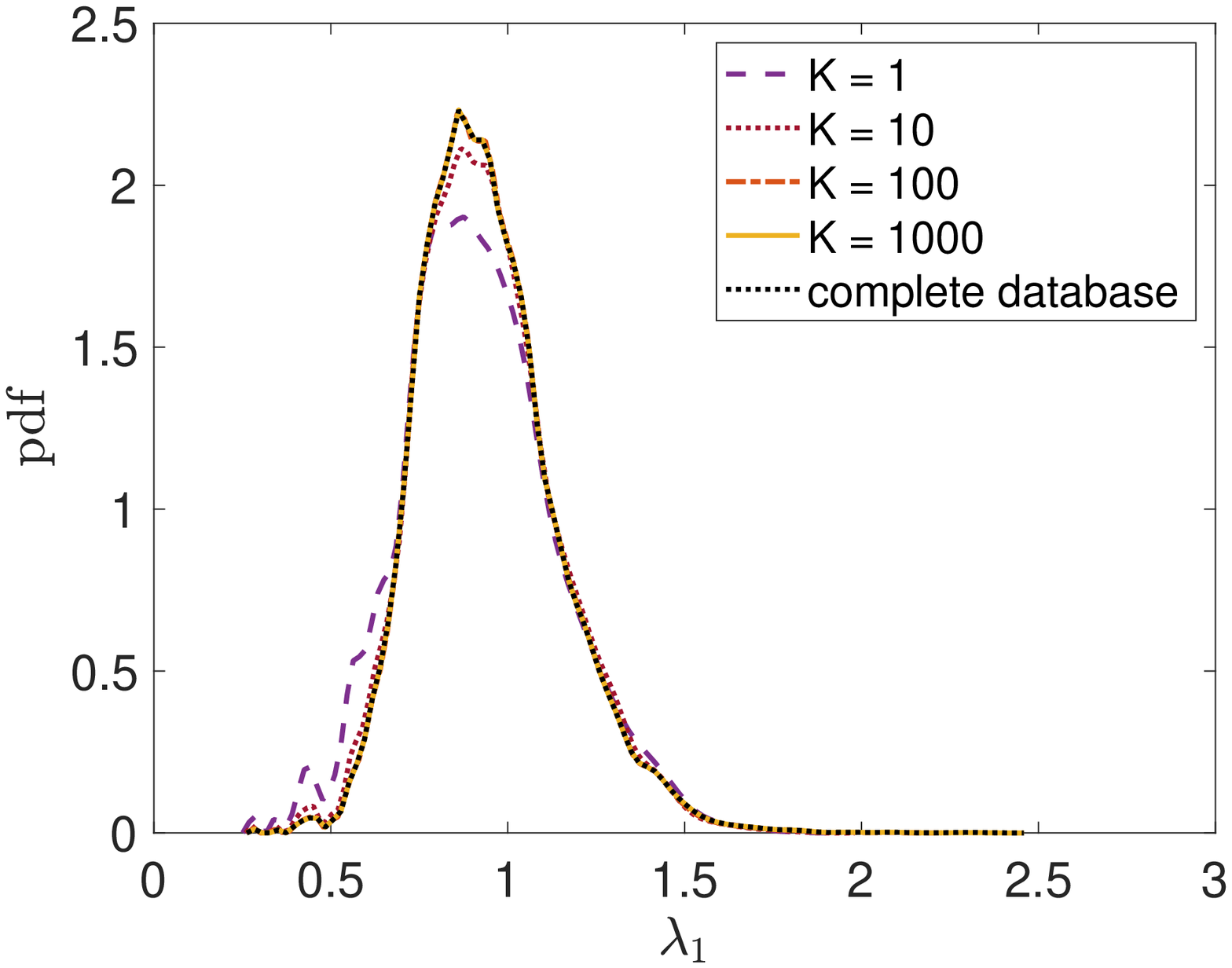}
              \label{lambda1smallT}      }  
       \caption{{\color{black}Subsampling the set of detected UPOs. 
      Panel \protect\subref{scatter_lambda1}: Scatter plot  period vs. $\lambda_1$ for the detected UPOs. 
             Panels \protect\subref{positive_LEsmallT}, \protect\subref{KSentropysmallT}, and \protect\subref{lambda1smallT}: Same as Figs. \ref{positive_LE}, \ref{KSentropy}, and \ref{first_LE_density}, respectively, but restricted to UPOs with $T\leq6.4$. }}
  \label{databasesmall} 
\end{figure}

In a chaotic system one expects to find that the number of UPOs with prime period smaller or equal than $T$ grows as $\propto \exp(h_{top}T)$, where $h_{top}$ is the topological entropy \cite{cvitanovic_2005}. As opposed to our previous study \cite{maiocchi_2022}, it is clear that our set of UPOs is far from being complete, as longer-period orbits are clearly underrepresented, {\color{black}see the curve referring to the complete database in} Fig. \ref{distribution_periods_shad_UPOs}. The difficulty in computing long-period UPOs has been widely discussed in the literature \cite{miller2000finding,cvitanovic_2005,gritsun_2008}. While clearly incomplete, as further discussed below, the dataset of detected UPOs can provide extremely valuable information on the properties of the L96 model.  The detected UPOs are characterised by vastly different instabilities properties, which provide a clear evidence of the heterogeneity of the attractor of the L96 model. The UDs number  varies from 2 to 9 across the UPOs (Fig. \ref{positive_LE}): this indicates a very serious violation of hyperbolicity via UDV \cite{Lai1997,Sauer1997,Sauer2002,Pereira2007,lucarini_2020}. Additionally, the Kolmogorov-Sinai entropy varies between $\approx0.5$ and $\approx10.0$ (Fig. \ref{KSentropy}) and the first LE varies between $\approx0.3$ and $\approx1.8$ (Fig. \ref{first_LE_density}).  

{
{\color{black}Figures \ref{distribution_periods_shad_UPOs}-\ref{scatter_lambda1} show that the periods of the detected UPOs are approximately integer multiples of a fundamental period $T_f\approx2.1$, which is associated with the least unstable UPO. 
Figure \ref{scatter_lambda1} additionally shows that the detected longer-period UPOs we find are significantly skewed towards low instability; it is extremely difficult to detect long-period, highly unstable UPOs. Indeed, we find that $\lambda_1^{max,T}\times T\approx const.$, where $\lambda_1^{max,T}$ is the largest first Lyapunov exponent detected among the UPOs with period $T$. This can be explained by considering that the UPOs detection algorithm is aimed at controlling errors that grow $\propto\exp\left(\lambda_1 T\right)$ for a UPO with period $T$\cite{gritsun_2008}. The scatter plot of the UPOs period vs. $h_{KS}$ is qualitatively similar; see Fig. 1a) in the Supplementary Material. Indeed - see Figs. \ref{KSentropysmallT} and \ref{lambda1smallT} - if we restrict the statistics of $\lambda_1$ to the orbits with $T\leq6.4$, where the cutoff corresponds to the first peak of the distribution in Fig. \ref{distribution_periods_shad_UPOs}, which marks the departure from the exponential growth of the number of UPOs with respect to their period, we obtain distributions of $h_{KS}$ and $\lambda_1$ that are shifted towards the right with respect to what has been obtained using the whole dataset of UPOs. Even when considering such set of low-period UPOs one finds considerable UDV - see Fig. \ref{positive_LEsmallT}. In agreement with the previous observations, the distribution of the UDs number is skewed to higher values with respect to what found when considering the whole dataset, compare with Fig. \ref{positive_LE}}.

It is worth exploring whether one find a clear localization in phase space of the anomalously stable and anomalously unstable UPOs. {\color{black}In} this regard, it is helpful to provide a visual representation of the UPOs. This can be achieved by considering the 3D projected space over the first three normalised time-dependent moments $C_1, C_2, C_3$, defined as:
\begin{equation}
C_k=\frac{\left(\sum_{j=1}^{{\color{black}J}} X_j^k\right)^{1/k}}{\left(\langle \sum_{j=1}^{{\color{black}J}} X_j^2\rangle\right)^{1/2}}, \quad k=1,2,3.\label{cs}
\end{equation}
where $\langle \bullet \rangle$ indicates the expectation value computed according to the invariant measure of the system. 

Figure \ref{UPOs} represents all UPOs of the database in such a projected space.  We will come back to this representation in Sect. \ref{statistics}. One notes that {\color{black}less unstable} UPOs are localised in the  part of the projected space closest to the origin.  Such visual impression is further supported by Figs. \ref{M1}, \ref{M2}, and \ref{M3}, which portray the statistics of the $C_k$'s stratified according to the UD of the UPOs.  The average value of $C_k$ computed over UPOs with a given number of UD increases monotonically with UD. In the case of $C_2$, this corresponds to the intuition that higher instability is associated with higher energy {\color{black} \cite{lorenz_1996,Lorenz2005,Vissio2020}. Indeed, one can interpret the periodic variations in the value of the $C_k$'s as describing the life-cycle of the unstable waves defined by the individual UPOs. 

Additionally, we have that the region of the phase space covered by lower-period - and thus anomalously unstable - UPOs misses the part of the attractor whose projection is closer to the origin in the $(C_1,C_2,C_3)$ space, compare Fig. \ref{UPOssmallT} with Fig \ref{UPOs}. Further evidence is shown in Figs. 1b-1d) of the Supplementary Material, where we show that mean values of the $C_k$'s stratified according to the UD of the UPO are shifted to higher values when only lower-period UPOs are considered. Figure \ref{density} portrays the projection of the invariant measure of the system in the $(C_1,C_2)$ and shows the region where the presence of lower-period UPOs is scarce. Hence, when the chaotic trajectory is in the region of phase space corresponding to the black circle in Fig. \ref{UPOssmallT} or the rectangular region in Fig. \ref{density}, there will be only few and nearby lower-period UPOs, so that the best local approximation to the trajectory will necessarily come from higher-period ones.}



\begin{figure}
\centering
\subfloat[]{%
     \includegraphics[width=8cm]{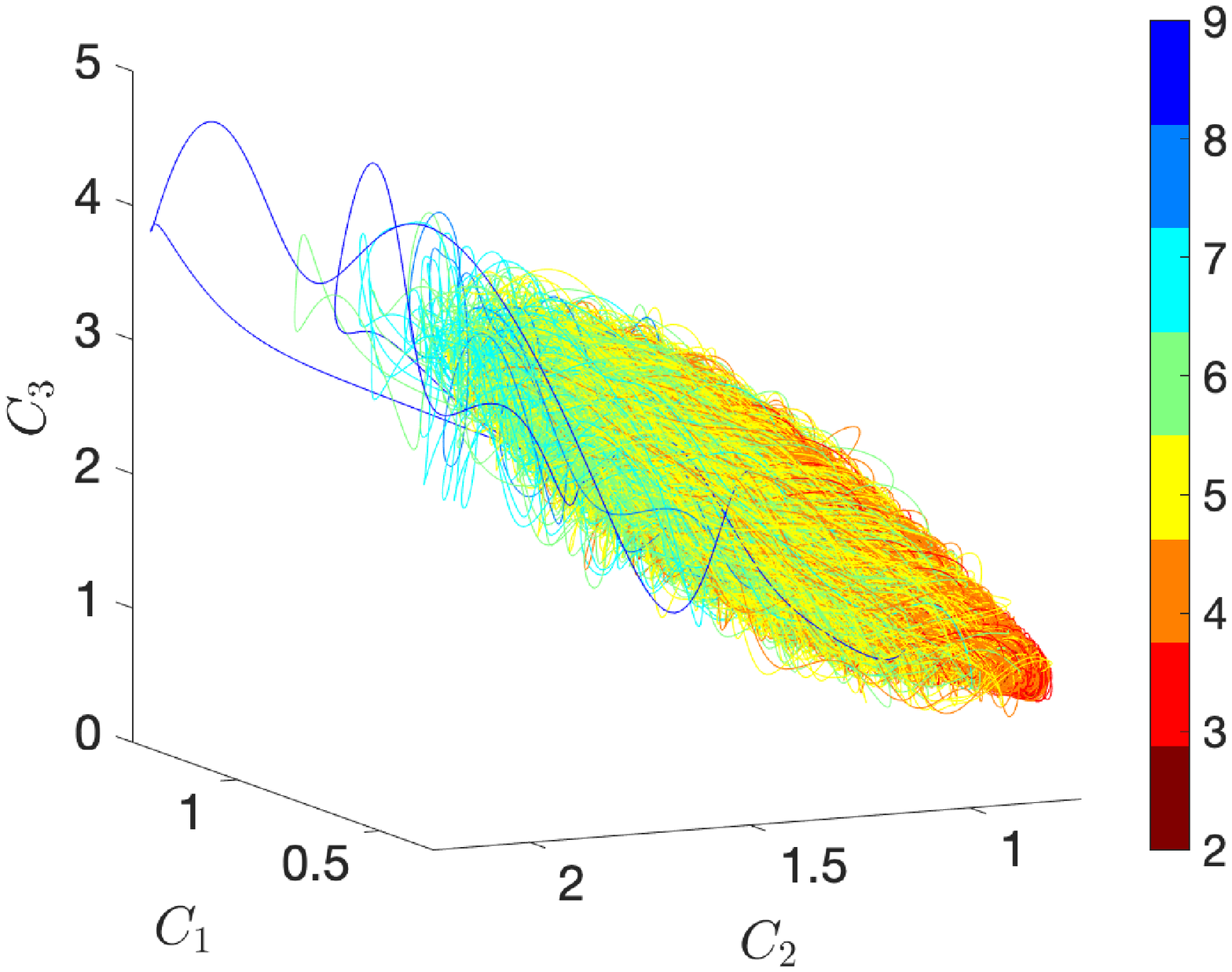}
           \label{UPOs}
    }
    \subfloat[]{%
     \includegraphics[width=8cm]{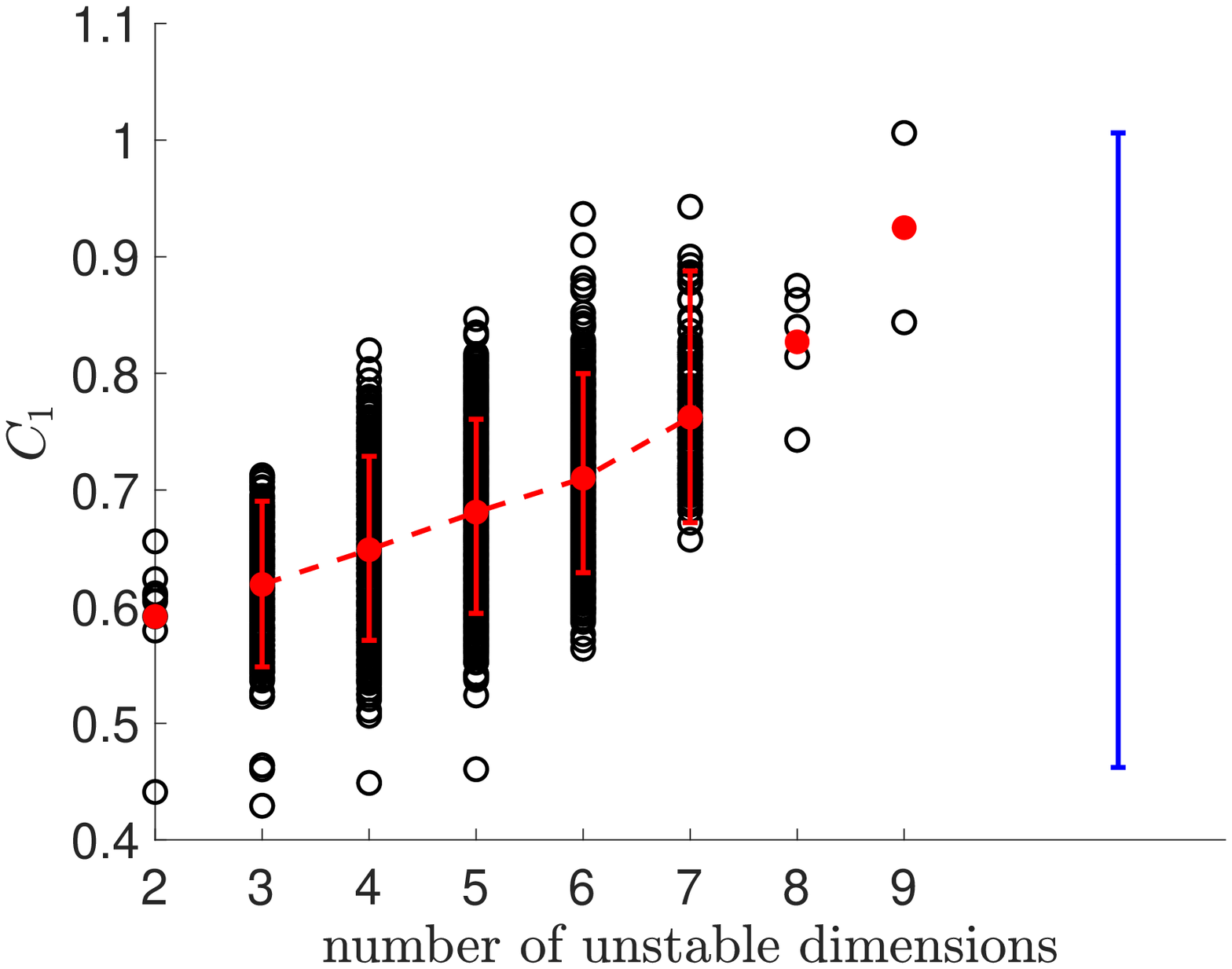}
           \label{M1}
    }\\
     \subfloat[]{%
   \includegraphics[width=8cm]{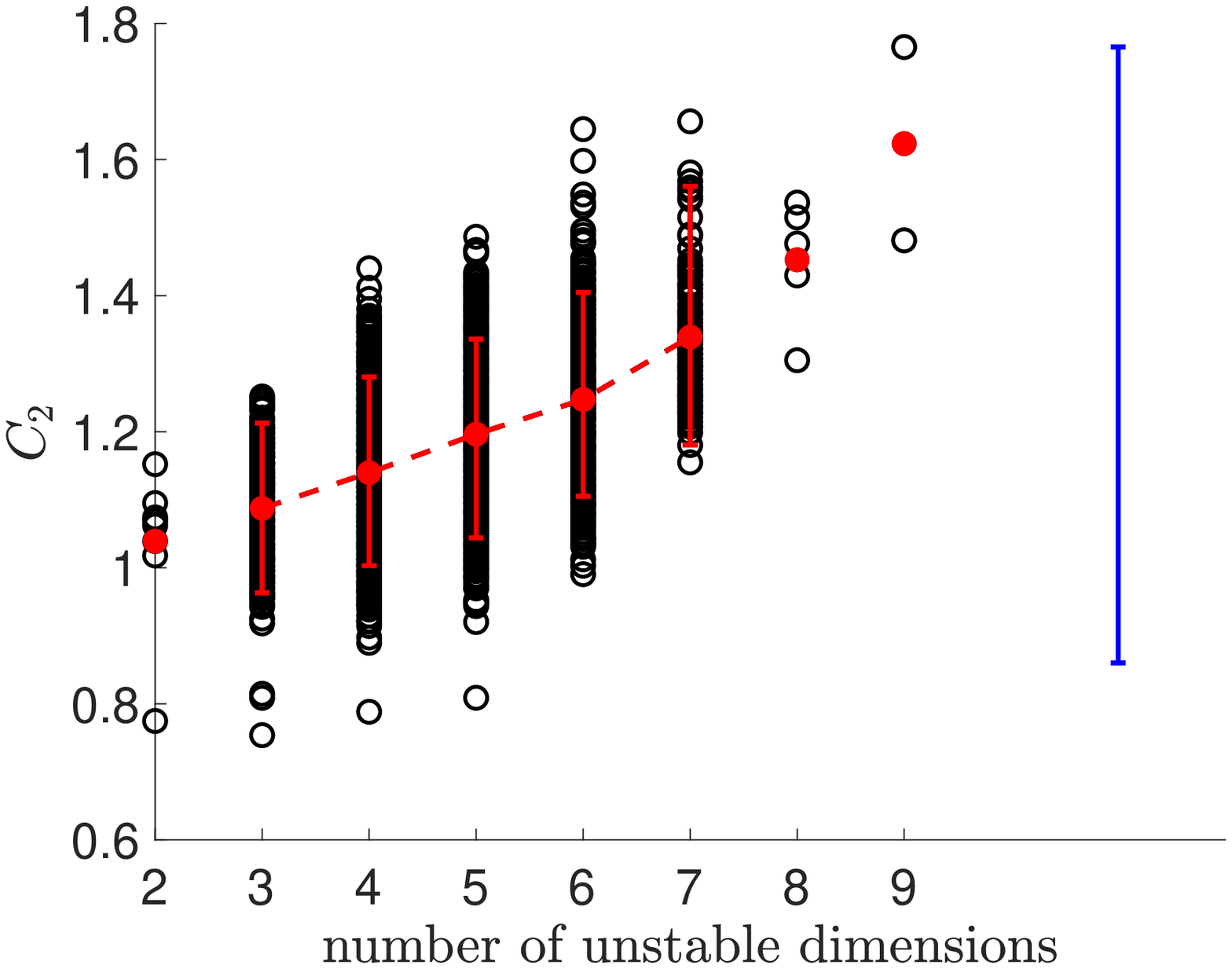}
              \label{M2}
}
     \subfloat[]{%
   \includegraphics[width=8cm]{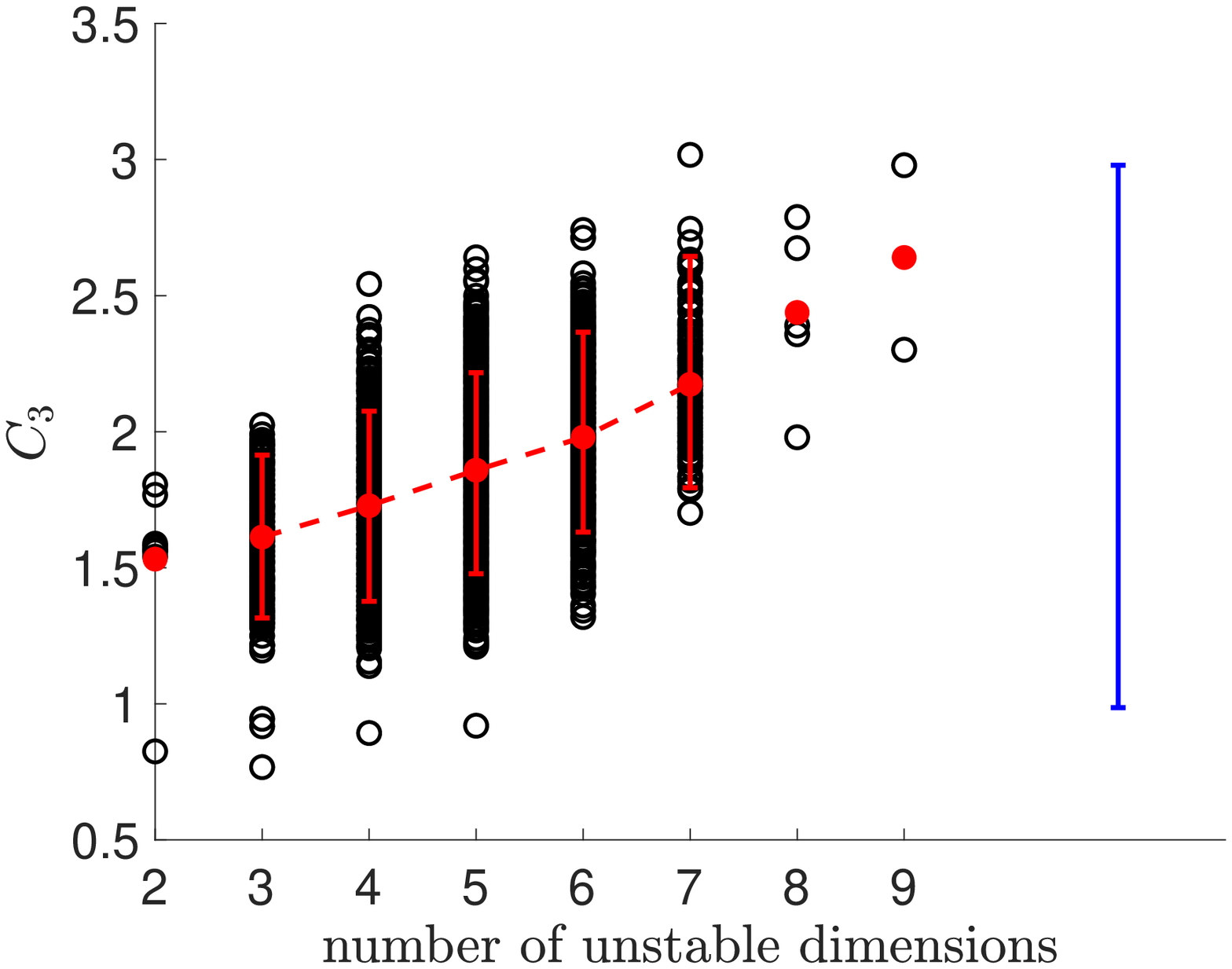}
              \label{M3}
}\\
 \subfloat[]{%
   \includegraphics[width=8cm]{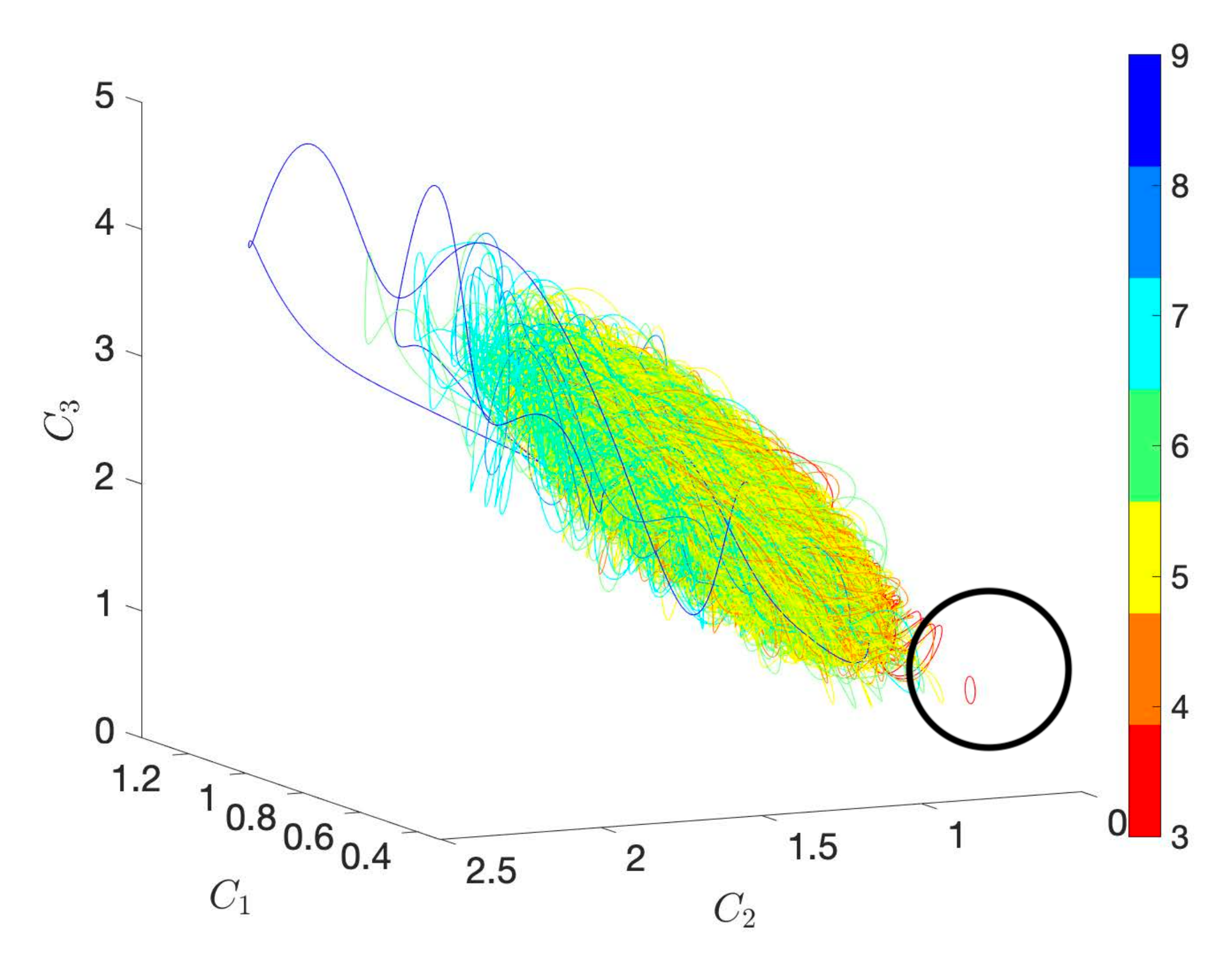}
              \label{UPOssmallT}
} \subfloat[]{%
   \includegraphics[width=8cm]{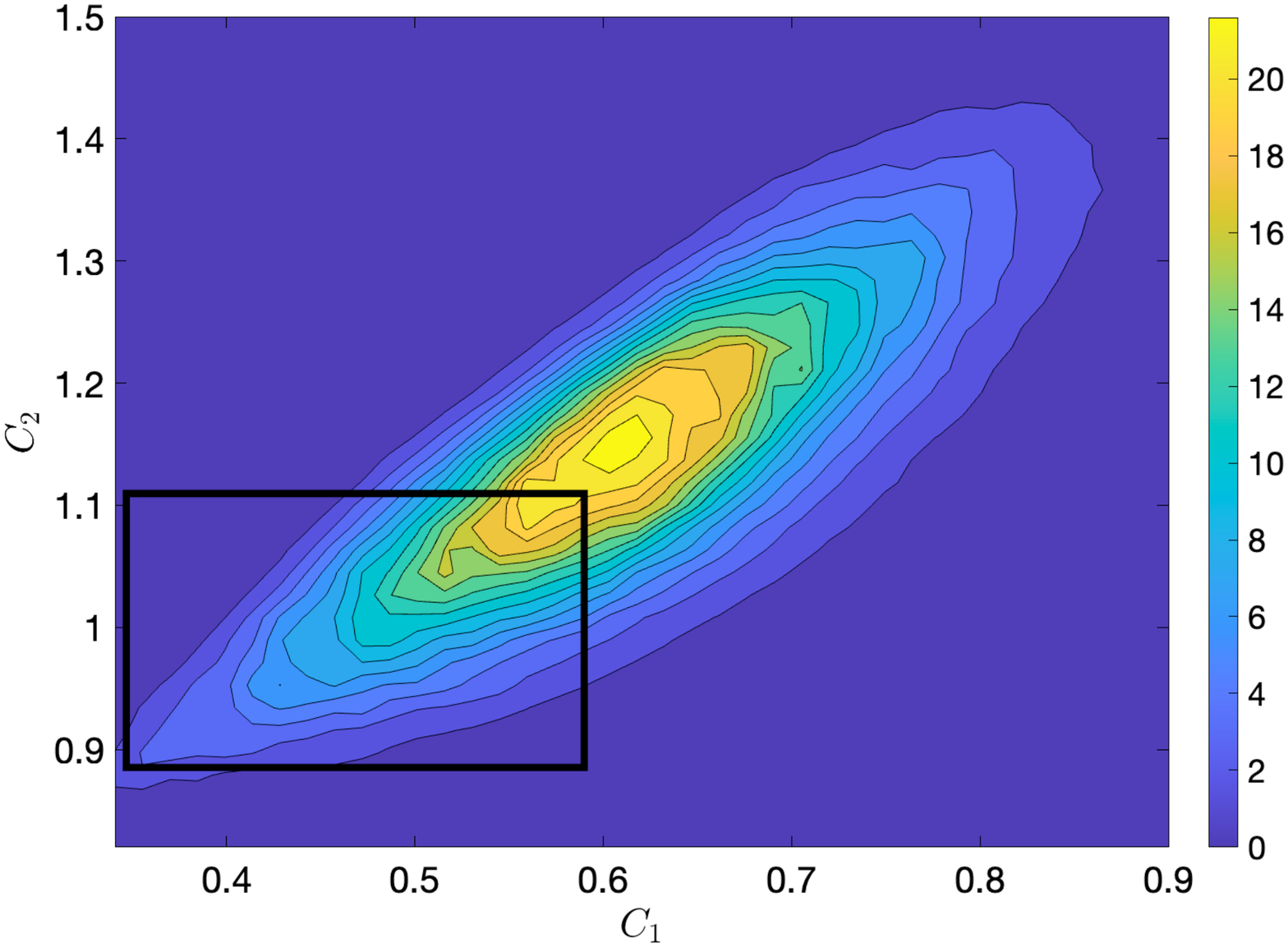}
              \label{density}
}
       \caption{Panel \protect\subref{UPOs}: UPOs of the system projected on the normalised moments $C_1, C_2, C_3$. The colour indicates the dimension of the unstable subspace. Panel \protect\subref{M1}: scatter plot of the first moment $C_1$ averaged along all the UPOs with the same number of UD vs their number of UD (black dots). The bars indicate the range between the 5th and 95th percentile for each UDs value. In blue the range between 5th and 95th percentile value of the corresponding statistics calculated along the chaotic trajectory. Panel \protect\subref{M2}: same as Panel \protect\subref{M1}, but for $C_2$. Panel \protect\subref{M3}, same as Panel \protect\subref{M1}, but for $C_3$. {\color{black}Panel \protect\subref{UPOssmallT}: same as Panel \protect\subref{UPOs}, but for UPOs with $T\leq6.4$. The circle indicates the region where major discrepancy is found with respect to Panel \protect\subref{UPOs}. Panel \protect\subref{density}: Projection of the invariant measure of the system in the $(C_1,C_2)$ space. The rectangle depicts the region that is  sparsely covered by UPOs with $T\leq6.4$; compare with the circle in panel \protect\subref{UPOssmallT}.}
  \label{energy_numberUD} }
\end{figure}

\subsection{Shadowing by Unstable Periodic Orbits}
\label{section: shadowing}
As discussed in the introduction, UPOs can  be used to approximate the forward trajectory of chaotic dynamical systems. 
 We say that a UPO is shadowing a chaotic trajectory if the UPO is sufficiently close to the chaotic trajectory and co-evolves with the trajectory for some period of time. 
{\color{black}Here we proceed as \cite{maiocchi_2022} and first introduce the concept of "rank shadowing".} 

{\color{black}Let $\mathcal{U}=\{U_k\}_{k=1}^{N_{UPO}}$ be the set of UPOs of the database where the $k^{th}$ UPO is defined as $U_k=\{ u_k(s)\} _{s=1}^{T_k/dt}$, with $T_k$ being its prime period and $dt=0.01$ the time step. We have a total of $N_{UPO}\approx 3\times10^5$ UPOs.
We consider a chaotic trajectory $\mathcal{X}_{chaotic}$ consisting of the set of points $\mathcal{X}_{chaotic} = \{  x(t) \}_{t=1}^{N_{max}}$ with output given every  $dt$ and with length $T_{max} = N_{max}\cdot dt=3\times10^4$. We assume that the chaotic trajectory lives on the attractor of the system, \textit{i.e.} transients have been discarded.
We define as distance between the UPO $U_{\bar{k}}$ and the chaotic trajectory $\mathcal{X}_{chaotic}$ at time $t$ as $d_k(t)=\min_s| u_{\bar{k}}(s)-x(t)|$. At each time $t$ we can rank the UPOs according to their distance form $x(t)$. $U_l$ is the tier 1 UPO at time $t$ if $l=\arg \min_{k=1,\ldots,N_{UPO}}(d_k(t)) $. Furthermore, $U_p$ is the tier K UPO at time $t$ if $d_p(t)$ is the $K^{th}$ smallest value among all $d_k(t)$'s, $k=1,\ldots,N_{UPO}$.

\begin{figure}[h]
  \subfloat[]{
\includegraphics[width=9cm]{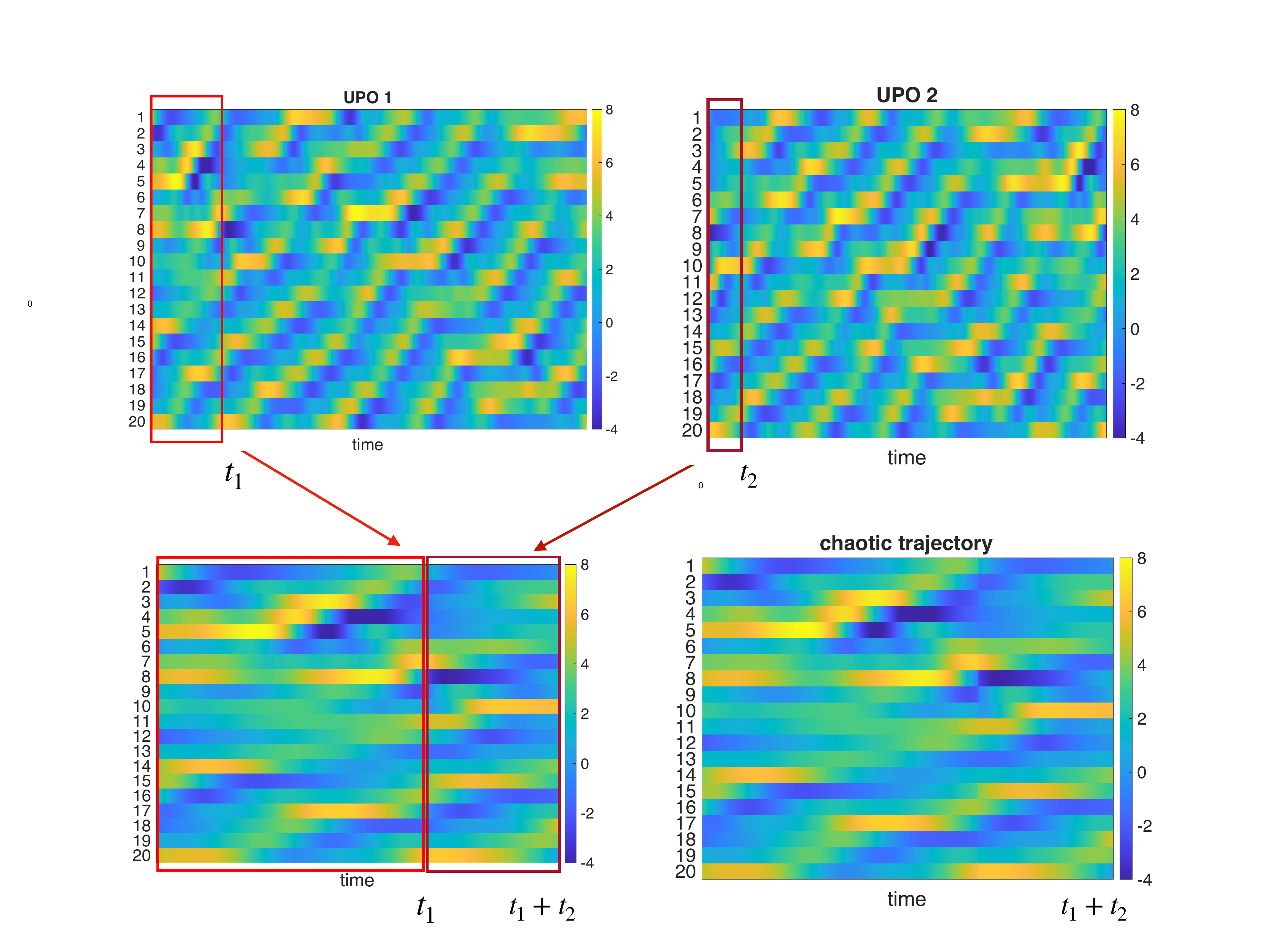}
\label{waves_panelA}}
  \subfloat[]{
\includegraphics[width=8cm]{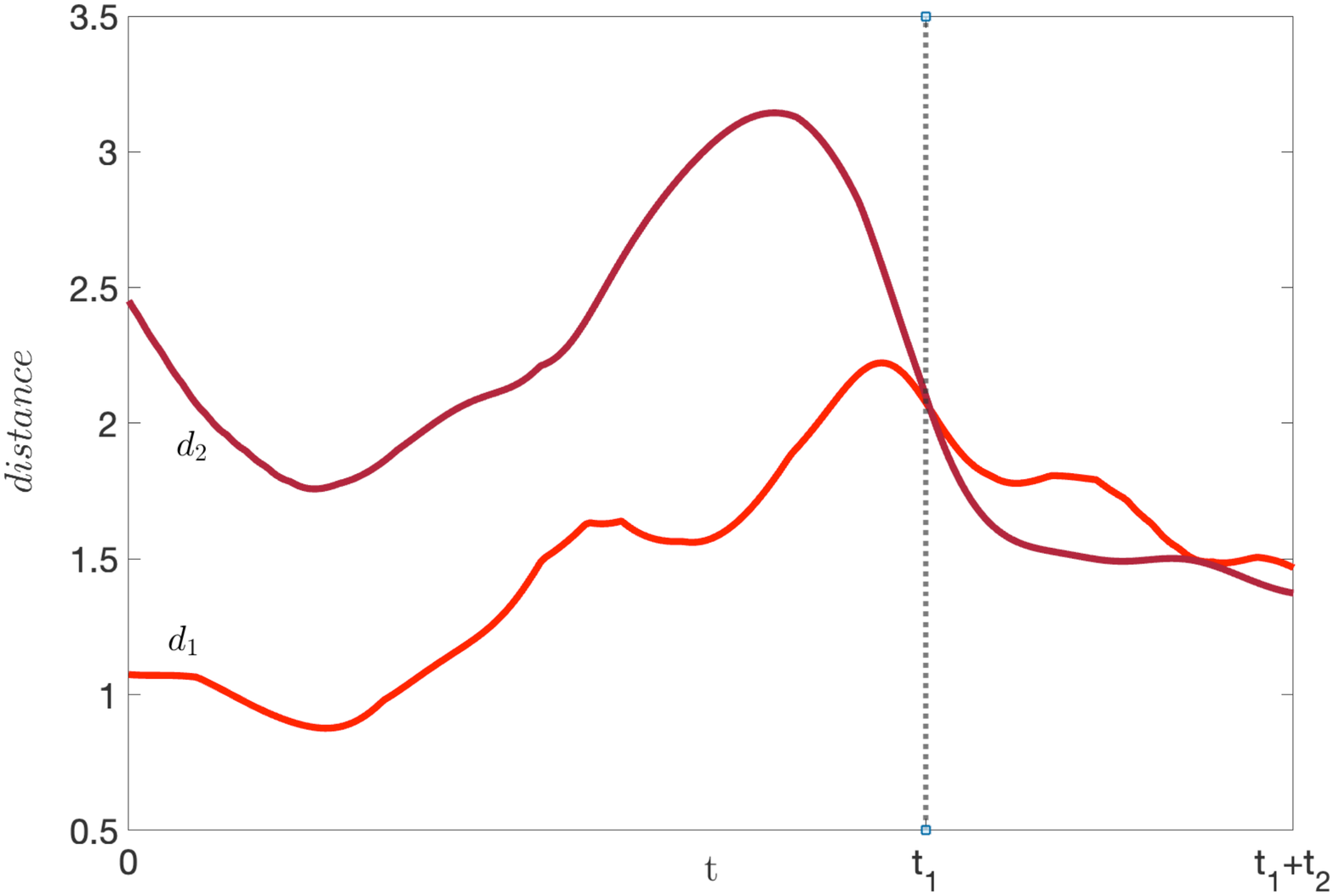}\label{waves_panelB}}
\caption{\label{waves} Panel \protect\subref{waves_panelA}: Subsequent shadowing of a segment of chaotic trajectory performed by two different UPOs (here UPO1 and UPO2). UPO1 (UPO2) has period  $T_1=10.8748
$ ($T_2=10.7626$) and possesses $5$ ($4$) positive LEs. {\color{black}In each of the four subpanels the x-axis indicates the time, the y-axis indicates the index $j$ of the $X_j$, $j=1,\ldots,20$ variables of the Lorenz '96 system, and the colorbar indicates the value of each variable. The space-time diagram of UPO1 and UPO2 over their full period is reported in upper row.} The chaotic trajectory (lower row, right hand side) is shadowed by UPO1 (bright red) for a time duration $t_1=1.78$ and then by UPO2 (dark red) for a time duration $t_2=0.87$. {\color{black}The two portions of the UPOs performing the shadowing are reported in the lower row, left hand side.} Panel \protect\subref{waves_panelB}: Time evolution of distance $d_1$ and $d_2$ between the chaotic trajectory and UPO1 and UPO2, respectively.}
\end{figure}
When constructing the time-dependent ranking of the UPOs in terms of proximity to the chaotic trajectory, we discover that only around $26\%$ of the total is selected at least once as first tier orbit. Instead, the fraction of UPOs that belong at least once to the first $K = 10, 100, 1000$ tiers increases up to $79\%, 99.6\%$, and $99.9\%$, respectively, of the complete database. {\color{black}The four panels of Fig. \ref{database} show that the $K=1$ tier shadowing orbits are preferentially of lower instability (and, hence, feature typically a longer period, see Fig. \ref{scatter_lambda1}) compared to the whole database because they repel less intensely  nearby trajectories. As we include orbits belonging to higher tiers of proximity, the statistical properties rapidly converge to those of the whole database. If we consider only the (much more homogeneous) lower-period UPOs, there is little dependence of the level of instability of the UPOs on their tier of proximity, see Figs. \ref{positive_LEsmallT}-\ref{lambda1smallT}.} As confirmed below, even orbits belonging to the $K = 10, 100, 1000$ tiers are very often rather close to the chaotic trajectory. 

We will consider two definitions of shadowing orbits. The first is
a stricter definition, which considers as shadowing UPOs only the sequence of tier 1 UPOs. We also consider a looser definition, which allows one to slightly prioritise persistence over proximity. In fact, at each time step, a UPO might still provide a very good local approximation to the trajectory even if it is not anymore the nearest one.  In particular, if $U_l$ is the closest UPO to the trajectory at time $t$, we say that $U_l$ ceases to shadow the trajectory at time $t+pdt$ if at that time $U_l$ is not anymore one of the $K$ closest UPOs, or, in other terms, it does belongs to one of first $K$ tiers. 
We then collect the time series of the distances $d_l(t+jdt)$, $j=0,\ldots,p-1$ and we say that $U_l$'s shadowing duration  (or $U_l$'s persistence) is $pdt$. At time $t+pdt$ the tier 1 UPO is selected as the next shadowing UPO. The strict definition of shadowing is obtained by setting $K=1$.

An example of shadowing (case $K=1$) is presented in Fig. \ref{waves}. Panel \protect\subref{waves_panelA} shows how a short portion of duration $t_1+t_2=2.65$ of the chaotic trajectory  is subsequently shadowed by two UPOs featuring different period and different UD. The juxtaposition of the two pieces of the shadowing UPOs of duration $t_1=1.78$ and $t_2=0.87$, gives a time-dependent field that is visually very similar to {\color{black}the} considered portion of the chaotic trajectory. Panel  \protect\subref{waves_panelB} shows the time evolution of the distance between the trajectory and the two UPOs. At time $t=t_1$ there is a transition in the shadowing, as the second UPO becomes the closest one to the chaotic trajectory. 
} 

By construction, choosing larger values of $K$ in the definition of shadowing leads to an increase in the average distance between the orbit and the shadowing UPOs, and, at the same time, to an increase in the persistence of each shadowing UPOs. 
Figures \ref{distance_tiers} and  \ref{persistence_tiers} show the distribution of the  distance $d$  and of the persistence $\pi$ of the shadowing UPOs when considering the strict (tier 1, in black) and the looser definition of shadowing, with $K\in\{10,100,1000\}$. 
The average distance increases from $1.81$ to $2.70$ as $K$ increases from 1 to 1000 (see Table \ref{table_prob}). One notices that the quality of the shadowing is in general rather good and extremely similar for $K=1$ and $K=10$: the  $95\%$ quantile of shadowing distances is approximately 2.99 and 3.21, which is well within the $0.5\%$ quantile of the typical distances distribution over the attractor (See Fig. \ref{typical_distances}) The mean persistence, in turn, increases from $0.22$ to $1.69$ time units. These values correspond to  average rectified distances ranging from $8$ to $63$. By comparing these numbers with the size of the attractor, the typical distances over the attractor (Fig. \ref{typical_distances}), and the average distance between the chaotic trajectory and the shadowing UPOs confirms that there is clear evidence of co-evolution. Assessing co-evolution is instrumental for constructing a framework that allows an accurate statistical and dynamical description of the chaotic flow \cite{krygier_2021}. 

\begin{figure}
\centering
     \subfloat[]{%
 \includegraphics[width=0.48\textwidth]{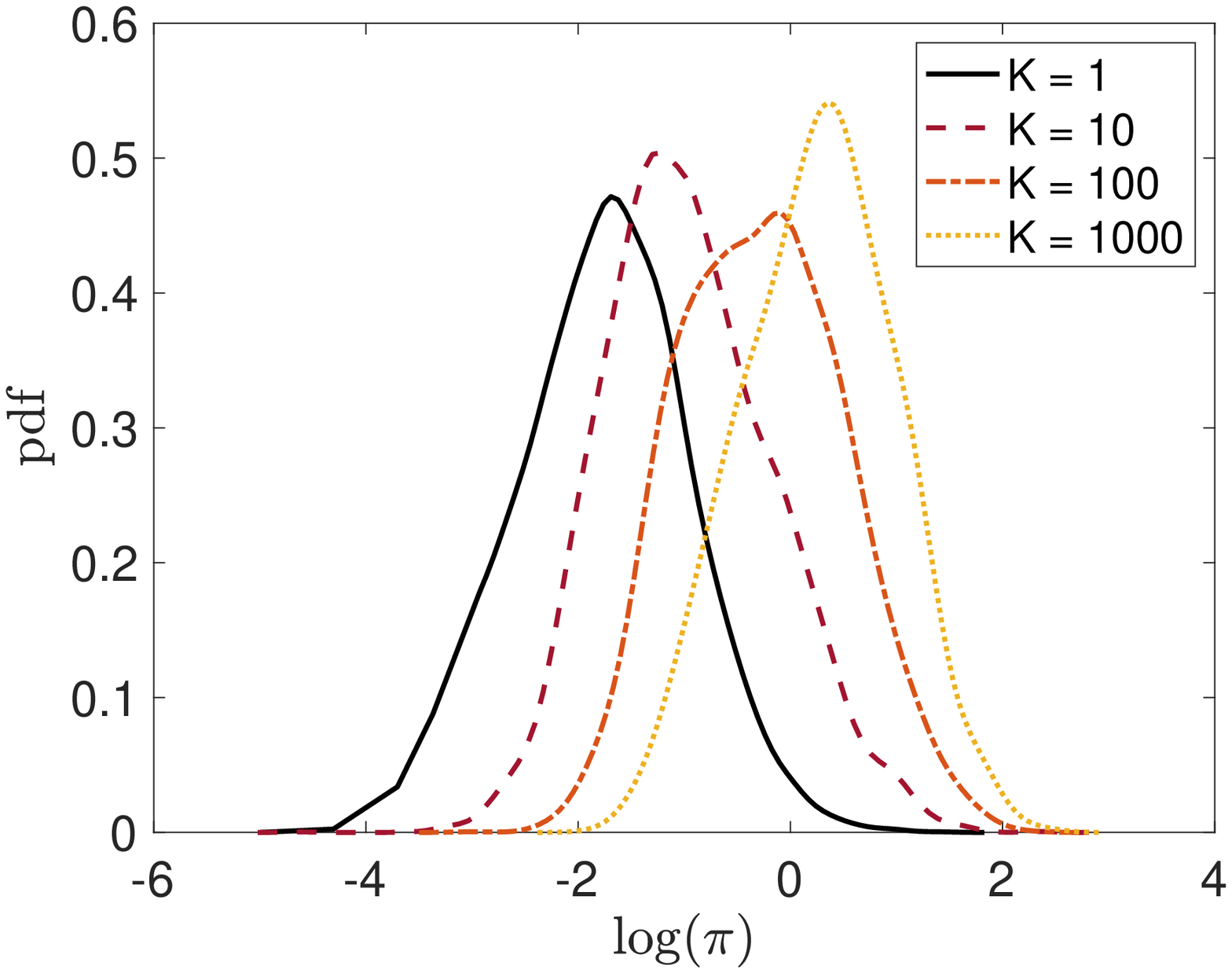}
 \label{persistence_tiers}
    }
     \subfloat[]{%
   \includegraphics[width=0.48\textwidth]{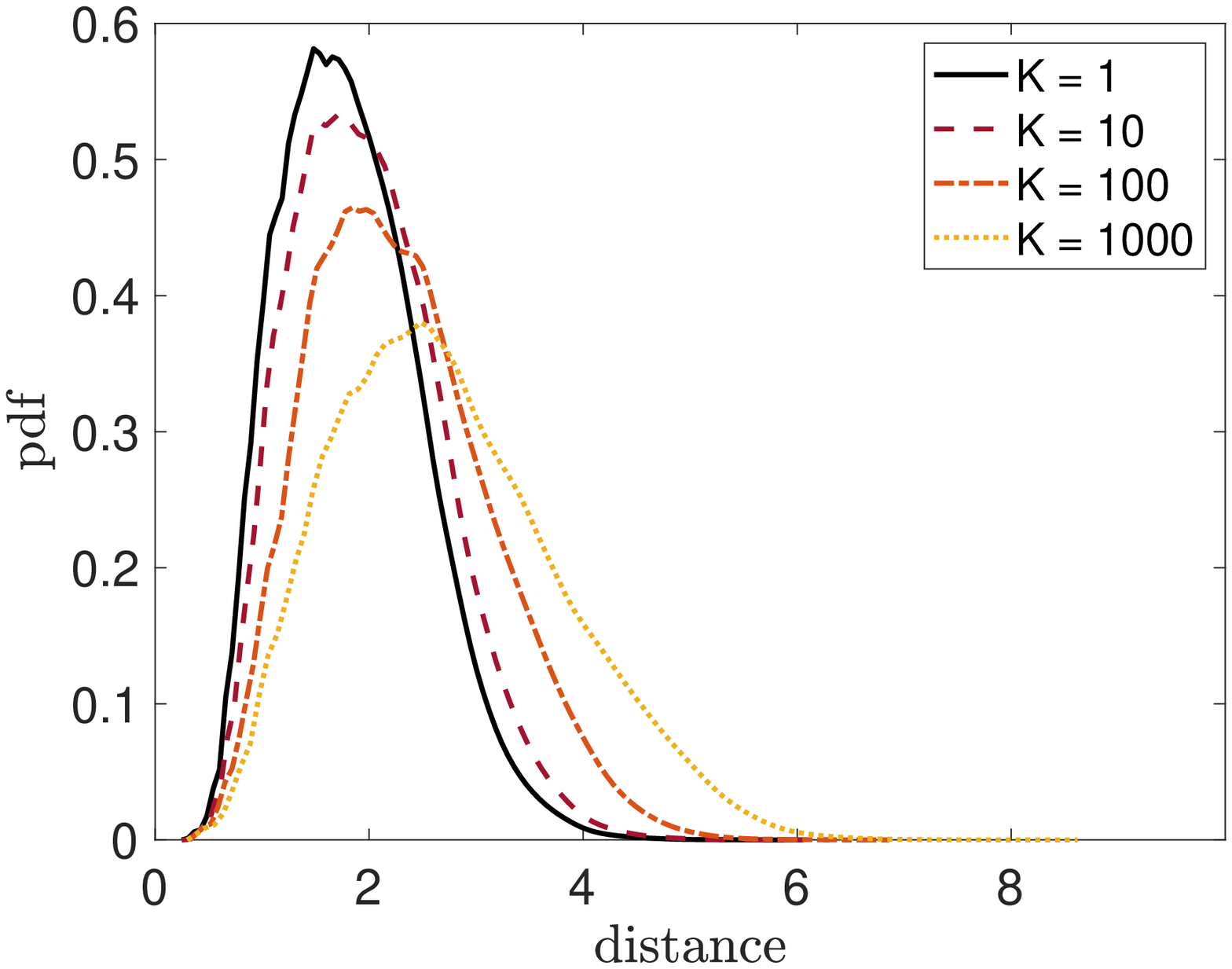}
              \label{distance_tiers}
}\\
     \subfloat[]{%
 \includegraphics[width=0.48\textwidth]{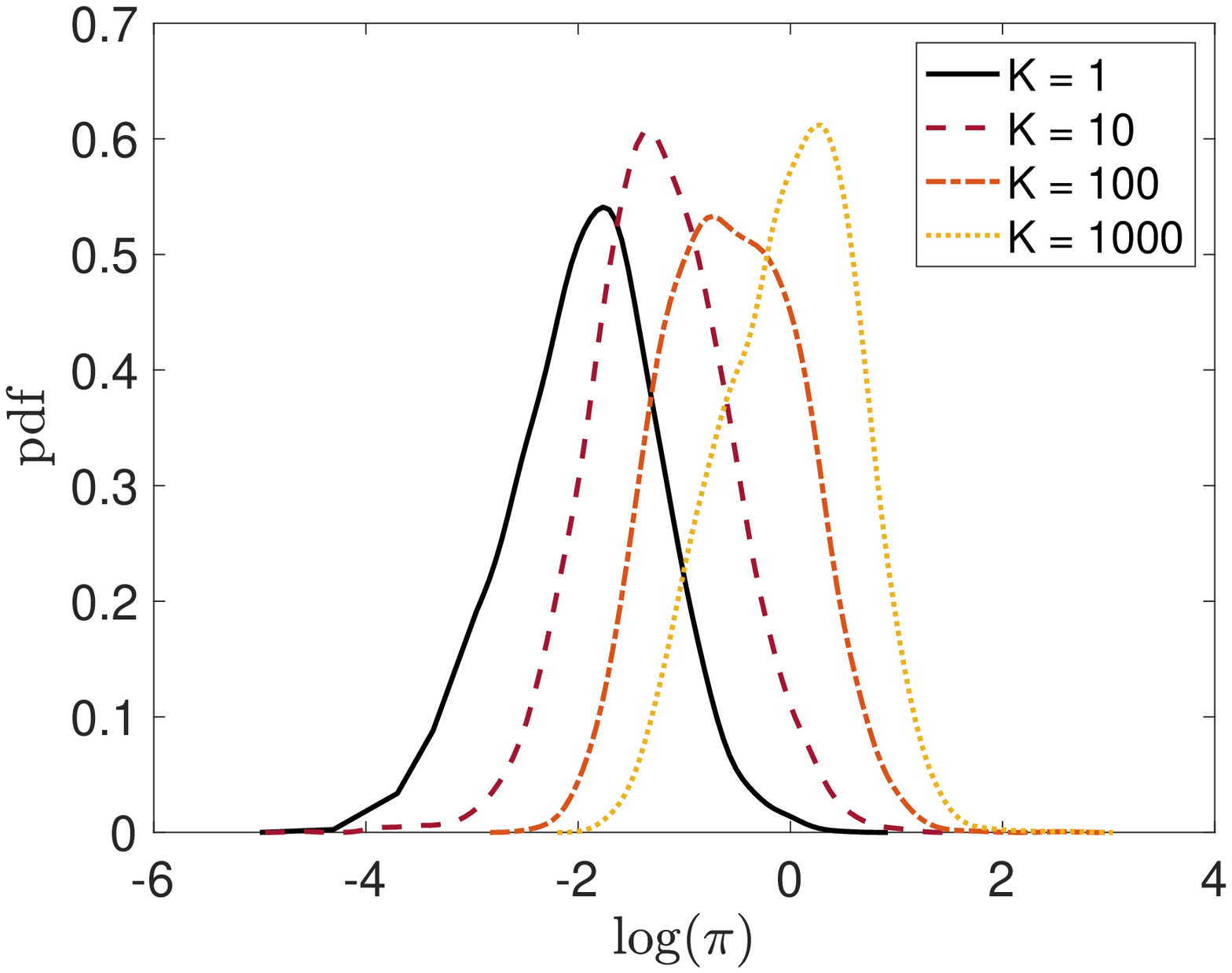}
 \label{persistence_tiers_smallT}
    }
     \subfloat[]{%
   \includegraphics[width=0.48\textwidth]{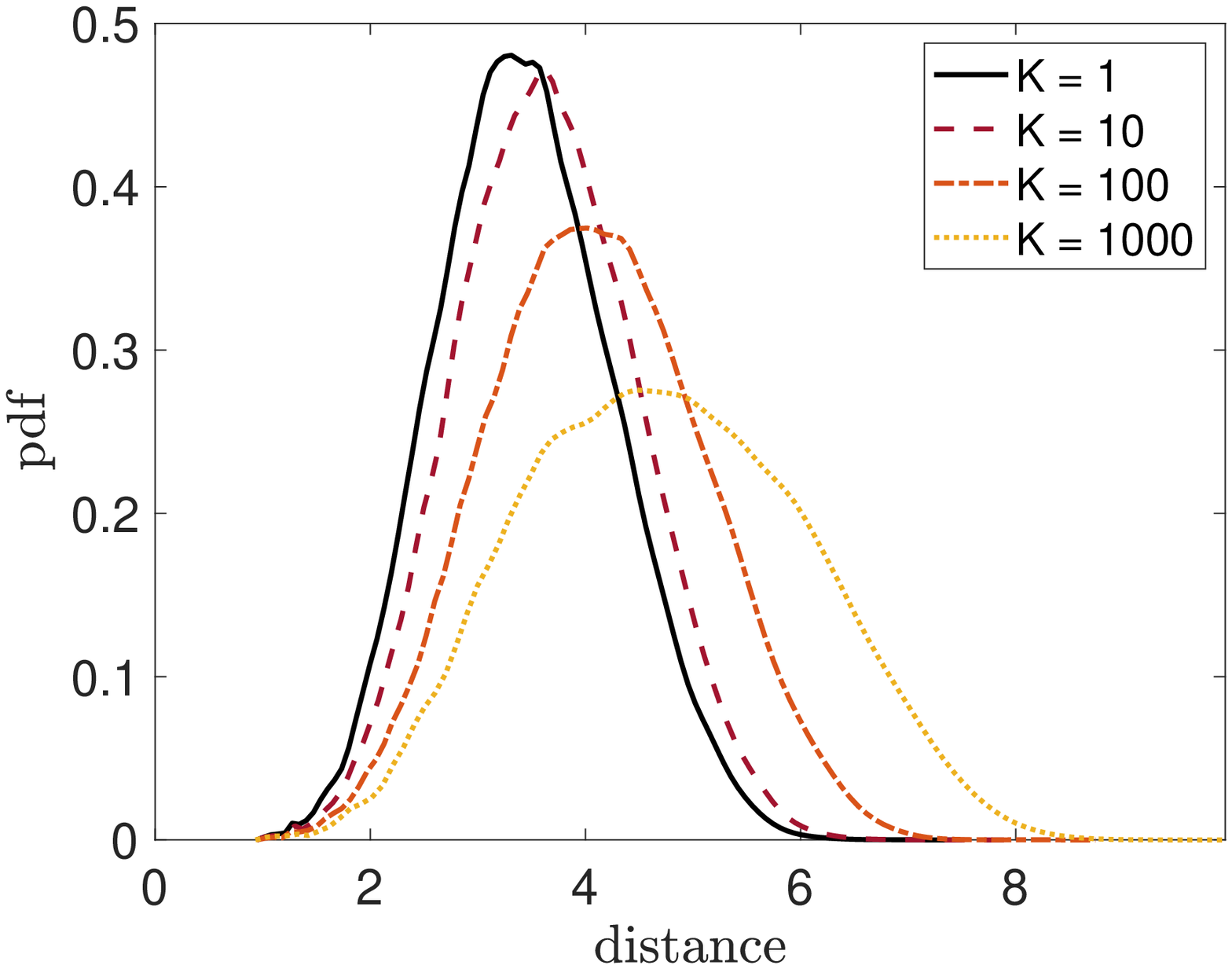}
              \label{distance_tiers_smallT}
}
        \caption{Panel \protect\subref{persistence_tiers}: 
       Probability distribution function of the $\log$ of the persistence $\pi$ of the tier 1 orbits (solid black line; mean persistence = 0.21), and the shadowing orbits with the looser definition with $K=10$ orbits (dashed red line, mean persistence 0.53), $K=100$ (dashed and dotted orange line, mean persistence 1.03) and $K=1000$ (dotted yellow line, mean persistence 1.67). Panel \protect\subref{distance_tiers}: Distribution of the distances from the chaotic trajectory when considering the looser definition of shadowing orbits that allows for fluctuations within the first K=10 tiers (red dashed line, mean distance 1.97), tier 100 orbits (dashed and dotted orange line, mean distance 2.26), tier 1000 orbits (dotted yellow line, mean distance 2.7) and again first tier orbits (black solid line, mean distance 1.81). Panel \protect\subref{persistence_tiers_smallT}: Same as Panel \protect\subref{persistence_tiers}, for $T<6.4$ UPOs. Panel \protect\subref{distance_tiers_smallT}: Same as Panel \protect\subref{distance_tiers}, for $T<6.4$ UPOs.   }
  \label{stability_propr_shad_UPOs} 
\end{figure}

{\color{black}Restricting our analysis to the lower-period UPOs unavoidably leads to an increase in the distance between the shadowing UPOs and the chaotic trajectory, compare Fig. \ref{distance_tiers_smallT} and Fig. \ref{distance_tiers}, and look at Table \ref{table_prob}. Note that, instead, the statistics of the log persistence changes negligibly, compare Fig. \ref{persistence_tiers_smallT} and Fig. \ref{persistence_tiers}, which indicates that the procedure is robust. Let's focus on the statistics of the distances between the tier 1 UPOs and the chaotic trajectory. The average distance increases by a rather substantial factor $\approx 1.9$: excluding the higher period UPOs amounts to losing not only about $80\%$ of all UPOs, but also the geometrically longest  and dynamically least unstable ones, so that the ability to cover the attractor  is substantially reduced. Additionally, if we restrict our analysis to the  $T<6.4$ UPOs, the average distance between the tier 1 UPOs and the chaotic trajectory increases by a factor $\approx2.5$ when considering the rectangular region in Fig. \ref{density}, whereas it increases by a factor $\approx1.8$ outside of it. This further supports what has been shown in Figs. \ref{UPOssmallT} and \ref{density} regarding the scarcity of the $T<6.4$ UPOs in the low-energy region of the attractor.}

\begin{table}[]
\centering
\begin{tabular}{l|llll|llll}
K                                     & 1    & 10   & 100  & 1000 & 1    & 10   & 100  & 1000 \\ \hline
95th percentile                       & 2.99 & 3.21 & 3.76 & 4.64 & 4.82 & 5.05 & 5.76 & 6.90 \\ \hline
mean\_\{dist\}                        & 1.81 & 1.96 & 2.26 & 2.70 & 3.44 & 3.64 & 4.09 & 4.74 \\
P\_\{\%\}(d\textless{}mean\_\{dist\}) & 0.09 & 0.10 & 0.12 & 0.14 & 0.17 & 0.18 & 0.21 & 0.25 \\ \hline
5th percentile                        & 0.87 & 0.95 & 1.05 & 1.17 & 2.15 & 2.29 & 2.48 & 2.68
\end{tabular}

\caption{\label{table_prob} \textcolor{black}{We report the distance statistic for both the complete (left hand side of the table) and reduced database of UPOs (right hand side of the table)}. The first and last rows indicate the $95th$ ($5th$ respectively) percentile of the distribution of distances of the shadowing orbits from the chaotic trajectory in the first tier and for $K=10,100,1000$. The central row reports the mean value of the distances distributions for each tier, and relative probability of achieving smaller distance across the attractor of the system. We can appreciate how such probability never exceeds $1\%$, confirming the closeness of the trajectory with the shadowing orbits.}
\end{table}




\section{Unstable Periodic Orbits and Unstable Dimension Variability}\label{Local}

\subsection{Finite-Time Lyapunov Exponents and Unstable Dimension Variability}
\label{section: Evidence of FTLE fluctuations}
In \cite{Sauer1997,Sauer2002,Pereira2007} it was shown how the UDV could be explained in terms of the presence of fluctuations between positive and negative values  of one specific FTLE (the one corresponding to the LEs with smallest absolute value) computed over a time scale $\tau$ also when considering very large values of $\tau$. {\color{black}See definitions and some essential information on FTLEs in App. \ref{finiteLyap}.} The presence of changeovers in the sign of the FTLE over such long time scales was proposed as evidence of the trajectory following closely UPOs having different UDs number. As mentioned earlier, in the system of interest here the UDV entails fluctuations between 2 and 9 of the number of UDs, hence one could expect to find that several FTLEs feature fluctuations between positive and negative values. 

This is indeed confirmed by our data. We consider here the first 10 LEs, ordered from the largest to the smallest. 
Note that for sufficiently large values of $\tau$ the distribution of all the FTLEs corresponding to nonzero LEs converges to a Gaussian with variance $\propto 1/\tau$, in agreement with previous studies \cite{Pazo2013,Laffargue2013,de2018exploring}. A different scaling is found for the FTLE corresponding to the vanishing LE (not shown).

Even considering very long averaging times, the support of the pdf of more than one FTLEs includes zero, meaning that one observes fluctuations about zero for the corresponding time series. Putting aside the vanishing LE, for which this property is trivial, this applies to 5 FTLEs for $\tau=10\tau_1$, which is already much longer than the period of the longest detected UPO), see Fig. \ref{FTLE10}. Clearly, the number of FTLEs fluctuating about zero decreases as one consider larger values for $\tau$. Nonetheless one finds four of such FTLEs when $\tau=30\tau_1$ (Fig. \ref{FTLE30}), and still two for the ultralong averaging time $\tau=100\tau_1$ (Fig. \ref{FTLE100}). The fluctuations about zero of the $6^{th}$ FTLE persist even for much long longer averaging times. 

Focusing on indicators of instability provided further information on the heterogeneity of the attractor. Panel \ref{KS} shows that the sum of the FTLEs corresponding to the four largest backward LEs - this provides the finite-time, local estimate of the Kolmogorov-Sinai entropy - have very large fluctuations, and the distribution has support extending to negative values up to averaging times of about $\tau=3\tau_1$. We also see - Panel \ref{LM} - that the largest (ordered) backward FTLE can have negative values for averaging times up to $\tau=3\tau_1$. This implies that the system features (temporary) return of skill. 
Finally, Panel \ref{UDV} shows the distribution of the number of positive FTLEs. Note that we have removed from the count the direction of the flow, whose corresponding FTLE obviously fluctuates between small positive and small negative values. We find confirmation that that the number of positive FTLEs has very large fluctuations even for very long averaging times $\tau$.

\begin{figure*}
     \subfloat[]{%
   \includegraphics[trim={2cm 1cm 3cm 0.5cm},clip,width=0.46\textwidth]{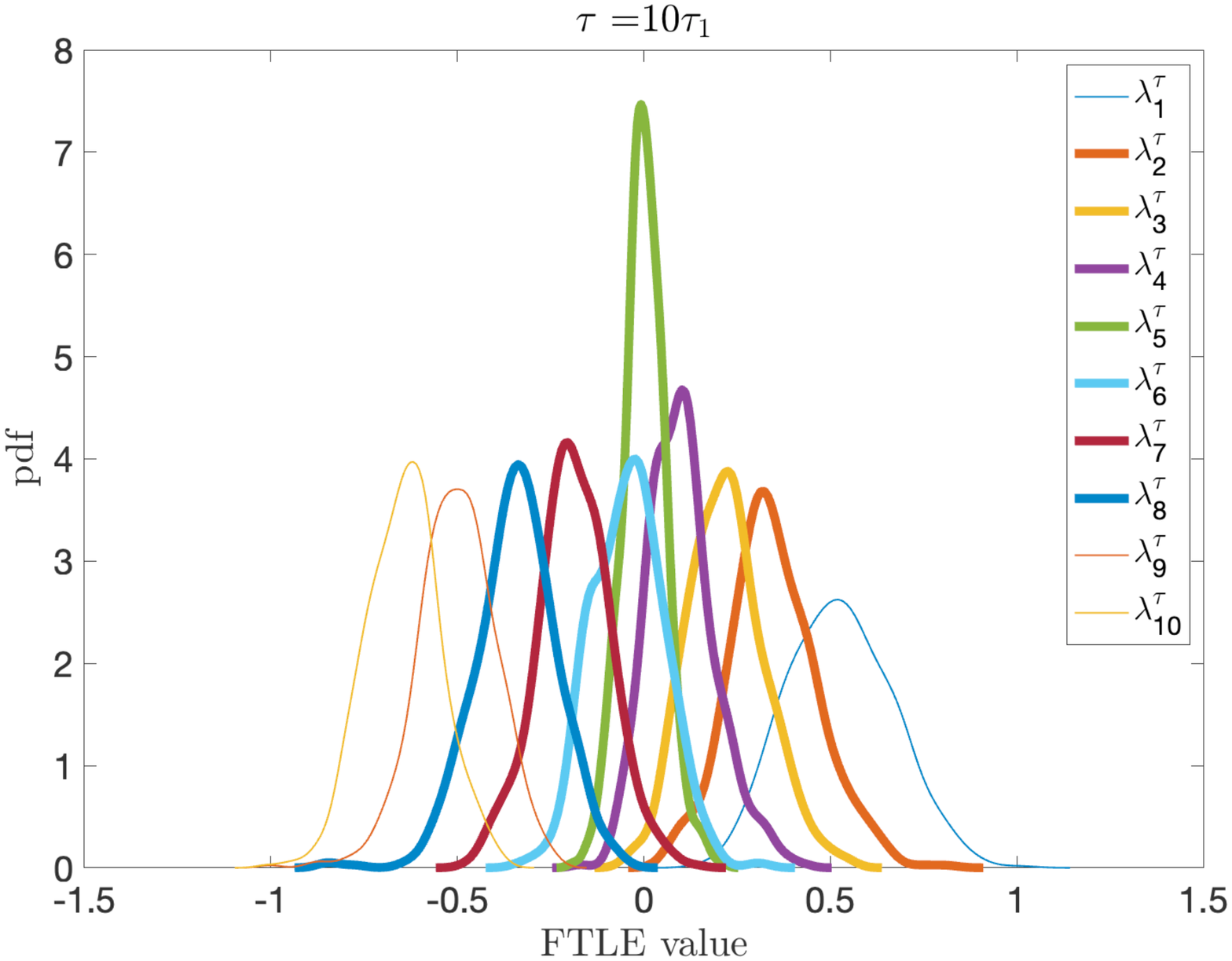}
              \label{FTLE10}
    }
     \subfloat[]{%
   \includegraphics[trim={2cm 1cm 3cm 0.5cm},clip,width=0.46\textwidth]{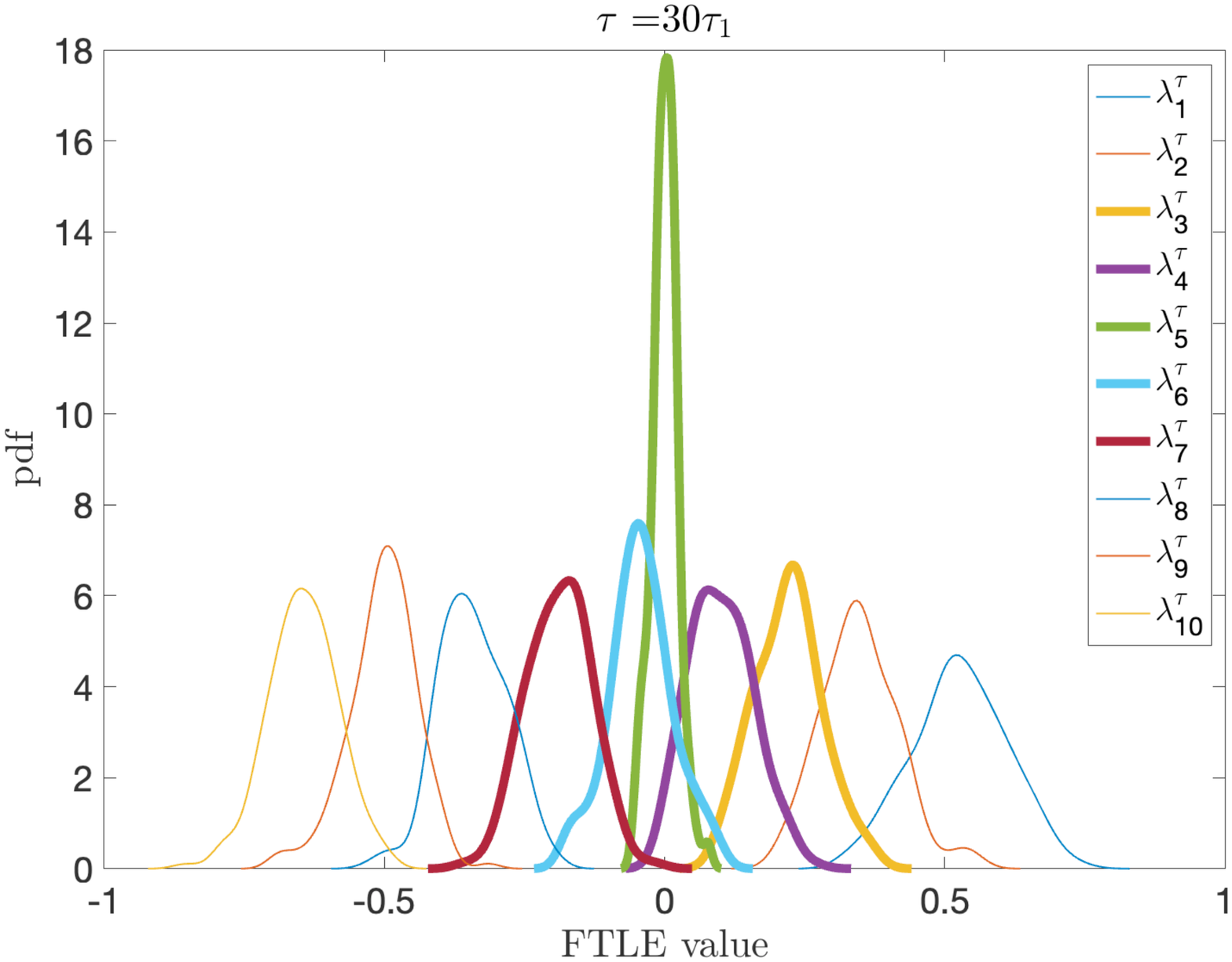}
              \label{FTLE30}
    }\\
     \subfloat[]{%
   \includegraphics[trim={2cm 1cm 3cm 0.5cm},clip,width=0.46\textwidth]{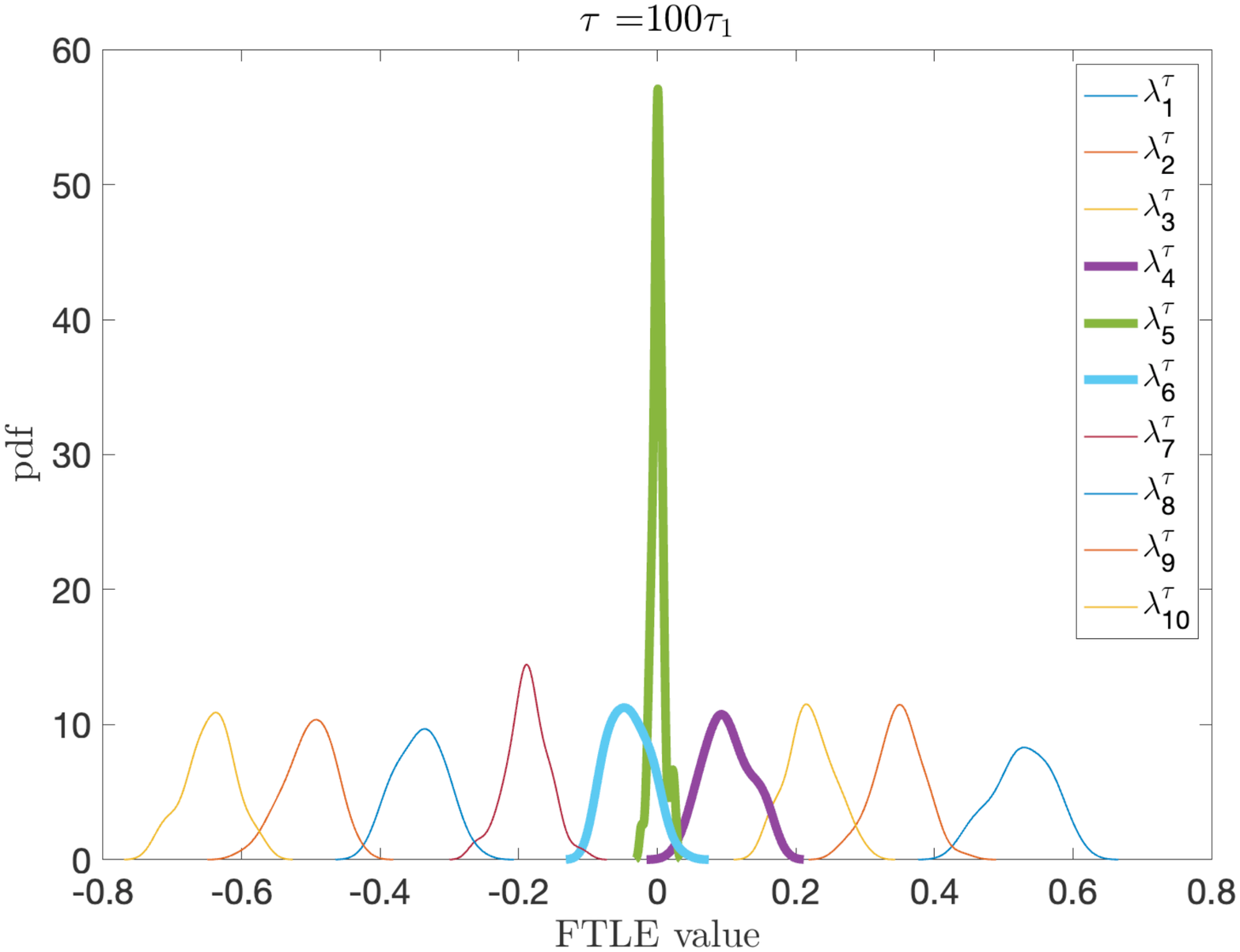}
              \label{FTLE100}
    }
        \subfloat[]{%
   \includegraphics[trim={2cm 1cm 3cm 0.5cm},clip,width=0.46\textwidth]{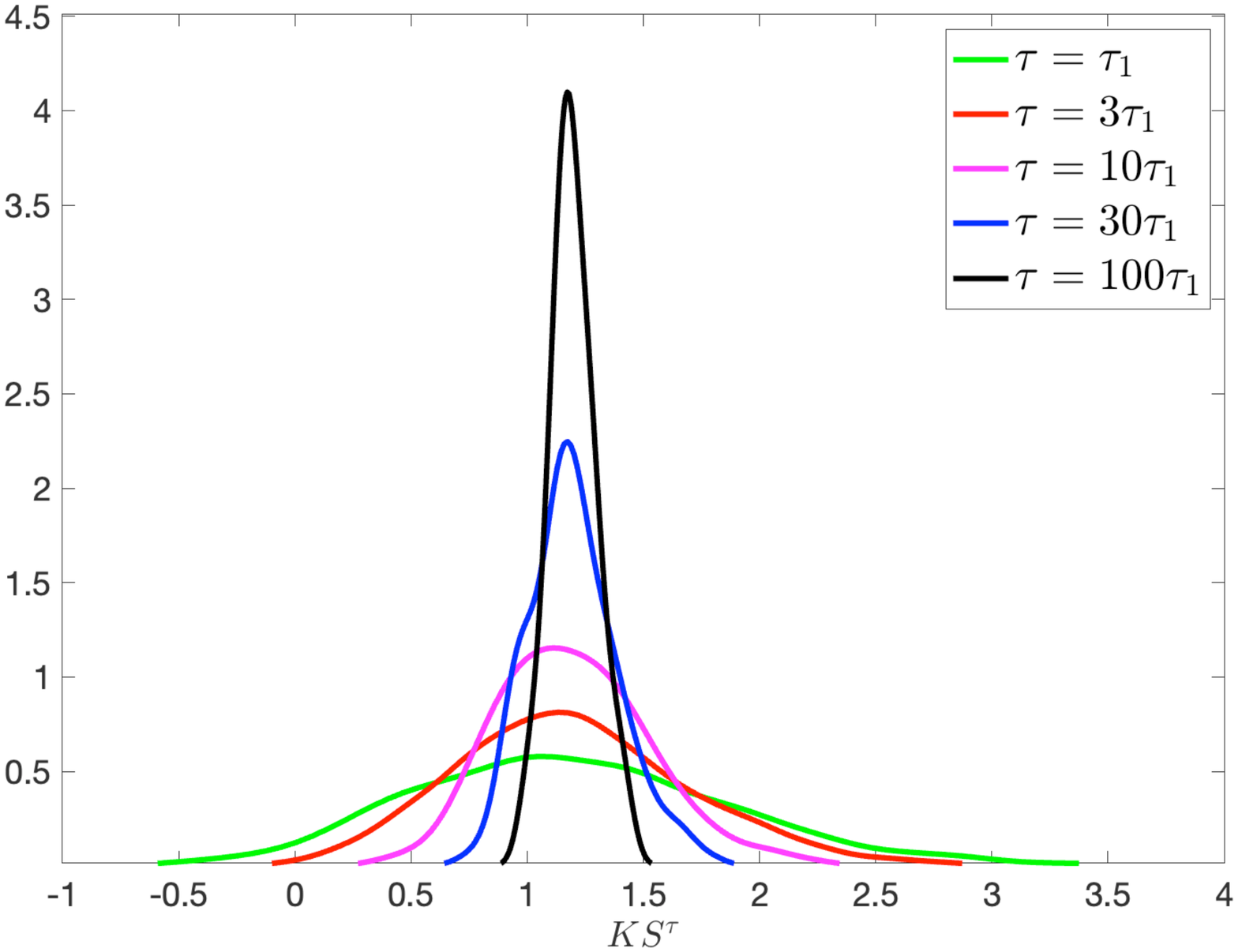}
              \label{KS}
    }\\
     \subfloat[]{%
   \includegraphics[trim={2cm 1cm 3cm 0.5cm},clip,width=0.46\textwidth]{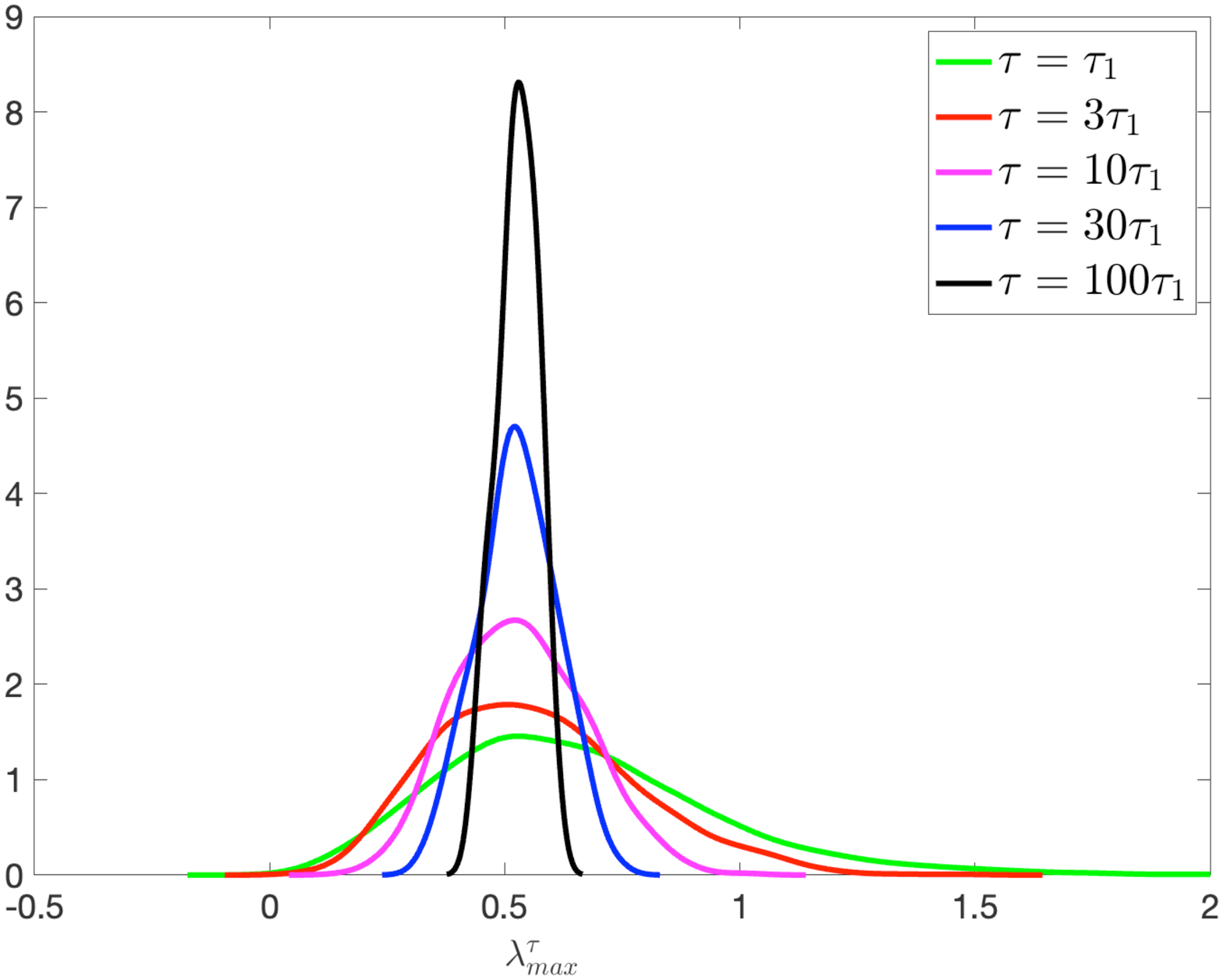}
              \label{LM}
    }
     \subfloat[]{%
   \includegraphics[trim={2cm 1cm 3cm 0.5cm},clip,width=0.46\textwidth]{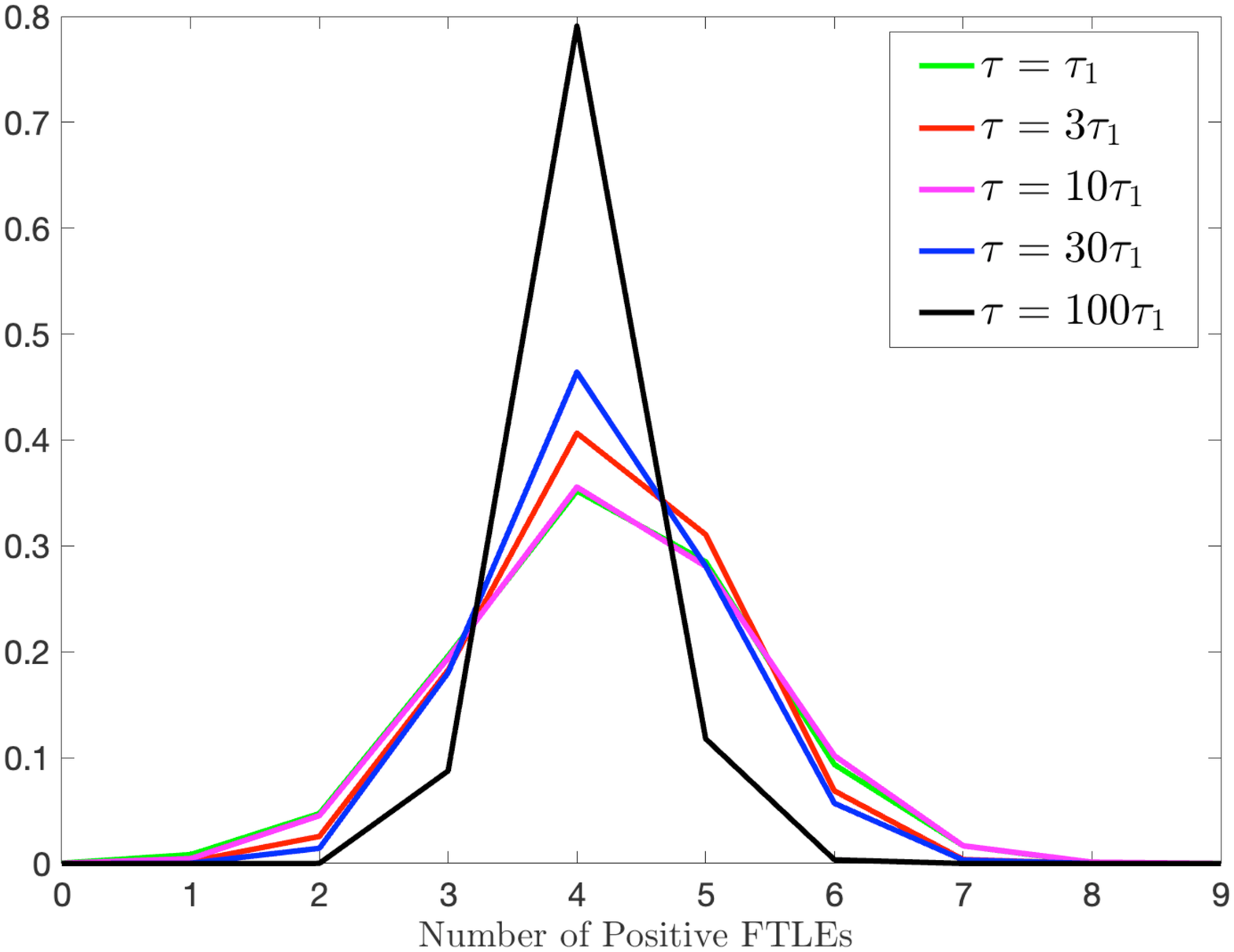}
              \label{UDV}
    }
       \caption{Evidence of heterogeneity of the tangent space. The $5^{th}$ LE corresponds to the direction of the flow. 
       Panel \protect\subref{FTLE10}: Distribution of the first 10 FTLEs with averaging time $\tau=10\tau_1$. Thick lines correspond to pdfs whose support include zero. Panel \protect\subref{FTLE30}: Same as \protect\subref{FTLE10}, with $\tau=30\tau_1$. Panel \protect\subref{FTLE100}: Same as \protect\subref{FTLE10}, with $\tau=100\tau_1$. Panel \protect\subref{KS}: Distribution of the sum of the first four backward FTLEs for different averaging times $\tau$. Panel \protect\subref{LM}: Distribution of the largest FTLE for different averaging times. Panel \protect\subref{UDV}: Distribution of the number of positive FTLEs for different averaging times $\tau$.}\label{FTLE}
\end{figure*}

\subsection{Finite-Time Lyapunov Exponents and Shadowing Unstable Periodic Orbits}
\label{section: FTLE fluctuations in terms of UPOs}
We now wish to provide an interpretation of such variability in terms of UPOs. Our intuition is that the local stability properties of the tangent space, measured in terms of the values of the FTLEs, are somehow encoded in neighbouring UPOs populating that same region of the attractor. 
In what follows, it is important to keep in mind that UPOs are large-scale structures in the phase space of the system and have a very long period compared to the typical shadowing times. 
With reference to the notation and framework defined in Section \ref{UPOs}, let's suppose that {\color{black}between the times $t_k$ and $t_k+\tau_k$ the chaotic trajectory $x(t)$ is being shadowed by the UPO $U_k$}, 
before being approximated by another UPO $U_h$ starting at time $t_h = t_k + \tau_k$. We then have a sequence of shadowing orbits $U_k$, each one associated to a  persistence time $\tau_k$. 
{\color{black}We then compute the spectrum of the FTLEs of the chaotic trajectory between $t_k$ and $t_k+\tau_k$}, and investigate the correlation with the corresponding LEs 
of the shadowing UPO $U_k$. Please note that each orbit might be considered more than once when looking at the correlations. Note also that the values of the time intervals $\tau_k$  change substantially along the trajectory, hence the various considered FTLEs are in general computed for different time horizons.


\begin{table}[]
\centering
\begin{tabular}{l|llll}
\hline
K                           & 1  & 10 & 100 & 1000 \\ \hline
$\lambda_{max}^\tau$          & 0.23 & 0.27 & 0.33  & 0.39   \\
$\lambda_1^\tau$        & 0.15 & 0.17 & 0.22  & 0.27   \\ \hline
$h_{KS,+}^\tau$ & 0.34 & 0.39 & 0.45  & 0.53   \\
$h_{KS}^\tau$        & 0.25 & 0.30 & 0.36 & 0.44   \\ \hline
\end{tabular}
\caption{Temporal correlation between the local properties of the chaotic trajectory and relative shadowing UPOs.\label{FTLEs_corre_table}}
\end{table}

The results are presented in Table \ref{FTLEs_corre_table}. 
There is a weak yet positive linear correlation between the first LE $\lambda_1$ of the shadowing UPOs and the corresponding first FTLE $\lambda_1^{\tau}$  when considering the orbits shadowing the trajectory in the first tier. The correlation is stronger if, instead, we consider the largest local LE $\lambda_{max}^{\tau}$ {\color{black}(i.e. the largest one obtained after reordering the local FTLEs)}. We need to remember that we are comparing two very different objects: a local property with a global structure. The link between local and global properties is clearer when we consider a measure of instability that encompasses the whole unstable manifold. The linear correlation between the sum of the first four FTLE ($KS^{\tau})$ and the Kolmogorov-Sinai entropy of the shadowing UPOs (first tier) is about 0.25, whilst a higher value (0.34) is found when considering the sum of the local positive FTLEs ($KS_+^{\tau}$). 
{\color{black}As we relax our definition of shadowing according to what described in Sect. \ref{section: shadowing}, we obtain that as we consider larger and larger values of $K$, the correlations between the corresponding local and global properties increases}. 
Such an increase in the correlation derives from the fact that higher values of $K$ result in longer shadowing times for the UPOs (Fig. \ref{persistence_tiers}), while maintaining good proximity to the trajectory (Fig. \ref{distance_tiers}).

\begin{figure*}
    \subfloat[]{%
      \includegraphics[trim={1cm 0 1cm 0},clip,width=0.45\textwidth]{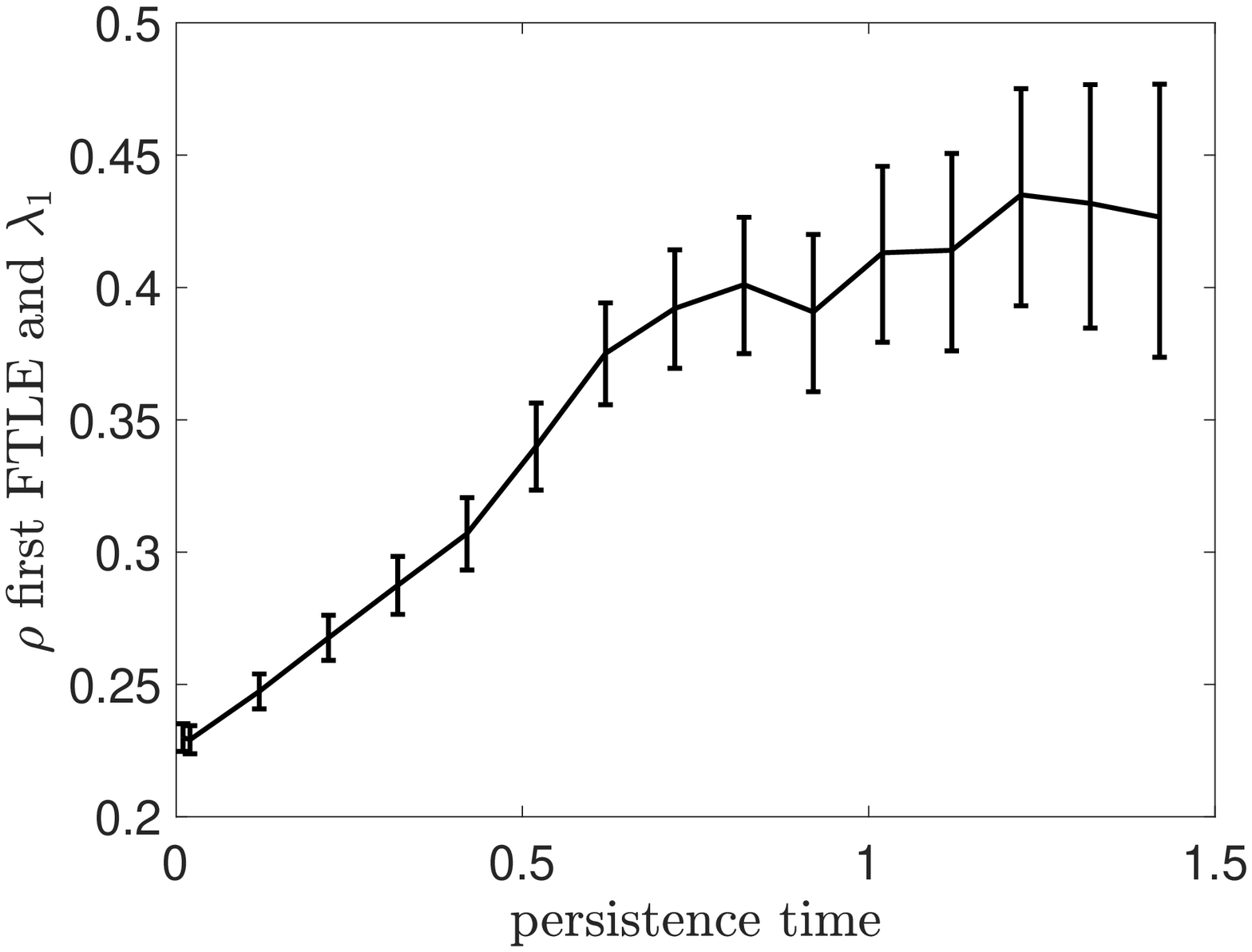}
           \label{corr_lambda1}
    }
    \qquad \qquad
     \subfloat[]{%
   \includegraphics[trim={1cm 0 1cm 0},clip,width=0.45\textwidth]{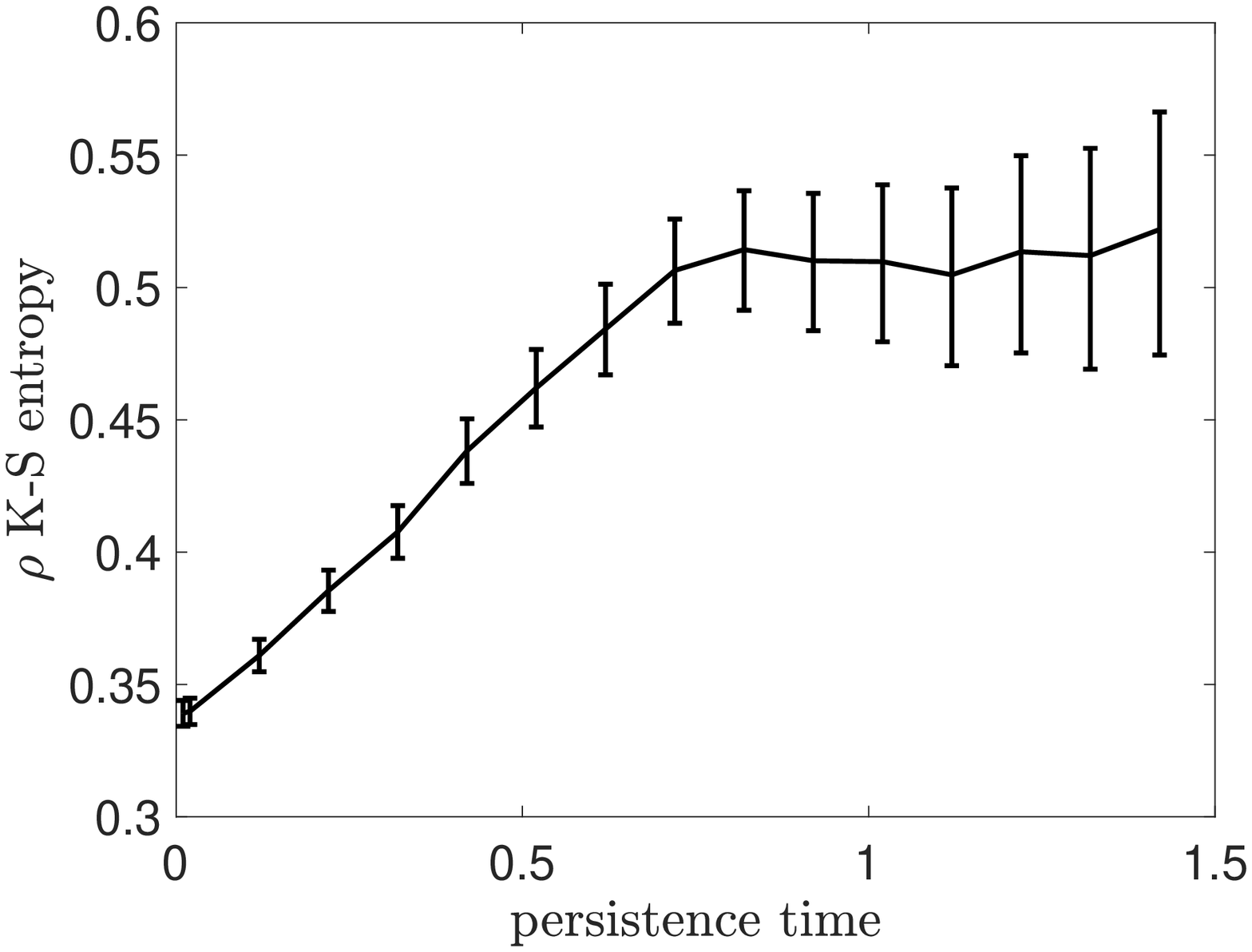}
              \label{corr_KS}
    }
    
        \subfloat[]{%
      \includegraphics[trim={1cm 0 1cm 0},clip,width=0.45\textwidth]{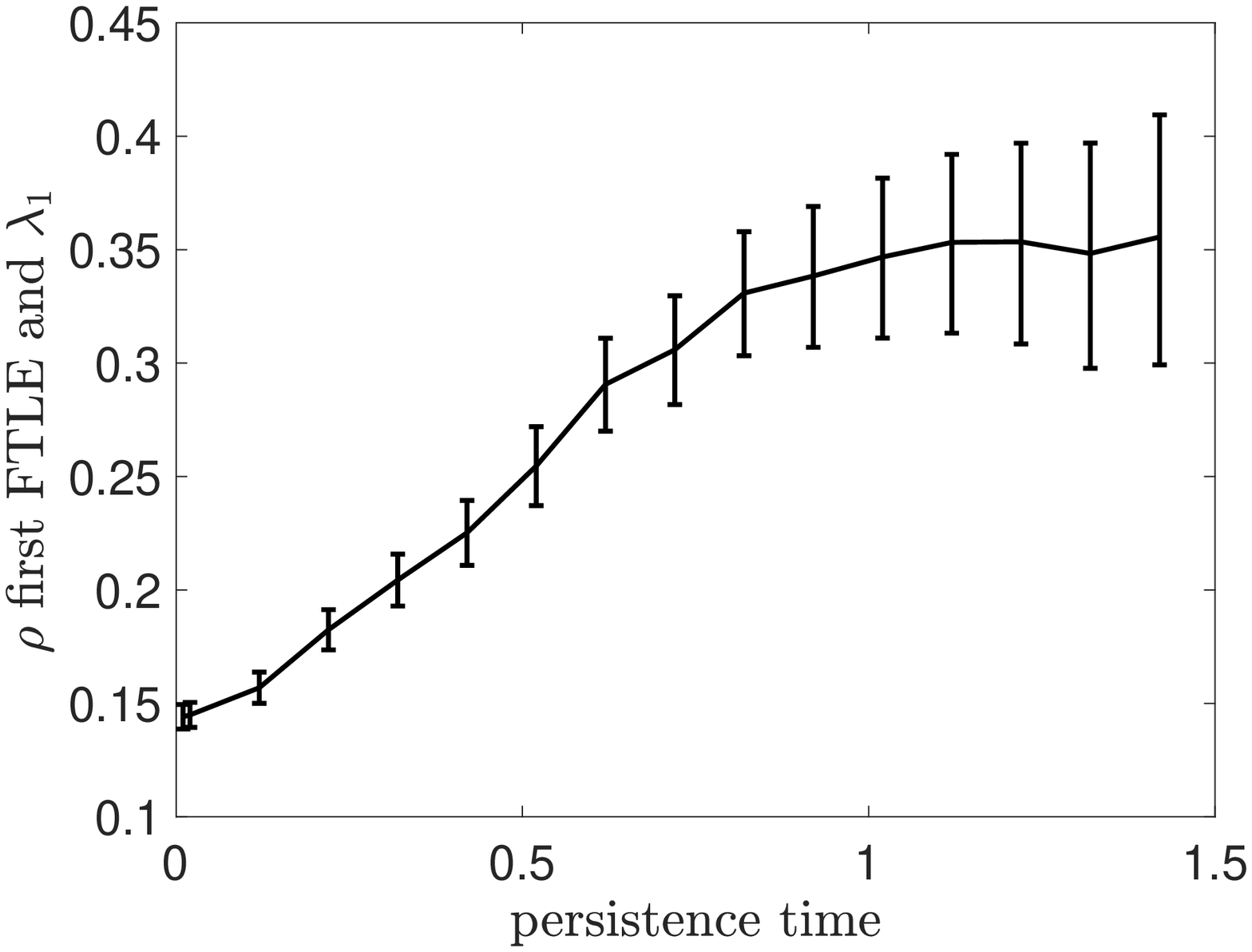}
           \label{corr_lambda1or}
    }
    \qquad \qquad
     \subfloat[]{%
   \includegraphics[trim={1cm 0 1cm 0},clip,width=0.45\textwidth]{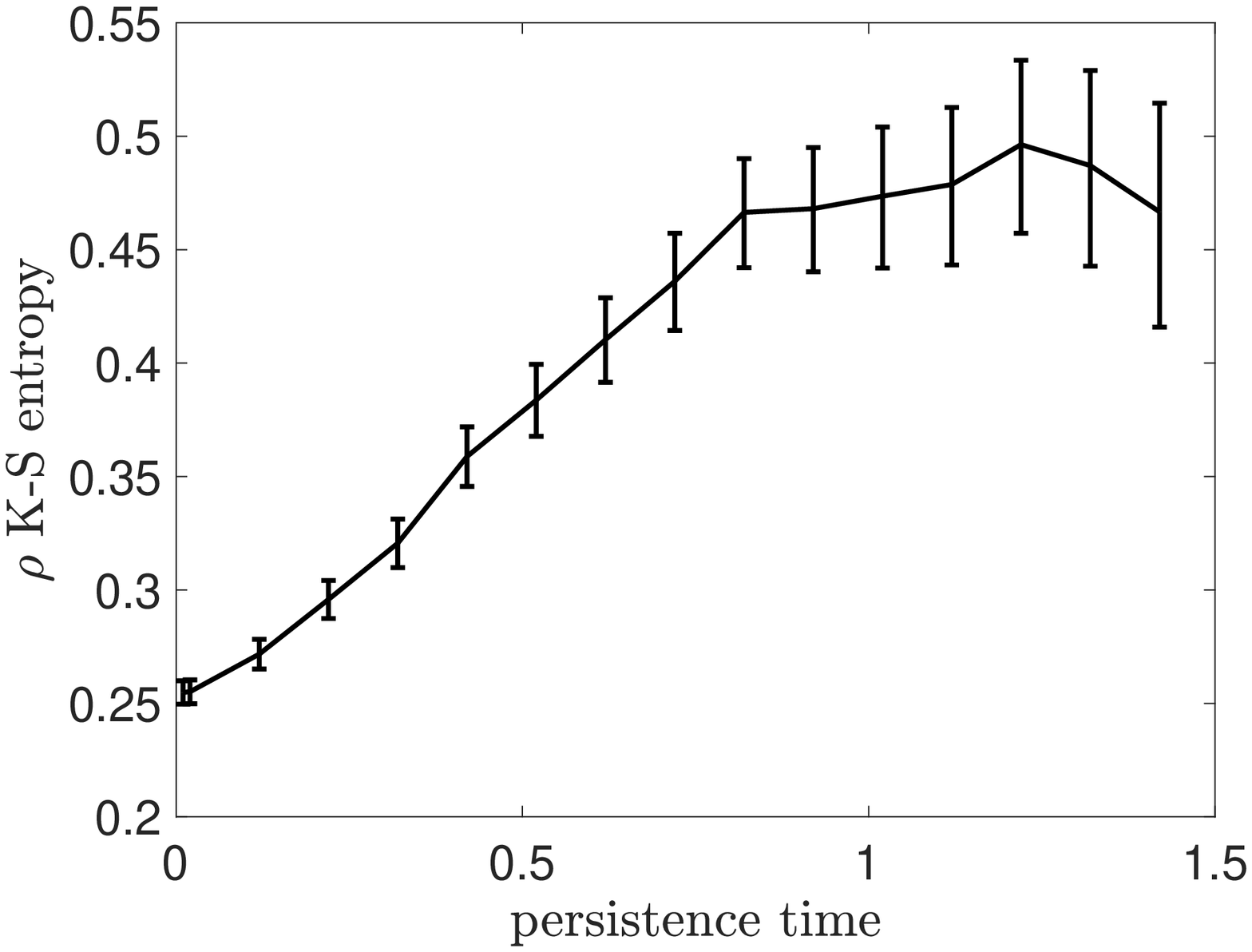}
              \label{corr_KSor}
    }
       \caption{{\color{black}Panel \protect\subref{corr_lambda1}: correlation between the first FTLE $\lambda_1^{\tau}$ and $\lambda_1$ of the corresponding shadowing UPO as a function of the minimum persistence time of the UPO. Panel \protect\subref{corr_KS}: correlation between $h_{KS}^{\tau}$ (sum of the first four FTLE) and $h_{KS}$ of the corresponding shadowing UPO as a function of the minimum persistence time of the UPO.} Panel \protect\subref{corr_lambda1or}: Same as \protect\subref{corr_lambda1}, for the largest FTLE $\lambda_{max}^\tau$. Panel \protect\subref{corr_KSor}: Same as \protect\subref{corr_KS}, for the local Kolmogorov-Sinai entropy $h_{KS,+}^{\tau}$(sum of the positive FTLE). The error bar corresponds to the $95\%$ confidence interval around the correlation value. {\color{black}The leftmost point of the diagrams corresponds to the values reported in Table \ref{FTLEs_corre_table} for $K=1$.}} 
  \label{correlation_persistence} 
\end{figure*}


{\color{black}It is possible to further explore such an aspect by looking in more detail into the properties of the shadowing UPOs chosen according to the strictest $K=1$ criterion. 
We then investigate how the correlation between the local instability properties of the chaotic trajectory and those of the shadowing UPOs change as we consider only UPOs that shadow the trajectory for longer and longer time durations. This allows us to restrict our analysis to what we may consider as the better performing UPOs.} 
Fig. \ref{correlation_persistence} shows that the link between the local instability properties of the trajectory and the UPOs steadily grows stronger as we consider {\color{black}UPOs that shadow for longer and longer times}. As an example, the linear correlation between the first LE of the UPO $\lambda_1$ and the first FTLE $\lambda_1^{\tau}$ of the chaotic trajectory increases from $0.15$ - full database - to around $0.35$ - when considering only UPOs that shadow for a time duration larger than 1. Of course, by definition, the number of UPOs considered for the statistics decreases as we demand more persistence (see Fig. \ref{persistence_tiers}), hence the uncertainty in our estimates grows as the UPOs dataset shrinks in size. Analogously, the linear correlations between the Kolmogorov-Sinai entropy of the shadowing UPO and $KS^{\tau}$ increases from  $0.25$ to around $0.50$. {\color{black}While these correlations might appear as relatively weak, we remark again that we are studying the link between local and global properties. Our results support the conjecture that the local stability properties of the flow can indeed be explained in terms of the properties of the shadowing UPOs. We also remark that, as shown in Fig. 2 of the Supplementary Material, the correlations shown in Fig. \ref{correlation_persistence} are substantially degraded if one considers the  database of $T<6.4$ UPOs, as a result of the much reduced ability to sample accurately the attractor of the system that we have already discussed earlier in the paper. }

\subsection{Unstable Dimension Variability and Quality of the Shadowing}

\label{section: shadowing quality}
The UDV is associated with another important phenomenon, namely the 
so-called shadowing breakdown \cite{dawson1994obstructions}. This refers to the fact that, whereas for hyperbolic system, the Anosov \cite{anosov1967geodesic} and Bowen's \cite{bowen_1975} shadowing lemma guarantees, roughly speaking, that each 
pseudo-orbit - e.g. the output of a numerical model of the system - stays uniformly close to some true trajectory, 
this does not necessarily holds true for non-hyperbolic systems. However, even though it is not possible to know in detail whether a computer simulated trajectory is representative of a true trajectory of the system, one can estimate for how long the numerical trajectory remains close to the true one \cite{Sauer1997,sauer1997long}. 
In general, it is expected that  the distance between the true trajectory and the pseudo-orbit increases when {\color{black}the trajectory goes through a glitch point, i.e when it performs a transition between regions featuring a different UDs number  \cite{dawson1994obstructions,sauer1997long}.} 

We wish to test whether a signature of the glitch points emerges as a deterioration of the ranked shadowing properties of the UPOs. 
We thus look at the statistics of distances between UPOs and the chaotic trajectory at the transitions points, i.e. when a new UPO takes over as best {\color{black}local approximation of the trajectory}. We distinguish between the case where the new shadowing UPO has the same number of positive LEs as the old one or not. If a change in the UD occurs, {\color{black}the trajectory goes through a glitch point, as in the case shown in Fig. \ref{waves}, where at time $t_1$ the UDs number of the shadowing UPO decreases from 5 to 4. Indeed, the distribution of distances shown in Fig. \ref{ddistribution} are statistically different at the $99\%$ confidence level as determined by the Kolmogorov-Smirnov and the average distance between the trajectory and the shadowing UPO is larger than in the case of glitch point. Clearly, since we are focusing on transition points, the quality of the shadowing is in both cases slightly lower than on average. } 




In order to better characterize the transitions points, we investigate the properties of the UPOs belonging to the first tiers {\color{black}- i.e. the best local approximations to the trajectory -} and check how similar they are in terms of number of UD. 
%
{\color{black}Figure \ref{variance100} shows that  } the variance in the number of UD  of the UPOs across the first $K=100$ tiers is  higher ($99.9\%$ confidence level) when considering glitch rather than non-glitch transition points. Additionally, the variance of UD is much smaller away from the transitions points. 
We propose the following interpretation. During each  shadowing window, the chaotic trajectory is surrounded by UPOs that are dynamically similar, so that low 
variability in the number of UD is found. 
When the trajectory approaches a transition point where the shadowing UPO changes, the UPOs around the chaotic trajectory become less homogeneous and the variability in the number of UD increases. 
At a glitch point, the hetereogeneity of the UPOs is higher because the system is {\color{black}performing a transition} between two qualitatively different regions of the phase space.

\begin{figure}[h]
    \subfloat[]{%
      \includegraphics[trim={0cm 0 0cm 0},clip,width=0.45\textwidth]{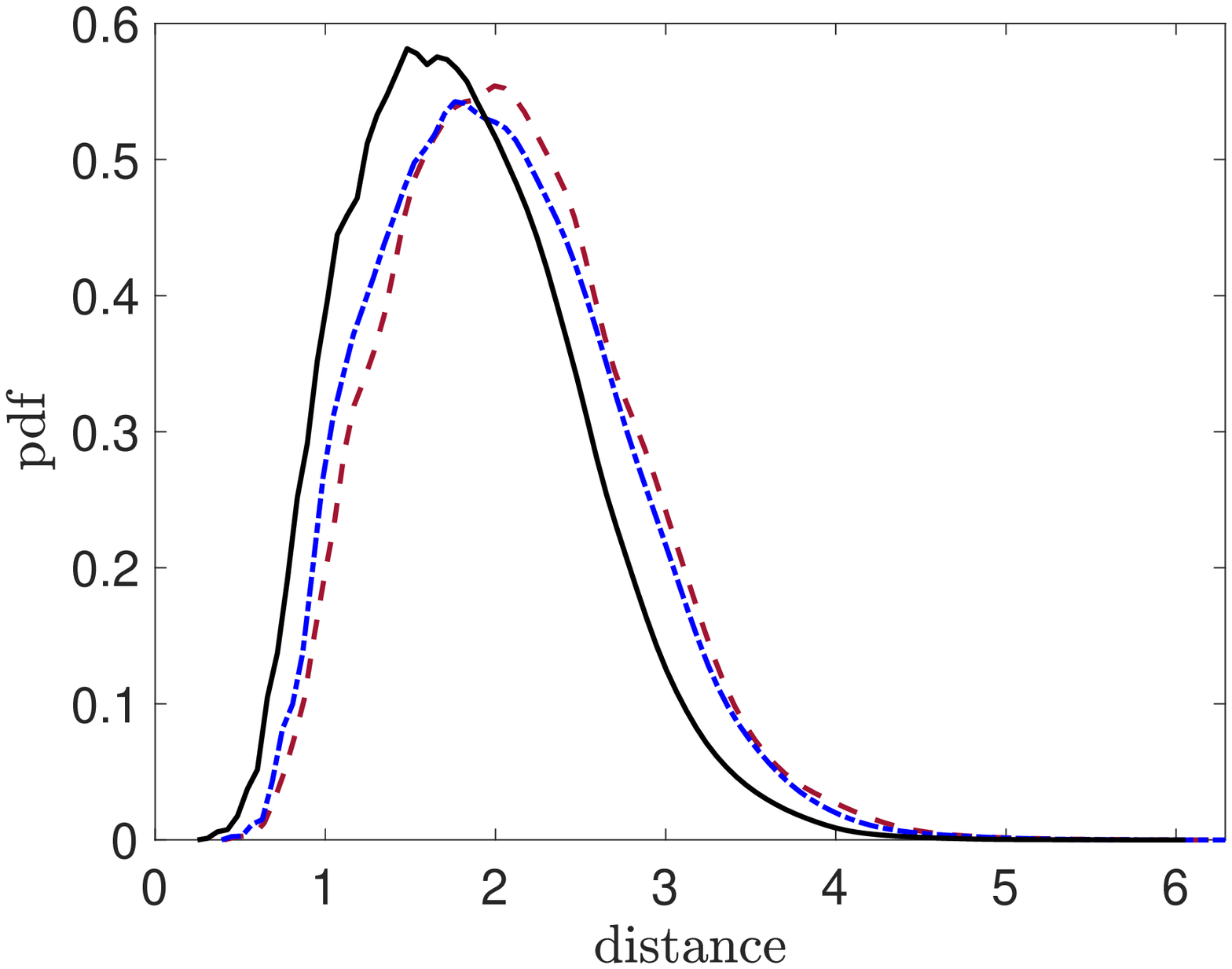}
           \label{ddistribution}
    }
    \qquad \qquad
     \subfloat[]{%
   \includegraphics[trim={0cm 0 0cm 0},clip,width=0.45\textwidth]{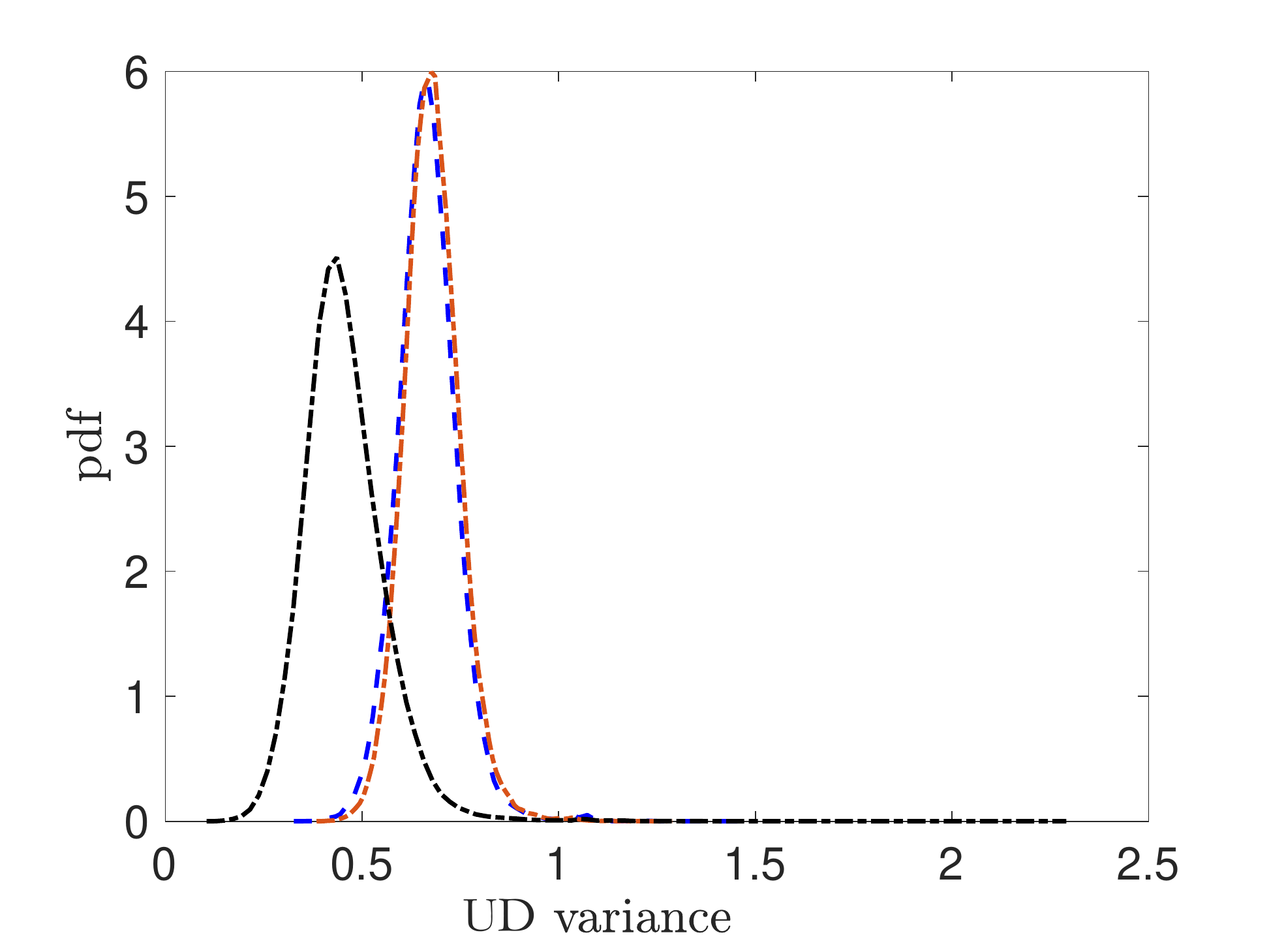}
              \label{variance100}
    }
\caption{Panel \protect\subref{ddistribution}: Probability distribution function for the distance of the first tier orbit limited to transition points: dark red dashed line refers to transitions associated to glitch points (mean distance 2.11), blue dotted line refers to transitions between orbits with the same UDs number  (mean distance 2.03). In black probability distribution function for the distance of the first tier orbit along the full chaotic trajectory (mean distance 1.81).  Panel \protect\subref{variance100}: Probability distribution function of the variance in the number of UD of the shadowing orbits within the first $K=100$ tiers. In black variance across the entire chaotic trajectory (mean variance 0.45), in blue variance associated to transition points with same UDs number (non-glitch points) (mean variance 0.66), in dark red variance in the UDs number at glitch points (mean variance 0.67). \label{fig: distance transitions}}
\end{figure}


\section{Coarse-Grained Dynamics via Finite-State Markov Chains}\label{statistics}
\subsection{A Statistical Analysis of Transition Points}
We take here a different angle for exploring the heterogeneity of the attractor of the system.  We 
propose a coarse grained representation of the dynamics by constructing 
 a finite state Markov chain process \cite{livi_2017}. {\color{black}Our approach differs from previous analysis \cite{froyland_2001, froyland_2003} because of the way we perform the partition of the phase space. Here, the neighbourhood of all UPOs with same number of UD are considered as a single state, and the mechanism of ranked shadowing outlined in Section \ref{section: shadowing} dictates the sequence of transitions from one state to another other.} 


We proceed as follows. 
We consider the $K=1$ shadowing algorithm, which  at each time-step $t$ selects the UPO $U_k$ of the full database that minimises the distance with the chaotic trajectory. We define the states $S=\{2,3,4,5,6,7,8\}$, where each $s_i \in S$ is the UDs number of the shadowing {\color{black}UPO. We say} that the system is in state $s_i$ at time $t$ if $s_i$ is the UDs number of the shadowing UPO at time $t$. Note that we have excluded the state corresponding to the few detected UPOs with 9 positive LEs because it makes little sense to include it in a statistical analysis.
The stochastic variable $\varsigma:\{1, ..., N_{max}\}\subset \mathcal{N}\rightarrow S$, describes the discrete Markov chain process as     $\varsigma(t)=s_i$, 
and the stochastic matrix describing the process can then be inferred in a frequentist way as 
{\color{black}\begin{equation}
    P_{i,j}^{dt} = \frac{\#\{ k : (\varsigma(k) = s_j)\wedge ( \varsigma (k+1)=s_i)\}}{\#\{ k : (\varsigma(k) = s_j\}}
\end{equation}}
where $\#$ {\color{black}denotes the cardinality and the denominator ensures that the matrix is properly normalised.}

Through the study of the spectral properties of the stochastic matrix describing the process it is possible to obtain information on the diffusion properties of the system. In particular, under the assumption of ergodicity, the first eigenvector $w_1$ corresponding to the unitary eigenvalue $\nu_1=1$ determines the unique invariant measure that will be attained exponentially fast and independently {\color{black}of the initial ensemble. Each subdominant eigenvector $w_k$ with corresponding eigenvalue $\nu_k$ sums to zero and describes a mode of the anomaly in the measure. The decay occurs on the timescale $\tau_k = -\frac{dt}{\ln\Re\{\nu^{(k)}\}}$,} where $\Re\{c\}$ indicates the real part of the complex number $c$. In particular, the process of relaxation to the invariant measure from a generic initial condition is dominated by the longest timescale $\tau_2 = -\frac{dt}{\ln\Re\{\nu^{(2)}\} }$.  

{\color{black}The first four subdominant eigenvalues are real.}  We have verified that $P^{dt}$ is ergodic and tested the Markovianity of the process by verifying that the eigenvalues of the matrix $P_{i,j}^{dt\times n}$ obtained by sampling the shadowing every $n$ time-steps has eigenvalues that scale with the $n^{th}$ power. {\color{black}Specifically, the relative normalised difference between the $10^{th}$ power of the eigenvalues of $P_{i,j}$ and the eigenvalues of the stochastic matrix obtained sampling the process every 10 time-steps assumes a value of about $0.5\%$.  We also verified that the  eigenvectors are very similar to those of $P_{i,j}^{dt\times n}$ ($w_k \cdot w_k^{dt\times n} \approx 1$ for $n=10$ $\forall k$}. We propose that such a coarse-grained representation of the dynamics characterises the statistics of the switching behaviour between clusters of UPOs with the same number of UD.

Figure \ref{fig:  transition matrix} shows the first five eigenvectors of $P^{dt}$, orderered according to the eigenvalue. The first eigenvector (in blue) returns the unique invariant measure of the system. Most often UPOs feature $UD=4$ and $UD=5$, in agreement with the distribution presented in Fig. \ref{positive_LE}. On the other hand, the modes corresponding to $\nu_2$ (in orange, $\tau_2 = 0.3982$) and $\nu_3$ (in yellow, $\tau_3=0.3134$) are responsible for the diffusion between anomalously stable and anomalously unstable UPOs. Indeed, both eigenvectors are characterised by negative components corresponding to $UD=3$ and $UD=4$, with positive values for larger UDs number. {\color{black}This means that if the initial ensemble is anomalously (un)stable, thus having an larger (smaller) number of members with high UDs number than dictated by the invariant measure, the anomaly will be damped mainly through these two modes.}   

The modes corresponding to $\nu_4$ and $\nu_5$ are responsible for the transitions between typical and atypical UPOs. In particular, the mode corresponding to  $\nu_4$ describes diffusion from UPOs with $UD=4$ to those with other values for UD. Finally, $\nu_5$ describes diffusion from UPOs with $UD=5$ to those with other values for UD.  It is also interesting to note that $\tau_4=0.2507>\tau_5 = 0.2138$, in agreement with the fact that the dominant mode of the invariant measure is given by $UD=4$.
\begin{figure}[h]
\centering
\includegraphics[width=10cm]{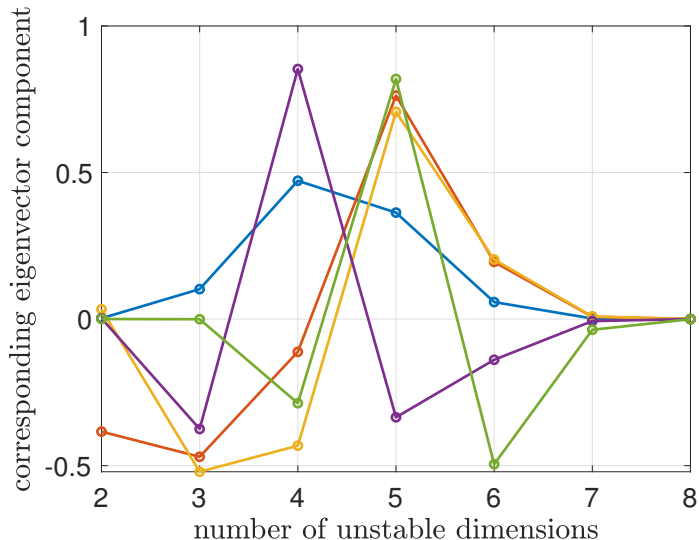}
\caption{\label{fig:  transition matrix} Eigenvectors of the transition matrix $P^{dt}$. For each eigenvector (represented by different colours) we represent the value of its different components corresponding to the different states of the system. $\nu_1=1$ (in blue) returns the invariant measure of the system and it is in agreement with the distribution presented in Fig. \ref{positive_LE}. The subdominant eigenvalues are $\nu_2=0.9752$ in orange, $\nu_3=0.9686$ in yellow, $\nu_2=0.9608$ in purple and $\nu_2=0.9543$ in green.  }
\end{figure}

Using Eq. (13) in \cite{gaspard2004}, it is possible to evaluate the KS entropy $h^P_{KS}$ of the finite state Markov chain described by the stochastic matrix $P_{i,j}^{dt}$. One obtains $h^P_{KS}\approx0.31$. {\color{black}This figure clearly indicate that we are investigating a system that creates information in the Shannon sense. Nonetheless, as a result of the procedure of coarse graining used for defining the stochastic matrix, the estimate of the KS entropy is lower than the value one can directly derive for the full system $h_{KS}\approx 1.23$, which is given by the sum of the first 4 LEs shown in Fig. \ref{global_LE}.}  


\subsection{Mixing between Large-Scale Regions of the Attractor with Different Instability}

\begin{figure*}
    \subfloat[]{%
      \includegraphics[trim={0cm 0cm .5cm 0cm},clip,width=0.46\textwidth]{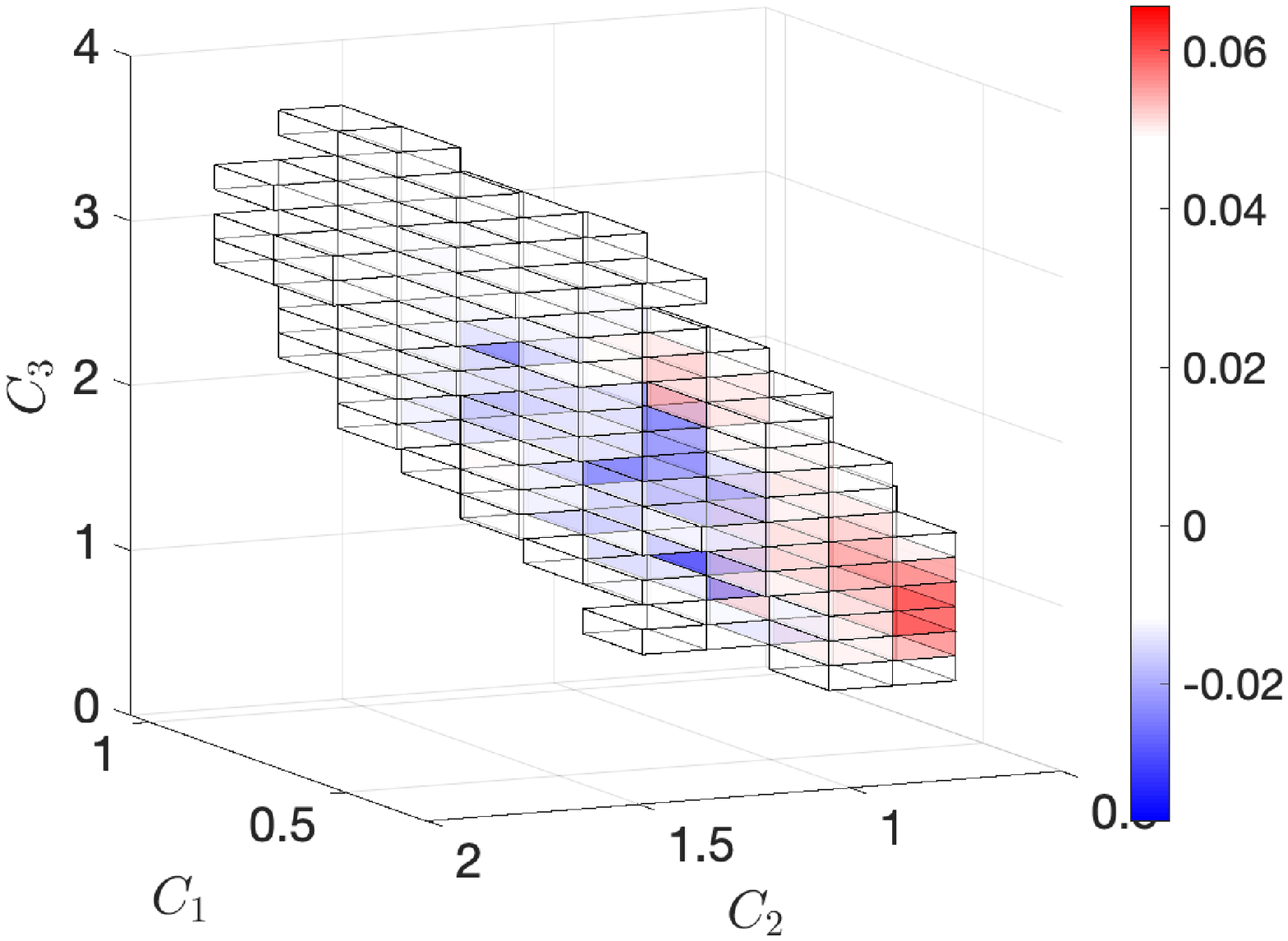}
           \label{l2}
    }
    \qquad \qquad
    \subfloat[]{%
   \includegraphics[trim={0cm 0cm .5cm 0cm},clip,width=0.46\textwidth]{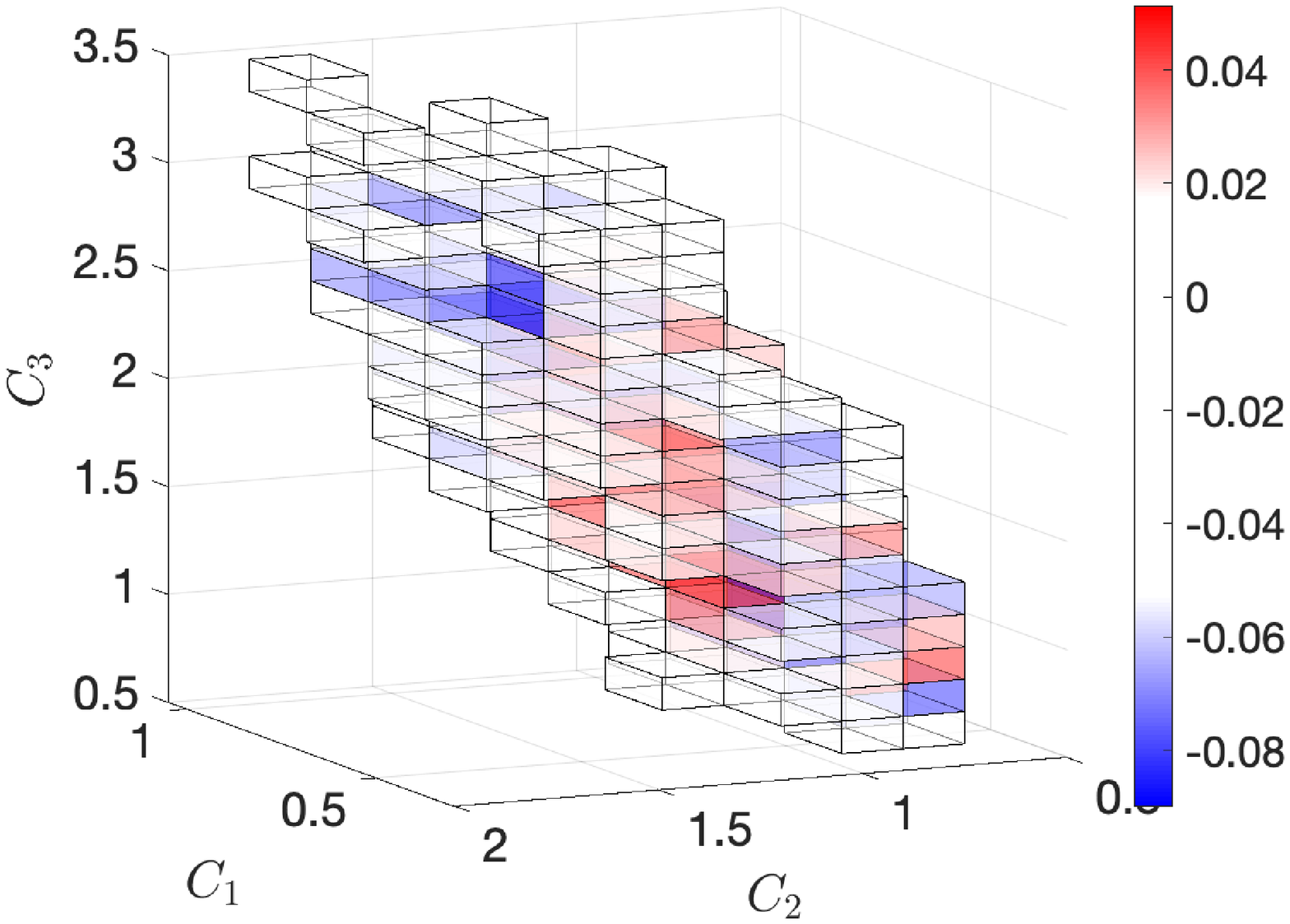}
              \label{l3}
    }\\
       \subfloat[]{%
   \includegraphics[trim={0cm 0cm .5cm 0cm},clip,width=0.46\textwidth]{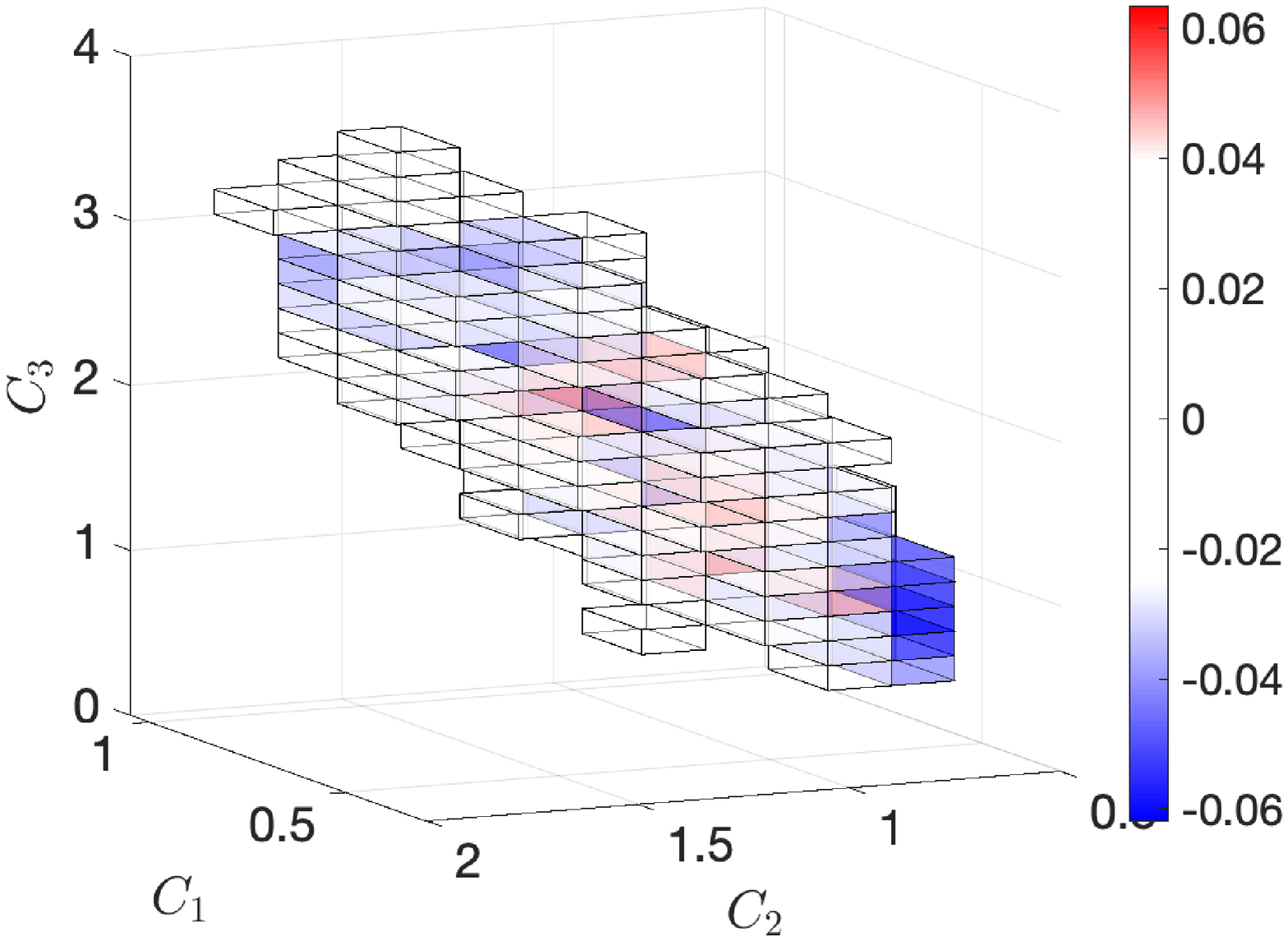}
              \label{l4}
    }
    \qquad \qquad
       \subfloat[]{%
   \includegraphics[trim={0cm 0cm .5cm 0cm},clip,width=0.46\textwidth]{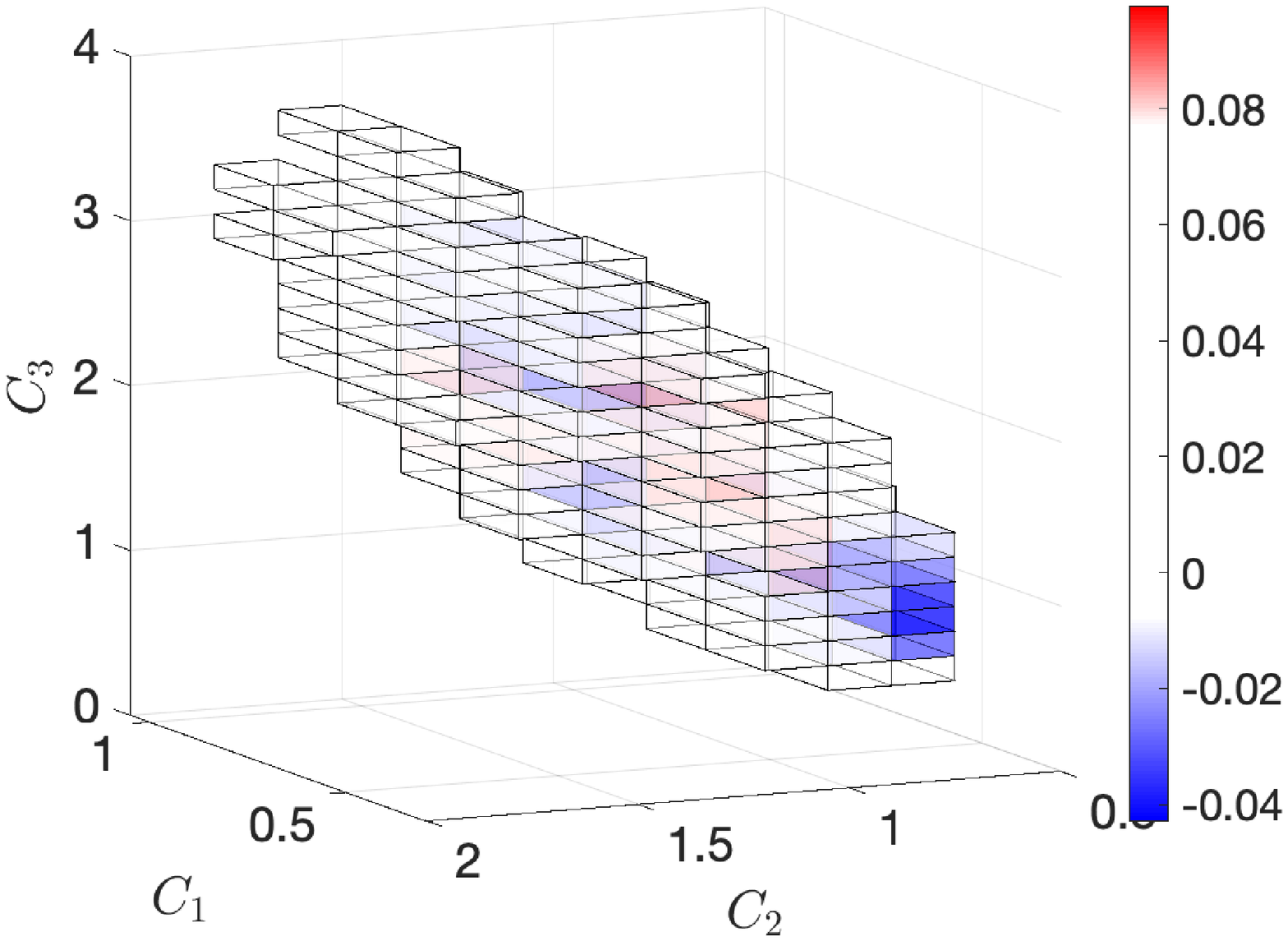}
              \label{l5}
    } 
 \caption{Projection of the subdominant eigenvectors (panel \protect\subref{l2}) $w^2$, (panel \protect\subref{l3}) $w^3$, (panel \protect\subref{l4}) $w^4$, (panel \protect\subref{l5}) $w^5$. of the stochastic matrix $Q^{dt}_{i,j}$ (see Sect. \ref{statistics}) in the space defined  by the normalised moments $(C_1,C_2,C_3)$ (see Eq. \ref{cs}).\label{cubes}}
 \end{figure*}
We now wish to investigate the relaxation {\color{black}of an ensemble} towards the invariant measure through a different {\color{black}coarse-graining} procedure. 
Following \cite{maiocchi_2022}, we construct a Markov chain where the states are the neighbourhoods of the shadowing UPOs. 
{\color{black}We define here the states $S=\{1,\ldots,P\}$, where each $i \in S$ is index of the shadowing UPO. The system is in state $i$ at time $t$ if the $i^{th}$ UPO is shadowing the chaotic trajectory at time $t$. 
The stochastic variable $\varsigma:\{1, ..., N_{max}\}\subset \mathcal{N}\rightarrow S$, describes the discrete Markov chain process as     $\varsigma(t)=i$, 
and the stochastic matrix describing the process can then be inferred in a frequentist way by performing the the $K=1$ shadowing using $P$ UPOs and evaluating: 
\color{black}\begin{equation}\label{Qm}
    Q_{i,j}^{dt} = \frac{\#\{ k : (\varsigma(k) = j)\wedge ( \varsigma (k+1)=i)\}}{\#\{ k : (\varsigma(k) = j\}}
\end{equation}
Ideally, one like to use the entire dataset of UPOs, so that $P=N_{UPO}$. Yet, given the very large number of UPOs considered in this study and the amount of available data, it is impossible to robustly evaluate the stochastic matrix by considering all UPOs. Hence, we choose as states of the system the neighbourhood of $P=1000$ UPOs. Such UPOs have been randomly chosen within the whole dataset in such a way to respect the distribution of periods as in the original database. In order to enforce robustness, we also verify that each UPO has been selected in the shadowing for at least the length of its period. We remark that, given the protocol above, the results presented below are weakly dependent on the specific choice of UPOs for $P=1000$ as well as on the chosen value for $P$ . 

We have verified that the process described by $Q_{i,j}^{dt}$ is ergodic, and tested its Markovianity (in particular, similarly to what we described earlier, we estimated an error of $0.4\%$ when considering sampling every $10$ time-steps. We also verified that the  eigenvectors are very similar to those of $Q_{i,j}^{dt\times n}$ ($w_k \cdot w_k^{dt\times n} \approx 1$ within less than $1\%$ for $n=10$ $\forall k$)). }
The first eigenvalues are $\nu_1=1, \nu_2 = 0.9887, \nu_3=0.9848,  \nu_4 = 0.9844,\nu_5 = 0.9831$ and the corresponding characteristic decay times are $\tau_2 = 0.8771,\tau_3 = 0.6511,\tau_4 = 0.6348,\tau_5 = 0.5850$. The KS entropy production rate of the stochastic matrix $Q_{i,j}^{dt}$ is $h^Q_{ks}\approx 0.15$: also in this case, {\color{black}one finds clear evidence of chaotic behaviour but underestimates the rate of creation of information of the system.} 


We want to represent the eigenvectors of the stochastic matrix in the projected space spanned by the normalised moments $C_1$, $C_2$, and $C_3$ introduced in Eq. \ref{cs} and used in Fig. \ref{energy_numberUD}. We proceed as follows.
We consider the compact set in $\mathbb{R}^3$ that contains the projected attractor of the system and take into account a partition in $5288$ cubes of side $0.15$. Each projected UPO of the system intersects a certain number of cubes, and each cube of the partition might contain contributions from different orbits. We measure the quantity of mass contained in each cube accordingly to the distribution given by the different subdominant eigenvectors. In order to do so, we set a fixed number of points $N$ to be represented in phase space, and we distribute such points to each UPO accordingly to the weight given by the corresponding component of the subdominant eigenvector $w^{(k)}$. For each UPO we represent the selected points equally spaced in time in the phase space and distinguish between negative and positive contributions. We finally quantify the total amount of mass contained in each cube of the partition by computing the algebraic sum of the points contained in it.

Figure \ref{cubes} presents the outcome of such a procedure for the eigenvectors corresponding to the first four subdominant eigenvalues. 
%
%
{\color{black}By looking at pattern of positive and negative values, one realizes} that the eigenvector corresponding to the first  subdominant eigenvalue $\nu_2$ - see Fig. \ref{l2} - is responsible for the transfer of mass between regions characterised by low values of the moments and the core of the attractor{\color{black}, compare with Figs. \ref{UPOs} and \ref{density}.} Considering that there is a clear relationship between the value of the moments and instability of the UPOs that preferentially populate the corresponding region of the attractor {\color{black}- see Figs. \ref{M1}-\ref{M3} - this is in good agreement with Fig. \ref{fig:  transition matrix}, where it is shown} that the first subdominant eigenvector describes the transfer of mass between UPOs with low instability and those with typical instability. The eigenvector corresponding to $\nu_3$ - see Fig. \ref{l3}  describes the transfer of mass between anomalously high and anomalously low values of the moments, also in qualitative agreement with the results shown in in Fig. \ref{fig:  transition matrix}. The eigenvectors corresponding to  $\nu_4$  and $\nu_5$ - see Figs. \ref{l4} and \ref{l5}, respectively - describe the transfer of mass between typical and atypical (anomalously high and anomalously low) values of the moments. Also in this case a good agreement is found with the corresponding eigenvectors in Fig. \ref{fig:  transition matrix}.

{\color{black}We remark that if we perform this analysis taking as starting point the reduced set of $T<6.4$ UPOs, unsurprisingly, the resulting subdominant eigenvector is rather different from the one depicted in Fig. \ref{l2}  because low-period UPOs are unable to sample accurately the low-energy region of the attractor, see Fig. 3 in the Supplementary Material.}



\section{Conclusions}\label{conclusions}
The use of  Lyapunov analysis is well accepted as a key tool for understanding the dynamical and statistical properties of complex systems \cite{pikovsky2016lyapunov}. In this paper we suggest that this effort should be complemented by taking a global approach based on the study of the skeletal dynamics associated with the UPOs \cite{cvitanovic_1988}. While {\color{black}their} computation is a challenging task \cite{chandler_2013}, {\color{black}UPOs indeed} have a great potential for clarifying  features and dynamical processes in complex systems \cite{cvitanovic_2013,lucarini_2020, kawahara_2001, suri2020capturing}.

As a step in the direction of better understanding the properties of  chaos in high-dimensional non-hyperbolic systems, we have here investigated strong violations of hyperbolicity in a specific version of the Lorenz '96 model \cite{lorenz_1996,Lorenz2005} with 20 degrees of freedom, taken as prototypical example of chaotic system with a nontrivial unstable manifold, i.e. with a dimensionality larger than one. We compute an extremely large set of UPOs immersed in the attractor, covering a large range of periods and geometrical lengths. While such a set is clearly incomplete, it allows for a rather accurate shadowing of the trajectory. We have verified that the system features UDV, where the UDs number ranges (at least) from 2 to 9. Fingerprints of the UDV are also found 
when performing a Lyapunov analysis of the system: several FTLEs fluctuate around zero also when very long averaging times are considered. Our results extend previous findings \cite{Sauer1997,Sauer2002,Pereira2007} because in our system we observe a much more pronounced UDV, as the heterogeneity of the UPOs is very high. 

We have also found that the local stability properties of the tangent space, measured in terms of their FTLES, are highly correlated to those of the UPOs, measured by their corresponding LEs, populating the same region of the phase space. This is a nontrivial result if one considers the fact that UPOs describe global features on the attractor.  The agreement becomes more evident as we focus our statistics on UPOs that shadow the trajectory for a long time.

We can rank the UPOs according to how close they are to the trajectory. While the closest orbit typically keeps its rank for a relatively short time, we find that  often the forthcoming rank 1 orbit has the same UDs number as the previous one. The less frequent transitions to the neighborhood of an UPO with a different UDs number - glitch points - are usually accompanied by a degradation of the quality of the shadowing. 

On slower time scales, the systems performs transitions between regions of the attractor associated with anomalously high and low instability, which are preferentially populated by UPOs with large vs low UDs number, respectively. Following \cite{maiocchi_2022}, we construct coarse-grained versions of the system based on using the neighbourhoods of the UPOs as building blocks and the dynamics as generator of the random process. It is then possible to see such transitions as slow relaxation processes of an arbitrary initial {\color{black}ensemble} towards the invariant measure, with the subdominant modes of the transfer operator being associated with slow fluctuations between high instability and low instability states. High (low) energy is typically associated with high (low) instability, in agreement with the thermodynamical understanding of the model. Regions associated with anomalous energy and corresponding anomalous instability can be though of as corresponding, by and large, to quasi-invariant sets of the system \cite{froyland_2014}. 

{\color{black}In a (relatively) high-dimensional chaotic systems like the one considered it is extremely hard if not practically impossible to compute all UPOs up to a given period T. The task of finding long-period UPOs becomes particularly daunting in the case of highly unstable orbits. Hence, our UPOs database is incomplete and biased. One might reasonably ask whether it is better to restrict our analysis to low-period UPOs, which are better sampled. Indeed, our results show the extreme importance of taking into account also the poorly sampled long-period UPOs. In fact, low-period UPOs cover poorly a specific region (the low-energy one) of the attractor. Hence, restricting our analysis to such orbits leads to a serious degradation of the quality of the shadowing and to losing key information on the dynamical and statistical properties of the system, in agreement with previous results \cite{lasagna_2020,yalniz_2020}. 

While we have focused on a rather specific model, our findings  wish to stimulate more general investigations on the link between the heterogeneity of the attractor  and the breakdown of hyperbolicity}. Such a heterogeneity has important implications in terms of robustness of the system, of the accuracy of the numerical models used to approximate its evolution, and of the efficacy of data assimilation strategies aimed at optimally merging observations and dynamics, as discussed in \cite{lucarini_2020} in the context of geophysical flows. 

There is a clear link between hyperbolicity of a system and applicability of linear response theory for computing the change in its statistical properties resulting from a perturbation to its dynamics \cite{ruelle_1998,ruelle_2009}; see also more recent developments in  \cite{Baladi08,Baladi14}. Hyperbolicity is also critical for the definition of rigorous algorithms aimed at implementing Ruelle's response formulas and evaluating separately the two contributions to the overall response coming from the stable and unstable components of the tangent space \cite{Ni2020,Chandramoorthy2020,SliwiakWang2022}. 
The serious breakdown of hyperbolicity discussed in this paper could reasonably be seen as a very  obstacle for the applicability of linear response formulas. Nonetheless, it has been amply shown that the linear response theory applies to a great degree of accuracy for the Lorenz '96 model, albeit in a slightly different configuration from the one considered here \cite{Lucarini2011}. As discussed in \cite{GhilLucarini2020}, linear response theory is very effective in providing accurate climate change projections using climate models of arbitrary level of complexity. There is no reasons to believe that such models are anywhere close to hyperbolicity. In fact the opposite seems to hold when the multiscale effects of atmosphere-ocean coupling is taken into consideration \cite{vannitsem2016statistical}. 

The clarification of the apparent mismatch between what is rigorously guaranteed by mathematical theorems and what is heuristically shown in multiple applications regarding the applicability of linear response theory in complex systems seems a topic of great scientific relevance both on theoretical grounds and for reasons of practical significance.


\section*{Acknowledgements}
The authors wish  to thank P. Cvitanovic, J. Dorrington, D. Faranda, G. Pavliotis, and J. Yorke for constant encouragement. {\color{black}The authors also acknowledge the constructive criticisms of  two anonymous reviewers.} VL acknowledges the support received from the EPSRC project EP/T018178/1 and from the EU Horizon 2020 project TiPES (grant no. 820970). CCM has been supported by an EPSRC studentship as part of the Centre for Doctoral Training in Mathematics of Planet Earth (grant number EP/L016613/1). AG was supported by the Moscow Center of Fundamental and Applied Mathematics at INM RAS (Agreement with the Ministry of Education and Science of the Russian Federation No.075-15-2022-286). {\color{black}YS was supported by JSPS Grant-in-Aid for Scientific Research (B) No. 21H01002.}

\section*{Data Availability}
{\color{black}The Supplementary Material of this publication, including extensive datasets and MATLAB codes, is available at Ref. \cite{Lucarini2023}}.

\appendix
\section{Mathematical framework}\label{appendix}
\subsection{Chaotic dynamical systems}\label{chaos}
Let us consider a continuous-time autonomous dynamical system $\dot{x}={f}({x})$ on a compact manifold $\mathcal{M} \subset \mathbb{R}^n$. We define a state at time $t \in  \mathbb{R}_{t>0}$ as ${x}(t) = S^t{x_0}$, where ${x_0} = {x}(0)$ is the initial condition and $S^t$ is the evolution operator. 
We assume that the system is dissipative (${\nabla} \cdot {f}<0$) with $\Omega \subset M$ compact attracting invariant set. The attractor supports a probability measure $\rho$, invariant and ergodic with respect  to $S^t$, specified in the following manner:
\begin{equation}
\langle \varphi \rangle = \int \rho(dx) \varphi (x) = \lim_{T\rightarrow \infty} \frac{1}{T} \int_0^T \varphi (S^t x_0) dt
\end{equation}
for any sufficiently regular function (observable) $\varphi:M \rightarrow \mathbb{R}$ and
for almost all initial conditions $x_0$ belonging to the basin of attraction of $\Omega$. 
\subsection{Lyapunov {\color{black}Exponents and Instability of the Flow}}\label{finiteLyap}
 We now want to characterise dynamical systems that feature sensitive dependence on initial conditions on the attractor. Following Pesin's theory \cite{Pesin1977}, this is most easily accomplished by using Lyapunov analysis \cite{Cencini2013,pikovsky2016lyapunov}. The  separation in time of infinitesimally nearby trajectories can be asymptotically quantified through specific global dynamical indicators \cite{pikovsky2016lyapunov}. 
We define the matrix
\begin{equation}
    \Lambda(x) = \lim_{t\to \infty} (J_t^\top(x)J_t(x))^{\frac{1}{2t}}
\end{equation}
where $J_t(x) = \nabla_x f (S^tx)$ is the {\color{black}tangent linear} matrix of the flow at time $t$ with initial condition $x\in \Omega$ and $\top$ indicates the transpose. It is possible to prove that the matrix {\color{black}$\Lambda(x)$} exists and its eigenvalues $\Lambda_i$ are real and constant almost everywhere with respect to the measure $\rho$. We call Lyapunov exponents (LE) of the systems the objects defined as $\lambda_i = \log(\Lambda_i)$. Usually they are ordered by size in descending order $\lambda_1 \geq \lambda_2 \geq \ldots \geq \lambda_n$ and a positive first LE $\lambda_1$ indicates that the system is chaotic. LEs are asymptotic quantities and refer to average properties over the attractor. The number of positive LEs is the number of UDs. The full Lyapunov spectrum allows to define the Kaplan-Yorke dimension, {\color{black}which is conjectured to correspond to the order 1 Renyi dimension \cite{ott_2002} of the attractor} as follows:
\begin{equation}
    D_{KY} = m + \frac{\sum_{i=1}^m \lambda_i}{|\lambda_{m+1}|}
\end{equation}
with $m$ being the highest index for which the sum of the largest $m$ LEs  is strictly positive. The quantity $D_{KY}$ can be thought as an approximate value of the number of excited degrees of freedom acting in the system \cite{FREDERICKSON1983185}.
The degree of chaoticity of a dynamical system can also be quantified via the Kolmogorov-Sinai entropy (approximated via Pesin's theorem in the case the invariant measure is of the Sinai-Ruelle-Bowen type) \cite{eckmann_1985} as:
\begin{equation}
    h_{KS} = \sum_{i=1}^n \lambda_i
\end{equation}
where $n$ is the number of positive LEs, which corresponds to the UDs number. The quantity $h_{KS}$ quantifies the production of information of the system. 

The local instability properties of the attractor are described by the  Finite Time Lyapunov Exponents (FTLE). These quantify the amount of stretching about the trajectory with initial condition $x$ on the attractor over a finite time interval $[0, t]$. They are local objects since their value depends on $x$ and $t$. They can be computed as the logarithm of the eigenvalues of the matrix $\Lambda(x,t) = (J_t^T(x) J_t(x))^{1/(2t)}$.
 One can also define as $h_{KS}^t= \sum_{i=1}^n \lambda_i^t$. The value of the $j^{th}$ FTLE can vary a lot along the attractor and in some cases, for a given $t$, the support of its probability distribution can include zero, meaning that the local FTLE can have a different sign compared to the corresponding $j^{th}$ LE. Correspondingly, the number of positive FTLEs can in general fluctuate and be different from $n$. Note also that $\lambda_1^t\leq\lambda_{max}^t$, where the latter is the largest among the $\lambda_j^t$'s and defines the local largest growth rate. Finally, one can define a local version of the Kolmogorov-Sinai entropy as $h_{KS,+}^t= \sum_{\{\lambda_i^t>0\}} \lambda_i^t\geq h_{KS}^t$.


\subsection{Unstable Periodic Orbits}\label{UPOsdef}
 A periodic orbit of period $T$ is an exact periodic solution of the evolution equation and it is defined as follows:
\begin{equation}
\label{periodicity}
S^T(x)=x.
\end{equation}
This representation is not unique. In fact, if equation \ref{periodicity} is satisfied, $S^{nT}(x)=x$ is verified as well $\forall n \in \mathbb{N}$ . In this work we will consider a periodic orbit to be identified by its prime period $T>0$ (we do not consider equilibria) and an initial condition $x_0$. The attractor of a chaotic dynamical systems is densely populated by unstable periodic orbits \cite{cvitanovic_2005}. It is known that a good understanding of the UPOs of the model plays a fundamental role in the characterisation of the system. As an example, it is possible to reconstruct the invariant measure of the system through the use of trace formulas, by considering the following expression for the average of any measurable observable $\varphi$: 
\begin{align}
\label{trace_formula}
\langle\varphi\rangle = \lim_{t \to \infty} \frac{\sum_{U^p,p\leq t} w^{U^p}\bar{\varphi}^{U^p}}{\sum_{U^p,p\leq t} w^{U^p}}
\end{align}
where $U^p$ is a UPO of prime period $p$, $w^{U^p}$ is its weight and $\bar{\varphi}^{U^p}$ is the average in time of the observable along the orbit. 
For uniformly hyperbolic dynamical systems this result is exact and the weight can be obtained, to a first approximation, by $w^{U^p}\propto \exp(-ph_{ks}^{U^p})$ \cite{grebogi_1988} , with $h_{ks}$ being the Kolmogorov-Sinai entropy of the system; see also discussion in \cite{Dhamala1999}.

\providecommand{\noopsort}[1]{}\providecommand{\singleletter}[1]{#1}%


\begin{thebibliography}{100}

\bibitem{Poincare1908}
H.~Poincar\'e, {\em Science et M\'ethode}.
\newblock Flammarion, Paris, 1908.

\bibitem{Ruelle1998}
D.~Ruelle, ``Henri poincaré's “science et méthode”,'' {\em Publications
  Mathématiques de l'IHÉS}, vol.~S88, pp.~179--181, 1998.

\bibitem{lorenz_1963}
E.~N. Lorenz, ``Deterministic nonperiodic flow,'' {\em Journal of atmospheric
  sciences}, vol.~20, no.~2, pp.~130--141, 1963.

\bibitem{Ruelle1971}
D.~Ruelle and F.~Takens, ``On the nature of turbulence,'' {\em Communications
  in Mathematical Physics}, vol.~20, no.~3, pp.~167--192, 1971.

\bibitem{Yorke1975}
T.-Y. Li and J.~A. Yorke, ``Period three implies chaos,'' {\em The American
  Mathematical Monthly}, vol.~82, no.~10, pp.~985--992, 1975.

\bibitem{hammerlindl_2019}
A.~Hammerlindl, B.~Krauskopf, G.~Mason, and H.~M. Osinga, ``Global manifold
  structure of a continuous-time heterodimensional cycle,'' {\em arXiv preprint
  arXiv:1906.11438}, 2019.

\bibitem{smale_1967}
S.~Smale {\em et~al.}, ``Differentiable dynamical systems,'' {\em Bulletin of
  the American mathematical Society}, vol.~73, no.~6, pp.~747--817, 1967.

\bibitem{anosov1967geodesic}
D.~V. Anosov, ``Geodesic flows on closed riemannian manifolds of negative
  curvature,'' {\em Trudy Matematicheskogo Instituta Imeni VA Steklova},
  vol.~90, pp.~3--210, 1967.

\bibitem{katok1995}
A.~Katok and B.~Hasselblatt, {\em Introduction to the Modern Theory of
  Dynamical Systems}.
\newblock Encyclopedia of Mathematics and its Applications, Cambridge
  University Press, 1995.

\bibitem{Ruelle_1989}
D.~Ruelle, {\em Chaotic evolution and strange attractors: the statistical
  analysis of time series for deterministic nonlinear systems}.
\newblock Lezioni Lincee, Cambridge, England: Cambridge University Press, 1989.

\bibitem{ruelle_1998}
D.~Ruelle, ``General linear response formula in statistical mechanics, and the
  fluctuation-dissipation theorem far from equilibrium,'' {\em Physics Letters
  A}, vol.~245, no.~3-4, pp.~220--224, 1998.

\bibitem{Bonatti_2011}
C.~Bonatti, ``Survey towards a global view of dynamical systems, for the
  c1-topology,'' {\em Ergodic Theory and Dynamical Systems}, vol.~31, no.~4,
  p.~959–993, 2011.

\bibitem{Pesin1977}
Y.~B. Pesin, ``{Characteristic} {Lyapunov} {exponents} {and} {smooth} {ergodic}
  {theory},'' {\em Russian Mathematical Surveys}, vol.~32, pp.~55--114, aug
  1977.

\bibitem{Benettin1980}
G.~Benettin, L.~Galgani, A.~Giorgilli, and J.-M. Strelcyn, ``Lyapunov
  characteristic exponents for smooth dynamical systems and for hamiltonian
  systems; a method for computing all of them. part 2: Numerical application,''
  {\em Meccanica}, vol.~15, no.~1, pp.~21--30, 1980.

\bibitem{Young2013}
L.-S. Young, ``Mathematical theory of lyapunov exponents,'' {\em Journal of
  Physics A: Mathematical and Theoretical}, vol.~46, p.~254001, jun 2013.

\bibitem{eckmann_1985}
J.-P. Eckmann and D.~Ruelle, ``Ergodic theory of chaos and strange
  attractors,'' {\em The theory of chaotic attractors}, pp.~273--312, 1985.

\bibitem{Nese1989}
J.~M. Nese, ``Quantifying local predictability in phase space,'' {\em Physica
  D: Nonlinear Phenomena}, vol.~35, no.~1, pp.~237--250, 1989.

\bibitem{Abarbanel1991}
H.~D.~I. Abarbanel, R.~Brown, and M.~B. Kennel, ``Variation of lyapunov
  exponents on a strange attractor,'' {\em Journal of Nonlinear Science},
  vol.~1, no.~2, pp.~175--199, 1991.

\bibitem{Gallez1991}
D.~Gallez and A.~Babloyantz, ``Lyapunov exponents for nonuniform attractors,''
  {\em Physics Letters A}, vol.~161, no.~3, pp.~247--254, 1991.

\bibitem{Aurell1997}
E.~Aurell, G.~Boffetta, A.~Crisanti, G.~Paladin, and A.~Vulpiani,
  ``Predictability in the large: an extension of the concept of lyapunov
  exponent,'' {\em Journal of Physics A: Mathematical and General}, vol.~30,
  pp.~1--26, jan 1997.

\bibitem{Cencini2013}
M.~Cencini and F.~Ginelli, ``Lyapunov analysis: from dynamical systems theory
  to applications,'' {\em Journal of Physics A: Mathematical and Theoretical},
  vol.~46, p.~250301, jun 2013.

\bibitem{pikovsky2016lyapunov}
A.~Pikovsky and A.~Politi, {\em Lyapunov exponents: a tool to explore complex
  dynamics}.
\newblock Cambridge University Press, 2016.

\bibitem{barreira2007}
L.~Barreira and Y.~Pesin, {\em Nonuniform Hyperbolicity: Dynamics of Systems
  with Nonzero Lyapunov Exponents}.
\newblock Encyclopedia of Mathematics and its Applications, Cambridge
  University Press, 2007.

\bibitem{bonatti_2004}
C.~Bonatti, L.~J. D{\'\i}az, and M.~Viana, {\em Dynamics beyond uniform
  hyperbolicity: A global geometric and probabilistic perspective}, vol.~3.
\newblock Springer Science \& Business Media, 2004.

\bibitem{zhang2012find}
W.~Zhang, B.~Krauskopf, and V.~Kirk, ``How to find a codimension-one
  heteroclinic cycle between two periodic orbits,'' {\em Discrete \& Continuous
  Dynamical Systems}, vol.~32, no.~8, p.~2825, 2012.

\bibitem{cvitanovic_1988}
P.~Cvitanovi{\'c}, ``Invariant measurement of strange sets in terms of
  cycles,'' {\em Physical Review Letters}, vol.~61, no.~24, p.~2729, 1988.

\bibitem{cvitanovic_1991}
P.~Cvitanovi{\'c}, ``Periodic orbits as the skeleton of classical and quantum
  chaos,'' {\em Physica D: Nonlinear Phenomena}, vol.~51, no.~1-3,
  pp.~138--151, 1991.

\bibitem{cvitanovic_2005}
P.~Cvitanovic, R.~Artuso, R.~Mainieri, G.~Tanner, G.~Vattay, N.~Whelan, and
  A.~Wirzba, ``Chaos: classical and quantum,'' {\em ChaosBook. org (Niels Bohr
  Institute, Copenhagen 2005)}, vol.~69, p.~25, 2005.

\bibitem{Gaspard2005}
P.~Gaspard, ``Chaos, scattering and statistical mechanics,'' {\em Chaos}, 2005.

\bibitem{kazantsev1998unstable}
E.~Kazantsev, ``Unstable periodic orbits and attractor of the barotropic ocean
  model,'' {\em Nonlinear processes in Geophysics}, vol.~5, no.~4,
  pp.~193--208, 1998.

\bibitem{chandler_2013}
G.~J. Chandler and R.~R. Kerswell, ``Invariant recurrent solutions embedded in
  a turbulent two-dimensional kolmogorov flow,'' {\em Journal of Fluid
  Mechanics}, vol.~722, pp.~554--595, 2013.

\bibitem{lucas2015recurrent}
D.~Lucas and R.~R. Kerswell, ``Recurrent flow analysis in spatiotemporally
  chaotic 2-dimensional kolmogorov flow,'' {\em Physics of Fluids}, vol.~27,
  no.~4, p.~045106, 2015.

\bibitem{van2006periodic}
L.~Van~Veen, S.~Kida, and G.~Kawahara, ``Periodic motion representing isotropic
  turbulence,'' {\em Fluid dynamics research}, vol.~38, no.~1, p.~19, 2006.

\bibitem{krygier2021exact}
M.~C. Krygier, J.~L. Pughe-Sanford, and R.~O. Grigoriev, ``Exact coherent
  structures and shadowing in turbulent taylor-couette flow,'' {\em arXiv
  preprint arXiv:2105.02126}, 2021.

\bibitem{cvitanovic2010geometry}
P.~Cvitanovi{\'c} and J.~Gibson, ``Geometry of the turbulence in wall-bounded
  shear flows: periodic orbits,'' {\em Physica Scripta}, vol.~2010, no.~T142,
  p.~014007, 2010.

\bibitem{kreilos2012periodic}
T.~Kreilos and B.~Eckhardt, ``Periodic orbits near onset of chaos in plane
  couette flow,'' {\em Chaos: An Interdisciplinary Journal of Nonlinear
  Science}, vol.~22, no.~4, p.~047505, 2012.

\bibitem{suri2020capturing}
B.~Suri, L.~Kageorge, R.~O. Grigoriev, and M.~F. Schatz, ``Capturing turbulent
  dynamics and statistics in experiments with unstable periodic orbits,'' {\em
  Physical Review Letters}, vol.~125, no.~6, p.~064501, 2020.

\bibitem{yalniz_2020}
G.~Yalniz and N.~B. Budanur, ``Inferring symbolic dynamics of chaotic flows
  from persistence,'' {\em Chaos: An Interdisciplinary Journal of Nonlinear
  Science}, vol.~30, no.~3, p.~033109, 2020.

\bibitem{Dhamala1999}
M.~Dhamala and Y.-C. Lai, ``Unstable periodic orbits and the natural measure of
  nonhyperbolic chaotic saddles,'' {\em Phys. Rev. E}, vol.~60, pp.~6176--6179,
  Nov 1999.

\bibitem{kawahara_2001}
G.~Kawahara and S.~Kida, ``Periodic motion embedded in plane couette
  turbulence: regeneration cycle and burst,'' {\em Journal of Fluid Mechanics},
  vol.~449, p.~291, 2001.

\bibitem{page2022recurrent}
J.~Page, P.~Norgaard, M.~P. Brenner, and R.~R. Kerswell, ``Recurrent flow
  patterns as a basis for turbulence: predicting statistics from structures,''
  {\em arXiv preprint arXiv:2212.11886}, 2022.

\bibitem{Lai1997}
Y.-C. Lai, Y.~Nagai, and C.~Grebogi, ``Characterization of the natural measure
  by unstable periodic orbits in chaotic attractors,'' {\em Phys. Rev. Lett.},
  vol.~79, pp.~649--652, Jul 1997.

\bibitem{Sauer1997}
T.~Sauer, C.~Grebogi, and J.~A. Yorke, ``How long do numerical chaotic
  solutions remain valid?,'' {\em Phys. Rev. Lett.}, vol.~79, pp.~59--62, Jul
  1997.

\bibitem{Sauer2002}
T.~D. Sauer, ``Shadowing breakdown and large errors in dynamical simulations of
  physical systems,'' {\em Phys. Rev. E}, vol.~65, p.~036220, Feb 2002.

\bibitem{Pereira2007}
R.~F. Pereira, S.~E. de~S.~Pinto, R.~L. Viana, S.~R. Lopes, and C.~Grebogi,
  ``Periodic orbit analysis at the onset of the unstable dimension variability
  and at the blowout bifurcation,'' {\em Chaos: An Interdisciplinary Journal of
  Nonlinear Science}, vol.~17, no.~2, p.~023131, 2007.

\bibitem{Kalnay2003}
E.~Kalnay, {\em {Atmospheric Modeling, Data Assimilation and Predictability}}.
\newblock Cambridge: Cambridge University Press, 2003.

\bibitem{Ghil2020}
M.~Ghil and V.~Lucarini, ``The physics of climate variability and climate
  change,'' {\em Rev. Mod. Phys.}, vol.~92, p.~035002, Jul 2020.

\bibitem{Palmer2000}
T.~N. Palmer, ``Predicting uncertainty in forecasts of weather and climate,''
  {\em Reports on Progress in Physics}, vol.~63, pp.~71--116, jan 2000.

\bibitem{Slingo2011}
J.~Slingo and T.~Palmer, ``Uncertainty in weather and climate prediction,''
  {\em Philosophical Transactions of the Royal Society A: Mathematical,
  Physical and Engineering Sciences}, vol.~369, no.~1956, pp.~4751--4767, 2011.

\bibitem{Carrassi2018}
A.~Carrassi, M.~Bocquet, L.~Bertino, and G.~Evensen, ``Data assimilation in the
  geosciences: An overview of methods, issues, and perspectives,'' {\em WIREs
  Climate Change}, vol.~9, no.~5, p.~e535, 2018.

\bibitem{Wu2020}
Y.~Wu, Z.~Shen, and Y.~Tang, ``A flow-dependent targeted observation method for
  ensemble kalman filter assimilation systems,'' {\em Earth and Space Science},
  vol.~7, no.~7, p.~e2020EA001149, 2020.
\newblock e2020EA001149 2020EA001149.

\bibitem{Chen2021}
Y.~Chen, A.~Carrassi, and V.~Lucarini, ``Inferring the instability of a
  dynamical system from the skill of data assimilation exercises,'' {\em
  Nonlinear Processes in Geophysics}, vol.~28, no.~4, pp.~633--649, 2021.

\bibitem{Nese1993}
J.~M. Nese and J.~A. Dutton, ``Quantifying predictability variations in a
  low-order ocean-atmosphere model: A dynamical systems approach,'' {\em
  Journal of Climate}, vol.~6, no.~2, pp.~185 -- 204, 1993.

\bibitem{Yoden1995}
S.~Yoden and M.~Nomura, ``Finite-time lyapunov stability analysis and its
  application to atmospheric predictability,'' {\em Journal of Atmospheric
  Sciences}, vol.~50, no.~11, pp.~1531 -- 1543, 1993.

\bibitem{Nicolis1995}
C.~Nicolis, S.~Vannitsem, and J.-F. Royer, ``Short-range predictability of the
  atmosphere: Mechanisms for superexponential error growth,'' {\em Quarterly
  Journal of the Royal Meteorological Society}, vol.~121, no.~523,
  pp.~705--722, 1995.

\bibitem{Vannitsem1997}
S.~Vannitsem and C.~Nicolis, ``Lyapunov vectors and error growth patterns in a
  t21l3 quasigeostrophic model,'' {\em Journal of the Atmospheric Sciences},
  vol.~54, no.~2, pp.~347 -- 361, 1997.

\bibitem{de2018exploring}
L.~De~Cruz, S.~Schubert, J.~Demaeyer, V.~Lucarini, and S.~Vannitsem,
  ``Exploring the lyapunov instability properties of high-dimensional
  atmospheric and climate models,'' {\em Nonlinear Processes in Geophysics},
  vol.~25, no.~2, pp.~387--412, 2018.

\bibitem{Pazo2013}
D.~Paz\'o, J.~M. L\'opez, and A.~Politi, ``Universal scaling of
  lyapunov-exponent fluctuations in space-time chaos,'' {\em Phys. Rev. E},
  vol.~87, p.~062909, Jun 2013.

\bibitem{Laffargue2013}
T.~Laffargue, K.-D. N.~T. Lam, J.~Kurchan, and J.~Tailleur, ``Large deviations
  of lyapunov exponents,'' {\em Journal of Physics A: Mathematical and
  Theoretical}, vol.~46, p.~254002, jun 2013.

\bibitem{lucarini_2020}
V.~Lucarini and A.~Gritsun, ``A new mathematical framework for atmospheric
  blocking events,'' {\em Climate Dynamics}, vol.~54, no.~1, pp.~575--598,
  2020.

\bibitem{gritsun_2008}
A.~Gritsun, ``Unstable periodic trajectories of a barotropic model of the
  atmosphere.,'' {\em Russian Journal of Numerical Analysis and Mathematical
  Modelling}, vol.~23, no.~4, 2008.

\bibitem{gritsun_2013}
A.~Gritsun, ``Statistical characteristics, circulation regimes and unstable
  periodic orbits of a barotropic atmospheric model,'' {\em Philosophical
  Transactions of the Royal Society A: Mathematical, Physical and Engineering
  Sciences}, vol.~371, no.~1991, p.~20120336, 2013.

\bibitem{Schubert2016}
S.~Schubert and V.~Lucarini, ``Dynamical analysis of blocking events: spatial
  and temporal fluctuations of covariant lyapunov vectors,'' {\em Quarterly
  Journal of the Royal Meteorological Society}, vol.~142, no.~698,
  pp.~2143--2158, 2016.

\bibitem{Kwasniok2022}
F.~Kwasniok, ``Data-driven modeling using non-markovian optimal mode
  decomposition.'' 2022.

\bibitem{Faranda2017}
D.~Faranda, G.~Messori, and P.~Yiou, ``Dynamical proxies of north atlantic
  predictability and extremes,'' {\em Scientific Reports}, vol.~7, no.~1,
  p.~41278, 2017.

\bibitem{maiocchi_2022}
C.~C. Maiocchi, V.~Lucarini, and A.~Gritsun, ``Decomposing the dynamics of the
  lorenz 1963 model using unstable periodic orbits: Averages, transitions, and
  quasi-invariant sets,'' {\em Chaos: An Interdisciplinary Journal of Nonlinear
  Science}, vol.~32, no.~3, p.~033129, 2022.

\bibitem{barrio_2015}
R.~Barrio, A.~Dena, and W.~Tucker, ``A database of rigorous and high-precision
  periodic orbits of the lorenz model,'' {\em Computer Physics Communications},
  vol.~194, pp.~76--83, 2015.

\bibitem{froyland_2014}
G.~Froyland and K.~Padberg-Gehle, ``Almost-invariant and finite-time coherent
  sets: directionality, duration, and diffusion,'' in {\em Ergodic Theory, Open
  Dynamics, and Coherent Structures}, pp.~171--216, Springer, 2014.

\bibitem{Smith1999}
L.~A. Smith, C.~Ziehmann, and K.~Fraedrich, ``Uncertainty dynamics and
  predictability in chaotic systems,'' {\em Quarterly Journal of the Royal
  Meteorological Society}, vol.~125, no.~560, pp.~2855--2886, 1999.

\bibitem{lorenz_1996}
E.~N. Lorenz, ``Predictability: A problem partly solved,'' in {\em Proc.
  Seminar on predictability}, vol.~1, 1996.

\bibitem{Lorenz2005}
E.~N. Lorenz, ``{Designing Chaotic Models},'' {\em Journal of the Atmospheric
  Sciences}, vol.~62, pp.~1574--1587, 05 2005.

\bibitem{miller2000finding}
J.~Miller and J.~Yorke, ``Finding all periodic orbits of maps using newton
  methods: sizes of basins,'' {\em Physica D: Nonlinear Phenomena}, vol.~135,
  no.~3-4, pp.~195--211, 2000.

\bibitem{vanKekem2018PhysD}
D.~L. {van Kekem} and A.~E. Sterk, ``Travelling waves and their bifurcations in
  the lorenz-96 model,'' {\em Physica D: Nonlinear Phenomena}, vol.~367, pp.~38
  -- 60, 2018.

\bibitem{vanKekem2018NPG}
D.~L. van Kekem and A.~E. Sterk, ``Wave propagation in the lorenz-96 model,''
  {\em Nonlinear Processes in Geophysics}, vol.~25, no.~2, pp.~301--314, 2018.

\bibitem{Wilks2005}
D.~Wilks, ``{Effects of stochastic parametrizations in the Lorenz '96
  system},'' {\em Quarterly Journal of the Royal Meteorological Society},
  vol.~131, no.~606, pp.~389--407, 2005.

\bibitem{Arnold2013}
H.~M. Arnold, I.~M. Moroz, and T.~N. Palmer, ``Stochastic parametrizations and
  model uncertainty in the lorenz system,'' {\em Philosophical Transactions of
  the Royal Society A: Mathematical, Physical and Engineering Sciences},
  vol.~371, no.~1991, p.~20110479, 2013.

\bibitem{Vissio2018}
G.~Vissio and V.~Lucarini, ``{A proof of concept for scale-adaptive
  parametrizations: the case of the Lorenz '96 model},'' {\em Quarterly Journal
  of the Royal Meteorological Society}, vol.~144, pp.~63--75, 2018.

\bibitem{Chattopadhyay2020}
A.~Chattopadhyay, P.~Hassanzadeh, and D.~Subramanian, ``Data-driven predictions
  of a multiscale lorenz 96 chaotic system using machine-learning methods:
  reservoir computing, artificial neural network, and long short-term memory
  network,'' {\em Nonlinear Processes in Geophysics}, vol.~27, no.~3,
  pp.~373--389, 2020.

\bibitem{Gagne2020}
D.~J. Gagne~II, H.~M. Christensen, A.~C. Subramanian, and A.~H. Monahan,
  ``Machine learning for stochastic parameterization: Generative adversarial
  networks in the lorenz '96 model,'' {\em Journal of Advances in Modeling
  Earth Systems}, vol.~12, no.~3, p.~e2019MS001896, 2020.
\newblock e2019MS001896 10.1029/2019MS001896.

\bibitem{Gelbrecht2021}
M.~Gelbrecht, V.~Lucarini, N.~Boers, and J.~Kurths, ``Analysis of a bistable
  climate toy model with physics-based machine learning methods,'' {\em The
  European Physical Journal Special Topics}, vol.~230, no.~14, pp.~3121--3131,
  2021.

\bibitem{Blender2013}
R.~Blender and V.~Lucarini, ``Nambu representation of an extended lorenz model
  with viscous heating,'' {\em Physica D: Nonlinear Phenomena}, vol.~243,
  no.~1, pp.~86 -- 91, 2013.

\bibitem{Sterk2017}
A.~E. Sterk and D.~L. van Kekem, ``Predictability of extreme waves in the
  lorenz-96 model near intermittency and quasi-periodicity,'' {\em Complexity},
  vol.~2017, p.~9419024, 2017.

\bibitem{Hu2019}
G.~Hu, T.~B\'odai, and V.~Lucarini, ``Effects of stochastic parametrization on
  extreme value statistics,'' {\em Chaos: An Interdisciplinary Journal of
  Nonlinear Science}, vol.~29, no.~8, p.~083102, 2019.

\bibitem{Trevisan2004}
A.~Trevisan and F.~Uboldi, ``Assimilation of standard and targeted observations
  within the unstable subspace of the observation--analysis--forecast cycle
  system,'' {\em Journal of the atmospheric sciences}, vol.~61, no.~1,
  pp.~103--113, 2004.

\bibitem{Brajard2020}
J.~Brajard, A.~Carrassi, M.~Bocquet, and L.~Bertino, ``Combining data
  assimilation and machine learning to emulate a dynamical model from sparse
  and noisy observations: A case study with the lorenz 96 model,'' {\em Journal
  of Computational Science}, vol.~44, p.~101171, 2020.

\bibitem{Wilks2006}
D.~S. Wilks, ``Comparison of ensemble-mos methods in the lorenz '96 setting,''
  {\em Meteorological Applications}, vol.~13, no.~3, p.~243–256, 2006.

\bibitem{Duan2016}
W.~Duan and Z.~Huo, ``{An Approach to Generating Mutually Independent Initial
  Perturbations for Ensemble Forecasts: Orthogonal Conditional Nonlinear
  Optimal Perturbations},'' {\em Journal of the Atmospheric Sciences}, vol.~73,
  pp.~997--1014, 02 2016.

\bibitem{Hallerberg2010}
S.~Hallerberg, D.~Pazo, J.~M. Lopez, and M.~A. Rodriguez, ``Logarithmic bred
  vectors in spatiotemporal chaos: Structure and growth,'' {\em Physical Review
  E}, vol.~81, no.~6, p.~066204, 2010.

\bibitem{Carlu2019}
M.~Carlu, F.~Ginelli, V.~Lucarini, and A.~Politi, ``Lyapunov analysis of
  multiscale dynamics: the slow bundle of the two-scale lorenz 96 model,'' {\em
  Nonlinear Processes in Geophysics}, vol.~26, no.~2, pp.~73--89, 2019.

\bibitem{AbramovM2008}
R.~V. Abramov and A.~J. Majda, ``New approximations and tests of linear
  fluctuation-response for chaotic nonlinear forced-dissipative dynamical
  systems,'' {\em Journal of Nonlinear Science}, vol.~18, no.~3, pp.~303--341,
  2008.

\bibitem{Lucarini2011}
V.~Lucarini and S.~Sarno, ``A statistical mechanical approach for the
  computation of the climatic response to general forcings,'' {\em Nonlinear
  Processes in Geophysics}, vol.~18, no.~1, pp.~7--28, 2011.

\bibitem{Lucarini2012}
V.~Lucarini, ``Stochastic perturbations to dynamical systems: A response theory
  approach,'' {\em Journal of Statistical Physics}, vol.~146, no.~4,
  pp.~774--786, 2012.

\bibitem{Gallavotti2014}
G.~Gallavotti and V.~Lucarini, ``Equivalence of non-equilibrium ensembles and
  representation of friction in turbulent flows: the lorenz 96 model,'' {\em
  Journal of Statistical Physics}, vol.~156, no.~6, pp.~1027--1065, 2014.

\bibitem{Vissio2020}
G.~Vissio and V.~Lucarini, ``Mechanics and thermodynamics of a new minimal
  model of the atmosphere,'' {\em The European Physical Journal Plus},
  vol.~135, no.~10, p.~807, 2020.

\bibitem{Grassberger1983}
P.~Grassberger and I.~Procaccia, ``Measuring the strangeness of strange
  attractors,'' {\em Physica D: Nonlinear Phenomena}, vol.~9, no.~1,
  pp.~189--208, 1983.

\bibitem{saiki_2007}
Y.~Saiki, ``Numerical detection of unstable periodic orbits in continuous-time
  dynamical systems with chaotic behaviors,'' {\em Nonlinear Processes in
  Geophysics}, vol.~14, no.~5, pp.~615--620, 2007.

\bibitem{barrio2015database}
R.~Barrio, A.~Dena, and W.~Tucker, ``A database of rigorous and high-precision
  periodic orbits of the lorenz model,'' {\em Computer Physics Communications},
  vol.~194, pp.~76--83, 2015.

\bibitem{krygier_2021}
M.~C. Krygier, J.~L. Pughe-Sanford, and R.~O. Grigoriev, ``Exact coherent
  structures and shadowing in turbulent taylor–couette flow,'' {\em Journal
  of Fluid Mechanics}, vol.~923, p.~A7, 2021.

\bibitem{dawson1994obstructions}
S.~Dawson, C.~Grebogi, T.~Sauer, and J.~A. Yorke, ``Obstructions to shadowing
  when a lyapunov exponent fluctuates about zero,'' {\em Physical review
  letters}, vol.~73, no.~14, p.~1927, 1994.

\bibitem{bowen_1975}
R.~Bowen, ``$\omega$-limit sets for axiom a diffeomorphisms,'' {\em Journal of
  differential equations}, vol.~18, no.~2, pp.~333--339, 1975.

\bibitem{sauer1997long}
T.~Sauer, C.~Grebogi, and J.~A. Yorke, ``How long do numerical chaotic
  solutions remain valid?,'' {\em Physical Review Letters}, vol.~79, no.~1,
  p.~59, 1997.

\bibitem{livi_2017}
R.~Livi and P.~Politi, {\em Nonequilibrium statistical physics: a modern
  perspective}.
\newblock Cambridge University Press, 2017.

\bibitem{froyland_2001}
G.~Froyland, ``Extracting dynamical behavior via markov models,'' in {\em
  Nonlinear dynamics and statistics}, pp.~281--321, Springer, 2001.

\bibitem{froyland_2003}
G.~Froyland and M.~Dellnitz, ``Detecting and locating near-optimal
  almost-invariant sets and cycles,'' {\em SIAM Journal on Scientific
  Computing}, vol.~24, no.~6, pp.~1839--1863, 2003.

\bibitem{gaspard2004}
P.~Gaspard, ``Time-reversed dynamical entropy and irreversibility in markovian
  random processes,'' {\em Journal of Statistical Physics}, vol.~117, no.~3,
  pp.~599--615, 2004.

\bibitem{cvitanovic_2013}
P.~Cvitanovi{\'c}, ``Recurrent flows: the clockwork behind turbulence,'' {\em
  Journal of Fluid Mechanics}, vol.~726, pp.~1--4, 2013.

\bibitem{lasagna_2020}
D.~Lasagna, ``Sensitivity of long periodic orbits of chaotic systems,'' {\em
  Physical Review E}, vol.~102, no.~5, p.~052220, 2020.

\bibitem{ruelle_2009}
D.~Ruelle, ``A review of linear response theory for general differentiable
  dynamical systems,'' {\em Nonlinearity}, vol.~22, no.~4, p.~855, 2009.

\bibitem{Baladi08}
V.~Baladi, ``Linear response despite critical points,'' {\em Nonlinearity},
  vol.~21, no.~6, p.~T81, 2008.

\bibitem{Baladi14}
V.~Baladi, ``Linear response, or else,'' in {\em ICM Seoul 2014, Proceedings},
  vol.~III, pp.~525--545, 2014.

\bibitem{Ni2020}
A.~Ni, ``Approximating linear response by nonintrusive shadowing algorithms,''
  {\em SIAM Journal on Numerical Analysis}, vol.~59, no.~6, pp.~2843--2865,
  2021.

\bibitem{Chandramoorthy2020}
N.~{Chandramoorthy} and Q.~{Wang}, ``{A computable realization of Ruelle's
  formula for linear response of statistics in chaotic systems},'' {\em arXiv
  e-prints}, p.~arXiv:2002.04117, Feb. 2020.

\bibitem{SliwiakWang2022}
A.~A. \'{S}liwiak and Q.~Wang, ``A trajectory-driven algorithm for
  differentiating {SRB} measures on unstable manifolds,'' {\em SIAM J. Sci.
  Comput.}, vol.~44, no.~1, pp.~A312--A336, 2022.

\bibitem{GhilLucarini2020}
M.~Ghil and V.~Lucarini, ``The physics of climate variability and climate
  change,'' {\em Rev. Mod. Phys.}, vol.~92, p.~035002, Jul 2020.

\bibitem{vannitsem2016statistical}
S.~Vannitsem and V.~Lucarini, ``Statistical and dynamical properties of
  covariant lyapunov vectors in a coupled atmosphere-ocean model—multiscale
  effects, geometric degeneracy, and error dynamics,'' {\em Journal of Physics
  A: Mathematical and Theoretical}, vol.~49, no.~22, p.~224001, 2016.

\bibitem{Lucarini2023}
V.~Lucarini, C.~C. Maiocchi, A.~Gritsun, and Y.~Sato, ``Supplementary material
  for the article {''Heterogeneity of the Attractor of the Lorenz '96 Model:
  Bridging the Gap between Lyapunov Analysis and Unstable Periodic Orbits''},''
  10.6084/m9.figshare.24242458.v1 2023.

\bibitem{ott_2002}
E.~Ott, {\em Chaos in dynamical systems}.
\newblock Cambridge university press, 2002.

\bibitem{FREDERICKSON1983185}
P.~Frederickson, J.~L. Kaplan, E.~D. Yorke, and J.~A. Yorke, ``The liapunov
  dimension of strange attractors,'' {\em Journal of Differential Equations},
  vol.~49, no.~2, pp.~185--207, 1983.

\bibitem{grebogi_1988}
C.~Grebogi, E.~Ott, and J.~A. Yorke, ``Unstable periodic orbits and the
  dimensions of multifractal chaotic attractors,'' {\em Physical Review A},
  vol.~37, no.~5, p.~1711, 1988.

\end{thebibliography}
\end{document}